%% file: main.tex
\documentclass[12pt]{article}
\usepackage{fontspec}
\usepackage{setspace}
\usepackage{graphicx,subcaption}
\usepackage{geometry}
\usepackage{float}
\usepackage{placeins}
\usepackage{wrapfig}

\usepackage{multirow}

\usepackage{amsmath} 
\usepackage{mathtools} 
\usepackage{unicode-math} 

\usepackage{hhline}
\usepackage{caption}
\usepackage{indentfirst}
\usepackage[english]{babel}
\usepackage[flushleft]{threeparttable}
\usepackage[titletoc,toc,title]{appendix}

\usepackage[colorlinks=true, allcolors=blue]{hyperref}
\usepackage[authoryear]{natbib}
\usepackage{csquotes}
\usepackage{rotating} 
\usepackage{lscape} 

\usepackage{supertabular}
\usepackage{adjustbox}
\usepackage{array}
\usepackage{tabularx}

\usepackage{float} 
\usepackage{booktabs}
\usepackage{pdflscape}  
\usepackage{multicol}
\usepackage{ragged2e}
\newcommand{\sym}[1]{\rlap{#1}} 

\usepackage[linesnumbered,ruled,vlined,longend]{algorithm2e}
\makeatletter
\newcommand{\RemoveAlgoNumber}{\renewcommand{\fnum@algocf}{\AlCapSty{\AlCapFnt\algorithmcfname}}}
\newcommand{\RevertAlgoNumber}{\algocf@resetfnum}
\makeatother

\SetKwProg{Fn}{Function}{}{}

\begin{document}

\title{Estimating Long Run Welfare Outcome in Rotating Panel with Grouped Fixed Effects: Application to Poverty Dynamics in Peru\thanks{We thank Development seminar participants at Cornell University and the University of Notre Dame for helpful comments. This material is based upon work supported in part by the National Science Foundation under Grant Number 2242507. Any opinions, findings, and conclusions or recommendations expressed in this material are those of the author(s) and do not necessarily reflect the views of the National Science Foundation. Comments and suggestions are welcome and may be emailed to \href{mailto:hz399@cornell.edu}{hz399@cornell.edu} and \href{mailto:slee76@nd.edu}{slee76@nd.edu}} }
\author{Hongdi Zhao\thanks{Hongdi Zhao is Ph.D. candidate in Applied Economics and Management at Cornell University.}, \quad Seungmin Lee\thanks{Seungmin Lee is a postdoctoral research associate at Pulte Instutite for Global Development in the University of Notre Dame.}}
\date{April 2026}

\maketitle
\maketitle
\thispagestyle{empty} 
\begin{abstract}
\noindent

Household poverty dynamics are often difficult to investigate due to lack of long-term panel data. Existing methods, such as pseudo-panel and synthetic panel, offer widely used solutions based on repeated cross-section designs, but they do not exploit within-household variation in rotating panel designs, which provide very useful information for estimating long-run dynamics. This paper applies grouped fixed effects (GFE) to estimate poverty mobility and persistence in a rotating panel setting, using Peru’s National Household Survey on Living Conditions and Poverty (ENAHO). Using observed transitions, we show that GFE-implied poverty transitions closely track the data. In a one-step-ahead validation that holds out each household’s final observed year, predicted transition shares remain close to realized transition shares, indicating that the method captures short-run entry and exit dynamics out of sample. When benchmarked against synthetic panel point estimates, the GFE approach delivers transition measures that are closer to observed transitions on average, while also providing an interpretable grouping structure that supports richer descriptions of poverty persistence and mobility.

	
	
\end{abstract}

\clearpage
\pagenumbering{arabic}


\newpage
\section{Introduction}
\label{sec:introduction}

Individual and household well-being change over time, implying that poverty evolves as well. While some poor households are only transiently poor, quickly recovering back to non-poor status, many other poor households are chronically poor, trapped in poverty over the course of their lives. Understanding poverty dynamics - change in status and duration of status - is crucial in policy design for multiple reasons \citep{baulch_economic_2000}. First, it could identify a vulnerable population that is consistently poor over time, providing an opportunity to understand the reasons they are chronically poor. For instance, if we find that households in one region are more chronically poor than those in other regions, it could be due to the lack of public goods and infrastructure in that region. Second, different policies are needed to address the distinct types of poverty, since the causes of chronic and transient poverty differ. Social assistance programs, unemployment insurance, and cash transfers are effective against short-term poverty, which is often caused by temporary hardship, while productive asset accumulation, high-quality service provision, and formalization are needed to fight chronic poverty, which stems from the lack of productive assets and infrastructure. Third, the duration of poverty itself could exacerbate a household's poverty status. The lack of productive assets can lead to chronic poverty, but chronically poor households may reduce their assets or investment in capital accumulation as a coping strategy, such as selling their livestock or withdrawing their children from school, trapping themselves in poverty for longer periods \citep{attanasio_going_2001,carter_economics_2006}. However, studying household-level poverty dynamics has often been hindered by the lack of reliable panel data that observe the same households over time. Household-level panel data are often unavailable or suffers from a small sample size due to its high implementation and maintenance costs.

Several econometric and statistical approaches have been developed to estimate dynamics under such data limitations. Pseudo-panel \citep{deaton_panel_1985} is an early attempt to overcome data limitations in which micro-level units (individuals and households) are grouped into cohorts sharing common characteristics (e.g., birth cohort, village, or district), and dynamics are tracked at the cohort level from repeated cross-sectional data. The synthetic panel is a more recent approach that allows for the estimation of micro-level welfare outcome dynamics using only two rounds of cross-sectional data \citep{dang_using_2014}. Based on a few assumptions on the population and its parameters, the synthetic panel estimates the boundaries (upper- and lower-bound) of poverty mobility (e.g., entry into and exit from poverty) where true mobility lies in between. The literature has shown that bounded estimates almost always capture true poverty mobility across both developed and developing countries \citep{dang_using_2014,cruces_estimating_2015,herault_how_2019}.

The synthetic panel method has motivated follow-up studies. \citet{dang2023measuring} provides point estimates via imposing additional parametric assumptions. However, unlike bounded estimates, point estimates of poverty mobility vary across countries and settings, often deviating from the true point estimates. Since point estimates are imputed from cohort-level information, the validity of the final point estimates is sensitive to how cohorts are defined, for which there is no solid guidance. \citet{lucchetti_application_2025} developed the Least Absolute Shrinkage and Selection Operator - Predictive Mean Matching (LASSO-PMM) method, a semi-parametric method that does not require the parametric assumptions from \citet{dang2023measuring}. \citet{dalberto_estimating_2025} introduces a scenario-based approach that accounts for temporal data dependencies using a Bayesian framework and a matching method.

While researchers have developed these methods for short-term dynamics using repeated cross-sectional data, few studies have investigated long-term welfare estimation, which is crucial for identifying chronic deprivation. Although the methods used in prior papers could be theoretically feasible for welfare transition in the long run, the underlying assumptions of those approaches, such as the cohort-level autocorrelation in welfare outcomes, are less likely to hold over the longer term, thereby leading to biased and/or imprecise estimates of welfare dynamics. 

Rotating panel data is more readily available than long-term panel data. A rotating panel is a hybrid design that balances sample size and panel structure at a lower cost than a true panel. Under the rotating panel design, a subset of units (households or individuals) is periodically rotated out, and new units are rotated in, balancing the need for longitudinal insights with cost management. Due to its hybrid nature, the rotating panel design has been widely adopted internationally, including Brazil, Mexico, Peru and Vietnam.\footnote{Examples of rotating panel surveys are shown in Appendix Table \ref{Tab:RP_list}.} Since partial units overlap over a short period, typically from two to six years, they provide within-unit information that can be very useful in estimating unit-level welfare dynamics. Since the pseudo-/synthetic panel and its extensions are designed for pure repeated cross-sectional data, they treat each survey round as an independent draw from a stable population and recover mobility only through cohort-level information (i.e., the synthetic panel estimates dynamics through moments and assumptions about the joint distribution of unobservables). Therefore, if we apply either method to rotating panels, we necessarily discard within-household information that is observed, and we fail to utilize that overlap to discipline the correlation in unobserved shocks that drives mobility and persistence in the poverty transition measure. Overall, existing panel implementations may yield wide bounds or unstable point estimates precisely where short rotating panels provide an opportunity to do better. 

Although rotating panels combine the broad coverage of repeated cross-sections with some within-household follow-up, they have limitations for studying long-term poverty dynamics. First, the sample size of overlapping units is often small due to administrative costs, resulting in large standard errors in estimation \citep{van_den_brakel_dealing_2015}. Second, estimates can be biased if the units newly surveyed in a given period have systematically different characteristics from those surveyed in the same period, a phenomenon known as ``rotating panel bias'' \citep{bailar_effects_1975}. Third, most rotating panel data track units for only a short period, mostly due to administrative costs, which could impose restrictions on estimating dynamics. For example, one way to estimate dynamics is using a unit-level fixed-effects (FE) model to capture micro-level unobservables, and decompose welfare transitions into observables and unobservables and determine which one mainly drives welfare transition. However, estimating unit-level FE can be imprecise with modest T, and such effects does not have any economically interpretable implication. Even if we precisely estimate time-invariant fixed effects, they may not remain time-invariant over a longer period \citep{hill_limitations_2020,millimet_mis_2025}. 

In this paper, we employ Grouped Fixed Effects (GFE) estimator \citep{bonhomme2015grouped} (hereafter, BM2015) to address these limitations. Instead of estimating micro-level unobserved heterogeneity, the GFE estimator groups individual units into a pre-determined number of groups and estimates time-varying unobserved group-level heterogeneity. Compared to other estimators modeling unobserved heterogeneity in a way beyond classic two-way fixed effects (TWFE), such as correlated random effects \citep{mundlak_pooling_1978} and interacted fixed effects (IFE) \citep{bai_panel_2009}, BM2015 has the following advantage in our study setting. First, BM2015 allows heterogeneity to be time-varying, which is essential for investigating long-term welfare, where micro-level heterogeneity is very likely to be time-varying. Second, BM2015 models heterogeneity at the group level which can be precisely estimated and provide meaningful implication compared to micro-level FE which is often imprecisely estimated and has little implication. Third, it preserves intra-unit variations in welfare as well as covariates, allowing us to estimate the association between welfare and covariates with limited within-varation (e.g., educational attainment).  Individual units are assigned to a group through clustering; units with similar trends in heterogeneity (i.e., the outcome net of covariates) are grouped together. Therefore, each group exhibits a distinct trend in heterogeneity over time. Group assignments and model parameter estimation are conducted through either an iterative algorithm or a variable neighborhood search (VNS) algorithm. That is, the GFE estimator uses within-unit information by assigning units to groups in a data-driven way, addressing issues of sensitivity to the cohort definition and the loss of within-unit information from pseudo- and synthetic-panel methods.\footnote{\citet{bonhomme2022discretizing} extended BM2015 by relaxing certain assumptions, such as the discreteness of heterogeneity. In this paper, we use the framework from BM2015 because it provides a simple, transparent method for recovering distinct welfare trajectories from short panel overlap, which is the central objective of this paper. Applying \citet{bonhomme2022discretizing} to study poverty dynamics with the ENAHO data could be a potential extension to this paper.} 

BM2015 has been widely adopted for studying causal impacts on prediction. BM2015 applied the GFE method to examine the effects of income on democracy and found that, on average, income has little effect on democracy, but some countries were rapidly democratized at different times. \citet{oberlander_globalisation_2017} examined the effects of globalization on diet quality by grouping countries with similar trends in unobserved heterogeneity path into six groups. \citet{janys_mental_2024} estimated the causal effects of abortion on mental health by capturing the unobserved decision-making process driving risky behavior. \citet{aparicio-perez_disentangling_2025} investigated the impact of natural resources on economic growth by capturing the unobserved institutional quality that worked either positively or negatively on different groups of countries. In terms of predictive performance, the GFE estimator outperformed the standard fixed-effects estimator for predicting a rare event, such as a banking crisis \citep{pigini_grouped_2025}. In fact, GFE is one of the many examples where data-driven clustering outperforms predetermined grouping in forecasting future outcomes \citep{oh_dynamic_2023}. Since pseudo-panel and synthetic panel group units are based on predetermined characteristics, we hypothesize that the GFE estimator can better estimate poverty status than either of these.

In this paper, we study long-run poverty dynamics in Peru from 2007 to 2019 using the GFE estimator. By employing the GFE framework on Peru’s National Household Survey on Living Conditions and Poverty (ENAHO), which uses a panel design with a rotating sample up to 5 years, we leverage a rotating panel data structure by combining the strengths of short-panel identification with a data-driven grouping approach. We classify households into latent groups with distinct time patterns, and then leverage these estimated group-time components to recover poverty transition measures over horizons that exceed the observed panel length. By identifying unobserved heterogeneity and shock persistence from the observed overlap in the rotating panel, the GFE approach delivers more stable estimates of poverty dynamics than conventional synthetic panel implementations in the same setting. Moreover, the GFE method can estimate long-term poverty transitions by modeling group-specific time paths and a consistent structure for unobserved heterogeneity, whereas synthetic panels are generally designed for one-period-ahead (next-year) transitions. 

We find that by applying GFE estimator, our point estimates of poverty transition closely align with observed true poverty transitions in the data. Empirically, we show that GFE applied to ENAHO’s rotating panel structure can recover poverty-transition patterns that closely match observed transitions in the data. In a one-step-ahead validation that holds out each household’s final observed year, the GFE-implied transition shares path realized poor/non-poor transitions closely across end years. Moreover, when benchmarked against synthetic-panel point estimates, the GFE one-step approach yields smaller average discrepancies from observed transition shares while also delivering an interpretable latent-type structure that supports trajectory analysis beyond one-period transitions.

We further show that our main findings are robust to alternative sample definitions: removing the head of household's age restriction of the sample yields very similar optimal number of group selection and group structure, while restricting to the smaller subset of households observed for at least five rounds of survey leaves the optimal number of latent groups unchanged but makes the differences in group trajectories less stable with different sets of model specifications, highlighting the importance of adequate group--year support in rotating panels. We also test the model's ability to predict the medium-run poverty using households stayed in the data for at least four years, with the last two years as the testing data, and the prior years as the training data. We show that the model is able to predict the two years testing data with over 80\% accuracy rate. 

This paper makes the following contributions. First, we provide a framework for studying poverty dynamics in rotating panel settings that bridges the gap between true panels and repeated cross-sections. Using observed household overlap to estimate latent groups and their time profiles, we apply the GFE estimator, leveraging information that conventional synthetic panel methods cannot provide. Second, we show how GFE can be adapted to recover poverty transition measures over horizons longer than the observed panel lengths, offering a practical way to study medium- and longer-run mobility when long panels are unavailable. Third, to the best of our knowledge, this paper is the first to adopt GFE estimator to study poverty dynamics, providing a data-driven alternative to cohort-based synthetic panels that reduces sensitivity to arbitrary cohort definitions and yields more stable estimates of poverty dynamics in settings where only short panels are observed. Using Peru's ENAHO from 2007 to 2019, a rotating panel dataset,  we classify households into groups by using the GFE estimator such that the group-level change in unobserved variation in the expenditure best reflects the household-level change in that variation within the group. 

The remainder of the paper is organized as follows. Section~\ref{sec:empiricalstrategy} lays out the empirical framework and discussed how GFE is identified and estimated in rotating panels. Section~\ref{sec:implementation} describes the ENAHO rotating panel structures and our implementation of using GFE to estimate poverty transition on ENAHO data measures over both short and longer horizons. Section~\ref{sec:results} presents the main results on poverty persistent and mobility. Section~\ref{sec:robustness_check} shows robustness checks to alternative sample restrictions and section~\ref{sec:conclusion} concludes the paper with discussion on it's policy relevance.

\section{Empirical Strategy for GFE}
\label{sec:empiricalstrategy}

\subsection{Rotating Panels}

Datasets such as ENAHO follow a rotating panel design: households are interviewed for a limited number of consecutive years and then rotated out, while new households enter the sample each year. As a results, the dataset is neither a long true panel nor a sequence of independent cross sectional data. We present more detailed information about the ENAHO data, and its summary statistics in Section \ref{sec:implementation_data}. 

The partial overlap in datasets such as ENAHO is central to our empirical strategy. Even though each household is followed only for a few years, the repeated observations for a subset of households provide us direct information on persistence in expenditures and on time-varying shocks that affect similar households. At the same time, the continuing arrival of new households preserves the representativeness of each annual cross-section and provides rich variation in covariates and outcomes over time. We exploit this rotating panel structure in the next subsection by modeling households as belonging to a finite number of latent groups that share common time patterns in unobserved heterogeneity, and we estimate these group-time profiles using all available household-year observations.

\subsection{Grouped Fixed Effect (GFE) Estimator}

The GFE estimator provides a flexible way to model unobserved heterogeneity when full panel structures are not available. The key idea is that households can be classified into a finite number of latent ``types,'' where each type follows its own time pattern in unobserved components of welfare. Rather than assuming a single household fixed effect or imposing a parametric distribution of unobservables, the GFE estimator approximates continuous heterogeneity with a discrete set of groups and estimate both group membership and group-specific time profile directly from the data. 

Formally, we can model Inverse Hyperbolic Sine (IHS) transformed total household expenditure per-capita as a function of observed characteristics and a group-time component that captures time-varying unobserved heterogeneity common to households of the same latent type.\footnote{We use IHS-transformed expenditure to handle extremely high expenditure values, smoothing distribution.} Households are assigned to one of the $G$ groups, and each group has its own sequence of year-specific intercepts. This structure captures persistent differences across households (through group membership) while allowing the unobserved component to evolve over time (through group-time effects), which is particularly useful in contexts where macro shocks and structural change can shift the welfare distribution. 

The GFE estimator is particularly useful in rotating panels, where households are observed only for a limited number of years, such that estimating a separate household fixed effect for each unit is noisy and uninformative, and it does not provide a basis for extrapolating beyond observed years. The GFE estimator instead pools information across households that exhibit similar time patterns, using the overlap in the rotating panel to learn both the latent group and the evolution of group-level unobserved components. In this way, rotating panel overlap contributes directly to identifying the latent types and their dynamics. In the remainder of this section, we follow the GFE algorithm in BM2015, adapt it to the ENAHO rotating panel structure data with detailed description of our estimation procedure. We then explain how the estimated group-time paths are used to construct poverty transition measures over longer horizons.

\subsubsection{Model setup}

We model household welfare using a GFE estimator that allows for time-varying unobserved heterogeneity shared by households of the same latent type. Specifically, we estimate an expenditure equation of the form: 
\begin{equation}
\label{eq:basicmodel}
    y_{it} = x'_{it}\theta + \alpha_{g_{i} t} + \mu_{p_i} + \varepsilon_{it}
\end{equation}
where $y_{it}$ is a household welfare measure, such as the (log or IHS-transformed) per capita expenditure of household $i$ at time $t$; $x_{it} \in R^{K}$ is a vector of observed time-varying and time-invariant characteristics; $\theta \in R^{K}$ is a common vector of coefficients across all households; $\alpha_{g_{i} t}$ is a group-time fixed effect, capturing time-varying unobserved heterogeneity for group $g_{i}$ that household $i$ belongs to; $\mu_{p_i}$ is the time-invariant location fixed effects for individual $i$ in location $p$, such as province fixed effect, that nets out persistent location difference; and $\varepsilon_{it}$ is an idiosyncratic error term.

We observe units $i =1, \cdots, N$ living in $p = 1, \cdots, L$ over periods $t = 1, \cdots, T$ with outcome $y_{it}$ and regressors $x_{it} \in R^{K}$. Given Eq.~(\ref{eq:basicmodel}), let $d_{it} \in \{0, 1\}$ indicates observation (ENAHO is a rotating panel at the household and year level, so $d_{it} = 0$ if a household is not surveyed in year $t$). It is assumed that $x_{it}$ and $\varepsilon_{it}$ are contemporaneously uncorrelated given $d_{it}=1$.

Each unit belongs to one latent group $g_{i} \in \{1, \cdots, G\}$. $\theta \in \Theta$ (subset of $R^K$ parameter space) are common slopes across units, and units in the same group share the same time profile $\alpha_{gt} \in \mathcal{A}$ (e.g., all $i$ such that $g_i=1$ share the profile $\alpha_{1t}$. $\mathcal{A}$ is subset of $R^{G \times T}$ parameter space), and units in the same location $p$ share the same location-specific heterogeneity $\mu_{p} \in \Lambda$ (e.g., all $i$ such that $p_i=1$ share the heterogeneity $\mu_{1}$. $\Lambda$ is subset of $R^{L}$ parameter space) . The covariates vector $x_{it}$ may include strictly exogenous regressors and lagged outcomes. $x_{it}$ and $\alpha_{g_i t}$ are allowed to be arbitrarily correlated. We denote $\alpha$ as the set of all $\alpha_{gt}$'s, $\gamma$ as the set of all $g_{i}$'s, and $\mu$ as the set of all $\mu_{p_{i}}$. $\gamma \in \Gamma_g$ denotes a particular grouping of the $N$ units, where $\Gamma_g$ is the set of all groupings of $\{1,\cdots,N\}$ into at most $G$ groups. 

All members of group $g$ share the same unobserved trajectory. Neither the group membership $g_{i}$ nor the $\alpha_{g_{i}t}$ terms are observed prior. They are estimated jointly from the data. 

In our setting, local conditions such as geography, local prices, access to markets, and infrastructure can shift expenditure levels persistently and may otherwise be picked up by the latent groups. To separate these time-invariant location differences from the group-specific time profile, we include a province fixed effect $\mu_{p_i}$ for each household's province of residence. This inclusion of time-invariant heterogeneity is the first extension of the GFE framework described in BM2015, in which we use location-specific heterogeneity rather than unit-specific heterogeneity. Conceptually, the model decomposes expenditures into (i) observed covariates, (ii) a time-invariant location component, and (iii) a group-by-time component that captures shared time-varying unobserved heterogeneity among households of the same latent type. We discuss the option of adding province fixed effects instead of department (region) or districts fixed effects in more details in Section \ref{sec:model_specification}.

\subsubsection{Estimation procedure}
We estimate $(\theta, \alpha, \gamma, \mu)$ in Eq.~(\ref{eq:basicmodel}) using the GFE Algorithm 2 (Variable Neighborhood Search-VNS + Local Search) of BM2015, adapted to the unbalanced rotating panel structure of ENAHO.\footnote{To better adapt the algorithm to our setting, we do not use the replication code from BM2015. Instead, we implement the GFE algorithm in Python and adapt key steps to handle the partial household overlap and missing household-year cells implied by the rotating panel. Our code also allows us to add the time-invariant location fixed effects into the GFE estimator. Our implementation is written to be flexible and can be applied to other household surveys with similar rotating-panel designs with or without locational fixed effects.}

The GFE estimator chooses $(\theta, \alpha, \gamma, \mu)$ to minimize the sum of squared residuals over observed household year observations: 
\begin{equation}
\label{eq:GFE_estimates}
    (\hat{\theta}, \hat{\alpha}, \hat{\gamma}, \hat{\mu}) = \text{ argmin } _{(\theta, \alpha, \gamma, \mu) \in \Theta \times \mathcal{A} \times \Gamma_G \times \Lambda} \sum_{i=1}^N \sum_{t=1}^T d_{it} (y_{i t}-x_{i t}^{\prime} \theta - \alpha_{g_{i}t} - \mu_{p_i})^2
\end{equation}
where the minimum is taken over all possible groupings $\gamma = \{g_1,\cdots,g_N\}$ of $N$ households into $G$ groups, common parameters $\theta$, group-specific time effects $\alpha = \{\alpha_{gt}\}$, and location-specific effect $\mu = \{\mu_{p}\}$.

Directly minimizing this objective over all partitions $\gamma$ is computationally infeasible when $N$ is large. Following BM2015, we therefore use an iterative algorithm that alternates between (i) updating the group-time effects, common slopes, and location effects given group assignments, and (ii) updating group assignments given the current parameter estimates. In our setting, all updates are computed using only the observed household-year cells (those with $d_{it} = 1$). 

In each iteration, given a current grouping $\gamma$, we estimate $\theta$, $\alpha$ and $\mu$ by least squares on the pooled household-year sample with group-time indicators and location indicators. Given $(\theta, \alpha, \mu)$, each household is assigned to the group that yields the smallest sum of squared residuals over the years in which household is observed. Because the objective function is non-convex, the VNS + local search algorithm apply a multi-start strategy to reduce sensitivity to local minima.\footnote{The objective function is non-convex because it is a joint optimization over discrete group assignments and continuous parameters. Conditional on a fixed group assignment, the problem is just least squares, convex, and has a unique minimizer. However, since group assignment is discrete, joint optimization over the discrete group assignment and the continuous parameters renders the overall objective non-convex. For the multi-start, each start with a different seed.} We repeat this procedure until the objective function no longer decreases or the maximum number of iterations is reached. 

In practice, we implement the OLS update step by estimating $\theta$, group-by-time effects $\alpha_{gt}$, and the province-fixed effects $\mu_{p}$ using the observed household-year observations while absorbing province fixed effects. This implementation nets out time-invariant location differences so that group assignment is driven by within-province variation and by differences in households' time profiles. 

The step-by-step detailed explanation of the algorithm is described below.  

\begin{itemize}

\item \textbf{Step 1:} Let $(\theta^{(0)}, \alpha^{(0)}, \mu^{(0)}) \in \Theta \times \mathcal{A} \times \Lambda$ be an initial value, G be the number of groups pre-determined by the researcher, and set the tuning parameters for later steps.
    \begin{itemize}
        \item \textbf{Initializing $\theta^{(0)}$ and $\mu^{(0)}$}: We guess the initial $\theta^{(0)}$ and $\mu^{(0)}$ by regressing $y_{it}$ on $x_{it}$ with pooled data with province indicators using only rows with $d_{it}=1$.\footnote{Using OLS estimate as the initial guess is more stable than a random vector because we do not need to do a wild random guess. As random small-noise initializations are also allowed in BM2015 algorithm, We could use a random small noise vector from a normal distribution, or start with a zero vector.}
        \item \textbf{Initializing $\alpha^{(0)}$} (group-time fixed effects): Since group membership $g_i$ is not yet known, $\alpha^{(0)}$ must be initialized without groups. We initialize $\alpha^{(0)}$ as common time fixed effects estimated from the pooled observed household-year data after partial-ling out observed covariates and province fixed effects (using only observations with $d_{it} = 1$. We then set the initial time effect as the mean residual in each year: 
        $$
        \alpha_t^{(0)}=\frac{\sum_i d_{i t}\left(y_{i t}-x_{i t}^{\prime} \theta^{(0)}-\mu_{p_i}^{(0)}\right)}{\sum_i d_{i t}},
        $$
        where $\mu_{p(i)}^{(0)}$ denotes the estimated province fixed effect for household $i's$ province. Equivalently, $\alpha_t^{(0)}$ is the year-t average of the residualized outcome $y_{i t}-x_{i t}^{\prime} \theta^{(0)}-\mu_{p_i}^{(0)}$ over observed units. Then:
        $$
        \alpha_{g, t}^{(0)}=\alpha_t^{(0)} \quad \forall g \in\{1, \ldots, G\}
        $$
        Such that all the groups have identical time effects.
        \item \textbf{Initial grouping $\gamma_{init}$}: Once we have $(\theta^{(0)}, \alpha^{(0)}, \mu^{(0)})$, we can obtain the initial group assignment $\gamma_{init}$ that solves the following minimization problem:
        $$
        g_i^{(0)}=\underset{g \in\{1, \ldots, G\}}{\operatorname{argmin}} \sum_{t=1}^Td_{it}\left(y_{i t}-x_{i t}^{\prime} \theta^{(0)}-\alpha_{g t}^{(0)} - \mu_{p_i}^{(0)}\right)^2
        $$
        \item Setting up tuning parameters:
        \begin{itemize}
            \item $\texttt{n\_starts}$: the number of starting values we initialize. Since we follow the entire steps for each starting value, the entire steps will be repeated $\texttt{n\_starts}$ times.
            \item $\texttt{neighmax}$: the maximum neighborhood size controlling how many neighborhood jumps we allow in the following Step 3 before stopping the algorithm.\footnote{An neighborhood jump means that if we pick $n$ units, we move them to random new groups and re-run one update steps of Algorithm 1 in BM2015 to see if the objective improves. As $n$ increases from 1 to $\texttt{neighmax}$, we explore ``larger'' perturbations of the grouping. Larger values of the $\texttt{neighmax}$ would slower the program but provide more global exploration. We set the $\texttt{neighmax}$ to 5 to choose the best group $G$.}
            \item $\texttt{max\_local\_iters}$: the maximum number of local iterations that controls how many times we want to repeat the iterations to update $(\hat{\theta}, \hat{\alpha}, \hat{\gamma}, \hat{\mu})$ in Step 4.
            \item $\texttt{itermax}$: the maximum number of VNS iterations that controls how many times we want to repeat the entire VNS sequence\footnote{We set the $\texttt{itermax}$ = 10, the same order of magnitude used as the BM2015.}. The VNS iteration counter is initialized at $j=0$, and increased after each VNS cycle until $j$ reaches $\texttt{itermax}$
        \end{itemize}
    \end{itemize}

\item \textbf{Step 2:} Initialize neighborhood size.
    
Set the neighborhood size $n = 1$. $n = 1$ means that we start by exploring solutions that differ from the current best grouping by reassigning only one randomly chosen unit. 

\item \textbf{Step 3:} Neighborhood jump (``shake'').

Given the current grouping $\gamma^*$, randomly select $n$ units and relocate each selected unit to a (uniformly) randomly selected group, producing a new $\gamma'$. In our rotating-panel setting, our objective is always computed using only observed $(i,t)$ pairs with $d_{it}=1$, consistent with the unbalanced-panel formulation from BM2015.

After obtaining $\gamma'$, re-estimate $(\theta,\alpha, \mu)$ by OLS with group-by-time indicators, location indicators, and covariates conditional on $\gamma'$, using only observations with $d_{it}=1$ to obtain new parameter values $(\theta',\alpha',\mu')$.

\begin{equation}
\label{eq:GFE_estimaes_thetapr_alphapr}
    (\theta', \alpha', \mu') = \text{ argmin } _{(\theta, \alpha, \mu) \in \Theta \times \mathcal{A} \times \Lambda}\sum_{i=1}^N \sum_{t=1}^T d_{it} (y_{i t}-x_{i t}^{\prime} \theta - \alpha_{g_{i}t} - \mu_{p_i})^2
\end{equation}

\item \textbf{Step 4:} Capped refinement based on Algorithm 1.

Following the logic of Step 4 in BM2015, we apply a limited refinement procedure after the neighborhood jump rather than iterating Algorithm 1 to full convergence. In the implementation, this refinement consists of updating $(\theta,\alpha,\mu)$ once conditional on the shaken grouping $\gamma'$, followed by a capped sequence of greedy reassignment passes and a final OLS update. The cap is controlled by \texttt{max\_local\_iters}.

Our implementation follows the structure of Algorithm 2 in BM2015, but adapts the refinement step to the rotating-panel setting by using a capped practical variant rather than running Algorithm 1 to full convergence after each neighborhood jump.

\item \textbf{Step 5:} Greedy local search over single-unit reassignments. 

Starting from the grouping $\gamma'=\{g'_1,\dots,g'_N\}$, perform a \textit{local search} that systematically checks single unit moves $i: g'_i \mapsto g \in \{1,\dots,G\}$ for all $g \neq g'_i$, and accepts a move whenever it decreases the objective function, holding $(\theta',\alpha',\mu')$ fixed. This local search is run for at most \texttt{max\_local\_iters} passes and and stop when no further single-unit reassignment reduces the objective function. Denote the resulting locally optimal grouping as $\gamma''=\{g''_1,\dots,g''_N\}$. Given the updated grouping $\gamma''$, re-estimate
$$
(\theta'',\alpha'', \mu'') = \text{ argmin } _{(\theta, \alpha, \mu) \in \Theta \times \mathcal{A} \times \Lambda}\sum_{i=1}^N \sum_{t=1}^T d_{it} (y_{i t}-x_{i t}^{\prime} \theta - \alpha_{g''_{i}t} - \mu_{p_i})^2
$$
As noted by BM2015, ``local search allows to get around local minima that are close to each other, whereas random jumps aim at efficiently exploring the objective function while avoiding to get trapped in a valley.''

In our rotating panel set up, each candidate's move evaluates the following objective function: 
$$
\text{loss}_{i}(g)=\sum_{t=1}^T d_{it} \left(y_{i t}-x_{i t}^{\prime} \theta - \alpha_{g t} - \mu_{p_i} \right)^2
$$
so missing households' information in a given year would not affect the reassignment decision. 

\item \textbf{Step 6:} Acceptance rule.

If the objective function under $\gamma''$ is lower than the objective function under the current base $\gamma^*$, set $\gamma^* = \gamma''$ and return to \textbf{Step 2} (resetting $n = 1$ to restart the neighborhood exploration around the improved solution). Otherwise, set $n = n + 1$ and continue to Step 7.

\item \textbf{Step 7:} Increase neighborhood size up to $\texttt{neighmax}$.

If $n \leq \texttt{neighmax}$, go back to \textbf{Step 3} and perform another neighborhood jump of size $n$. If $n > \texttt{neighmax}$, proceed to Step 8. 

\item \textbf{Step 8:} Iterate VNS cycles up to $\texttt{itermax}$ and stop.

For $j=1,\ldots,\texttt{itermax}$, repeat Steps 2--7. After completing the $\texttt{itermax}$-th VNS cycle, stop and return the  best grouping $\gamma^*$ and associated parameters estimated.
\end{itemize}

\noindent \textbf{Multiple starting values}: Following the supplementary appendix in BM2015, Algorithm 2 typically runs for $\texttt{n\_starts}$ independent random initializations\footnote{BM2015 calls this parameter as $N_s$.}. $\texttt{n\_starts}$ is set to reduce sensitivity to initialization. We repeat Steps 1--8 for $\texttt{n\_starts}$ different initial values and also different seeds. We keep the solution with the lowest objective function value (i.e., which minimized squared errors.)

\begin{figure}[!htbp]
    \begin{center}
        \caption{GFE Estimation Flowchart}
        \label{fig:GFE_flowchart}
        \includegraphics[width=0.95\linewidth]{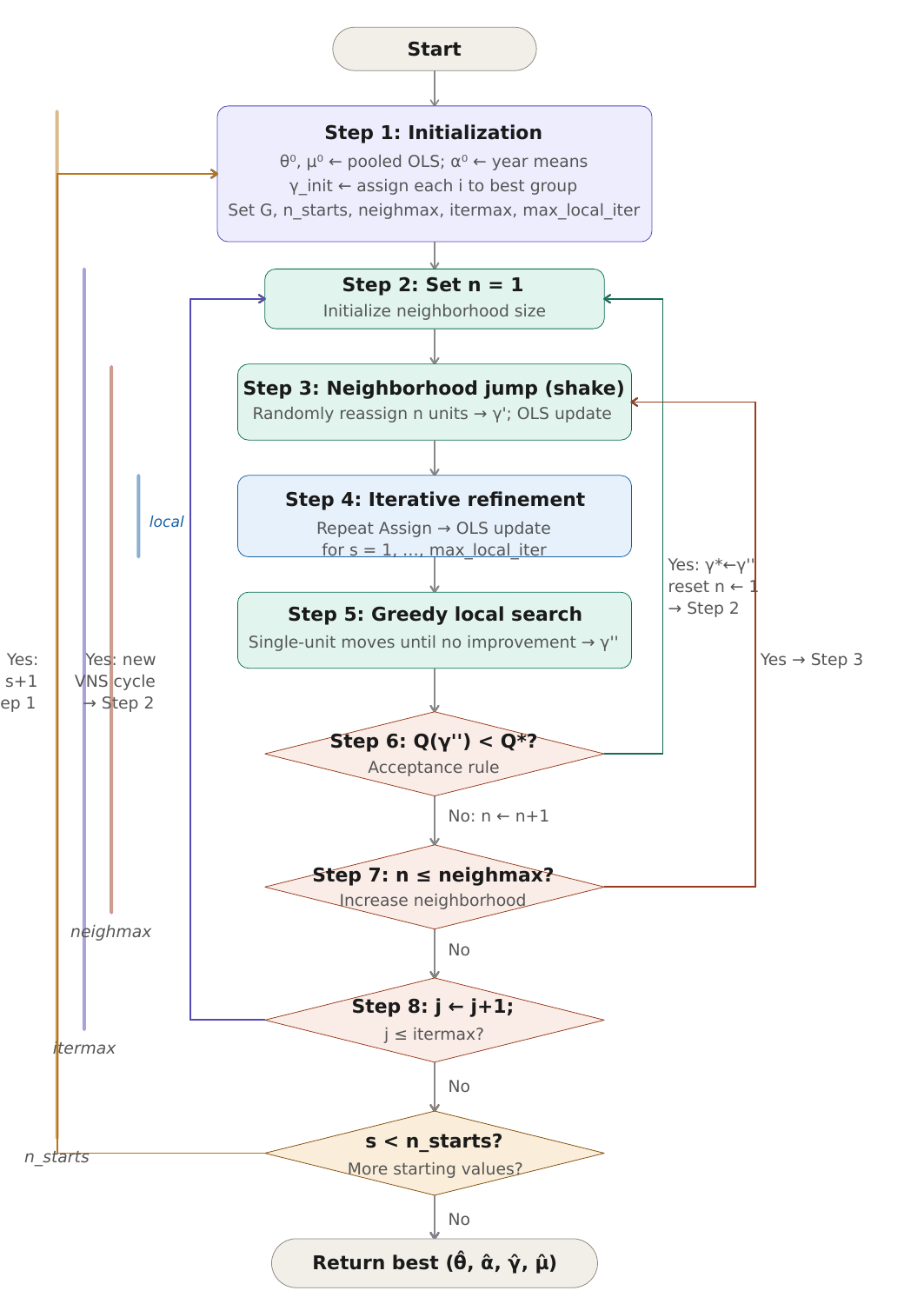}
    \end{center}
\end{figure}

Therefore, the choice of running parameters for the GFE Algorithm 2 in our case includes  $(\texttt{n\_starts}; \texttt{itermax}; \texttt{neighmax}; \texttt{max\_local\_iters})$. Setting up higher $\texttt{n\_starts}$ is computationally intensive. We start with $\texttt{n\_starts} = 3$ to find the better number of groups $G$ among all the $G$ we tried, and then we increase the $\texttt{n\_starts}$ to 10 for a few chosen better $G$ to check with the different start value to see if the algorithm yield the same results. See the Figure~\ref{fig:GFE_flowchart} for the flowchart the steps of the algorithm described above, and the pseudocode~(\ref{alg:gfe_vns}) in Appendix \ref{appendix:pseudocode}.

\subsection{Estimation in an Unbalanced/Rotating Panels}

A key practical feature of ENAHO data is that it is an unbalanced rotating panel: most households in this panel are observed for only a short consecutive period and are missing in other years. We therefore implement the GFE in a way that (i) uses all available household-year observations, and (ii) ensure that missing household-year observations do not mechanically affect either parameter updates or group assignments. 

Let our observed set to be $\Omega \equiv\left\{(i, t): d_{i t}=1\right\}$ where $d_{it} \in \{0, 1\}$ indicates whether household $i$ is surveyed (and has non-missing outcome and covariates) in year $t$. All least-squares updates and assignment steps are computed on $\Omega$ only, consistent with the masked objective in Eq.~(\ref{eq:GFE_estimates}). More precisely, conditional on a candidate grouping $\gamma = \{g_1, \dots, g_N\}$, the updated step estimates $(\theta,\alpha,\mu)$ by regression $y_{it}$ on $x_{it}$, group-by-time indicators, and province indicators using the pooled household-year sample restricted to $\Omega$. This provides us the estimates of common slopes $\theta$, the group-time effects $\{\alpha_{gt}\}$, and the province fixed effects $\{\mu_p\}$ based on the data observed cells only. 

Given $(\theta,\alpha,\mu)$, the assignment step also respect the rotating-panel structure, for each household $i$, we assign the group label by minimizing the sum of squared residuals over the years in which the household is observed:
$$
\hat{g}_i=\underset{g \in\{1, \ldots, G\}}{\operatorname{argmin}} \sum_{t=1}^Td_{it}\left(y_{i t}-x_{i t}^{\prime} \theta-\alpha_{g t} - \mu_{p_i}\right)^2
$$
Thus, years in which household $i$ is not interviewed ($d_{it} = 0$) contribute zero to the assignment criterion. This is important in practice: households with shorter observed windows are not penalized for missing years, and their group membership is inferred from the information they provide. Intuitively, the rotating panel still provides identifying variations because some households are observed in consecutive years. These overlaps link adjacent years and help us pin down both (i) household group membership and (ii) how each group's unobserved component evolves over time. At the same time, the continual entry of new households each year maintain the cross-sectional sample size, ensuring the precision with which we estimate the group-specific time effects in each year. 

After group assignments $g_i$, time effects $\alpha_{gt}$, covariates coefficients $\theta$, and province fixed effect coefficient $\mu_p$ are estimated, we can predict the outcome measured by IHS-transformed total expenditures per capita for any observed households:
\begin{equation}
\label{eq:predictedY}
    \hat{y}^{GFE}_{it} = x_{i t}^{\prime} \hat{\theta} + \hat{\alpha}_{\hat{g}_i t} + \hat{\mu}_{p_i} \qquad (i,t) \in \Omega
\end{equation}

These fitted values provide us a cleaned representation of household welfare that preserves (i) systematic variation explained by observables, (ii) persistent location differences capture by province fixed effects, and (iii) group-specific time patterns in unobserved heterogeneity. We can then use the estimated group-time paths to construct poverty transition measured over horizons that exceed a household's observed panel length, which we talk in detail in the next Section \ref{sec:from_estimateG_to_poverty_dynamics}.

\subsection{From Estimated Groups to Poverty Dynamics}
\label{sec:from_estimateG_to_poverty_dynamics}

The GFE provides us two useful measures that we can use directly to studying poverty dynamics in a rotating panel: (i) an estimated latent group assignment for each household, $\hat{g}_i$, and (ii) an estimated group-specific time path, $\hat{\alpha}_{gt}$ (net of observed covariates we used, and also province fixed effects).  The $\hat{\alpha}_{gt}$ summarizes how the expenditure trajectory evolves for households of latent type $g$, while $\hat{g}_i$ tells us which trajectory is most consistent with a given household's observed outcomes. 

Using these estimates, we predict a model-implied expenditure series that is similar to Eq.~(\ref{eq:predictedY}), but with missing years covariates $x_{it}$ filled by the researcher.
\begin{equation}
\label{eq:predictedY_fullsample}
    \hat{y}^{\mathrm{comp}}_{it} = x_{i t}^{\prime} \hat{\theta} + \hat{\alpha}_{\hat{g}_i t} + \hat{\mu}_{p_i}
\end{equation}

Essentially, Eq.~(\ref{eq:predictedY_fullsample}) defines a model-implied complete expenditure path $\hat{y}^{\mathrm{comp}}_{it}$ over the full time horizon, including years in which a household is not surveyed while Eq.~(\ref{eq:predictedY}) defines the in-sample fitted value for household-year cells that are actually observed in rotating panel, i.e., $(i,t) \in \Omega$ where $d_{it} = 1$. The predicted expenditure $\hat{y}^{\mathrm{comp}}_{it}$ replaces the missing household-year information with a group-consistent trajectory rather than treating the missing years information as attrition or requiring balanced panels. 

Constructing Eq.~(\ref{eq:predictedY_fullsample}) therefore requires the value of $x_{it}$ for every year $t$, even when the household is not observed. When the covariates are time-invariant, this extension is straightforward because the same $x_{it}$ can be carried across years. However, if some covariates vary over time, we could specify how $x_{it}$ is filled in for missing household-year cells. We could fill the missing by carrying forward the last observed value, anchoring to the first observed value, interpolating between observed waves, predicting $x_{it}$ from auxiliary models or external information, or other methods deemed appropriate by the researcher. Importantly, this covariate-completion step is an additional researcher input that affects the level and timing of the predicted series in Eq.~(\ref{eq:predictedY_fullsample}), while the group-time component and province fixed effect provide the structured dynamics and location adjustment learned from the observed rotating-panel overlap.

Once we have the fitted expenditures $\hat{y}^{\mathrm{comp}}_{it}$, we can translate them into poverty status using the year-specific poverty thresholds available in ENAHO data. We can compare $\hat{y}^{\mathrm{comp}}_{it}$ against the poverty line thresholds $z_{t}$ (total poverty line or food poverty line in year $t$), and define a poverty indicator as
$$
\hat{P}_{it} = 1\{\hat{y}^{\mathrm{comp}}_{it} < z_t\}
$$
and compute poverty dynamics over different horizons $h$ by comparing $\hat{P}_{i,t}$ and $\hat{P}_{i,t+h}$. This yields transition probabilities -- entry, exit, and persistence rates -- as well as transition matrices (poor, non-poor). Because the fitted series is available for all the year, we can also summarize long-run measures such as the share of time spent in poverty, and incidence of chronic poverty (e.g., poor in at least certain number of years within a window), and the distribution of poverty lengths (e.g., two years, three years, or more) from the predicted outcome. 


\subsection{Latent Group G Selection} 
\label{sec:latent_group_G_selection}

A key practical choice in the GFE framework is the number of latent groups, $G$. Larger $G$ increases flexibility by allowing more distinct time profiles of unobserved heterogeneity, but it also raises the risk of overfitting, especially in short, unbalanced panels where many households are observed for only a few years. We therefore select $G$ using a transparent model selection criterion, and we complement this with out-of-sample validation and qualitative checks on the stability and interpretability of the estimated group paths.

To select the number of latent group $G$, we combine Bayesian Information Criterion (BIC) used in BM2015 with Root Mean Square Error (RMSE). For a given $G$, the GFE estimator produces $(\hat{\theta}, \hat{\alpha}, \hat{\mu})$ and a minimized sum of squared residuals over observed household-year cells:
\begin{equation}
\operatorname{SSE}(G)=\sum_{i=1}^N \sum_{t=1}^T d_{i t}\left(y_{i t}-x_{i t}^{\prime} \hat{\theta}-\hat{\alpha}_{\hat{g}_i, t}-\hat{\mu}_{p_i}\right)^2
\end{equation}
where $d_{it} = 1$ indicates an observed household cell and $p_i$ denotes the household's time-invariant location (province). 

Let $NT_{obs} = \sum_{it} d_{it}$ be the number of observed household-year cell used in estimation. We compute the estimated error variance as 
\begin{equation}
\hat{\sigma}^2(G)=\frac{\operatorname{SSE}(G)}{N T_{\text {obs }}-g(G)}
\end{equation}
where $g(G) = GT + N + K + L$ is the effective parameter count. $GT$ counts the group-by-time effects $\alpha_{gt}$, $N$ counts the unit-level group assignments (one label per household), $K$ counts the common slope coefficients $\theta$, and $L$ counts the location fixed effects $\mu_p$ (one per province).

Our BIC is then computed as: 
\begin{equation}
\operatorname{BIC}(G)=\frac{\operatorname{SSE}(G)}{N T_{\mathrm{obs}}}+\hat{\sigma}^2(G) \frac{p(G)}{N T_{\mathrm{obs}}}\ln \left(N T_{\mathrm{obs}}\right) 
\end{equation}

The key difference between the BIC used in our application versus the one in BM2015 is that we additionally absorbed time-invariant location heterogeneity using province fixed effects. This adds $L$ additional parameters to the penalty term and slightly reduces the effective degree of freedom in $\hat{\sigma}^2(G)$. Aside from this adjustment, the criterion is computed in the same way, using only the observed household-year cells in the unbalanced rotating panel.

Because the GFE objective is non-convex in group assignments, different starting values can lead to different local optima. To reduce sensitivity to initialization, for each candidate $G$, we run the estimation algorithm multiple times with different random starts and retain the solution with the lowest objective function value, $SSE(G)$. We then compute $BIC(G)$ from the best solution. We choose the best $G$ by looking at the both RMSE and BIC values (for both RMSE and BIC, the lower value the better).

\section{Empirical Implementation}
\label{sec:implementation}

\subsection{Data: ENAHO 2007 - 2019}
\label{sec:implementation_data}

We use the National Household Survey on Living Conditions and Poverty (ENAHO, an acronym in Spanish) from 2007 to 2019 to investigate poverty dynamics in Peru.\footnote{We processed our data based on the programs from \citet{murillo_commodity_2024} with further adjustments.} ENAHO is a nationally-representative household survey that collects data on housing, household expenditure, education, health, employment, and income \citep{inei_peru_2017}. The main indicator for our study in the ENAHO is the poverty status (non-poor, poor) which is determined by whether household per capita expenditure is above the total poverty line (non-poor), below the total poverty line but above food poverty line (non-extremely poor) or below food poverty line (extremely poor)\footnote{We could not find the official reference how this poverty status in the data is determined, but we manually constructed the indicators comparing household per capita expenditure and the different poverty lines which matched the official indicators with 99.9\% accuracy.}. Every year, 30\% of the sample households are categorized as panel households which are repeatedly surveyed for up to six years. In our study period, 22.7\% of the panel households were surveyed no more than three times, and only 6\% of them were surveyed over 6 years. This rotating panel structure allows us to apply the GFE methods to study poverty dynamics overtime. For households who were in the data for at least three years, we can study their long term poverty dynamics using data from other households in the survey once they are grouped under the same group. The full dataset includes around 471,072 household-year observations from 375,897 unique households.

We further restrict our sample as follows. First, we restrict our study sample to households observed for at least three survey rounds. A minimum of two rounds is required for GFE estimation, and we use the final observed round for out-of-sample evaluation. Second, we restrict to households with a household head aged 25-55 in any given survey year. Because our analysis relies on identifying latent household groups with distinct time patterns, it is important that the set of households used for estimation reflects a relatively stable household structure. In rotating panel data, very young household heads are more likely to be entering newly formed households, while older heads are more likely to experience household dissolution, retirement-related changes, or mortality. All of which could generate non-random attrition and discrete shifts in household composition. These demographic transitions can mechanically alter observed expenditures and poverty status in ways that are difficult to distinguish from genuine welfare dynamics, and they can also complicate the interpretation of GFE ``types'' (groups).

To reduce these concerns and improve comparability of households across waves, we restricted our sample to households whose head is between 25 and 55 years old in any given survey year, an age range in which household formation and dissolution are less frequent and labor market attachment is more stable. After applying these sample restrictions (households surveyed for at least 3 years, and the head of household aged 25-55 in each survey year), the data sample we use for the GFE estimation includes 56,390 observations from 14,886 households. As a robustness check, we report results using the same three-wave requirement but without imposing the household-head age restriction with a sample of 104,304 observations from 27,034 households.

\begin{figure}[!htbp]
    \begin{center}
        \caption{The trend of poverty status by year}
        \label{fig:GFE_poverty_trend_by_year}
        \includegraphics[width=0.95\linewidth]{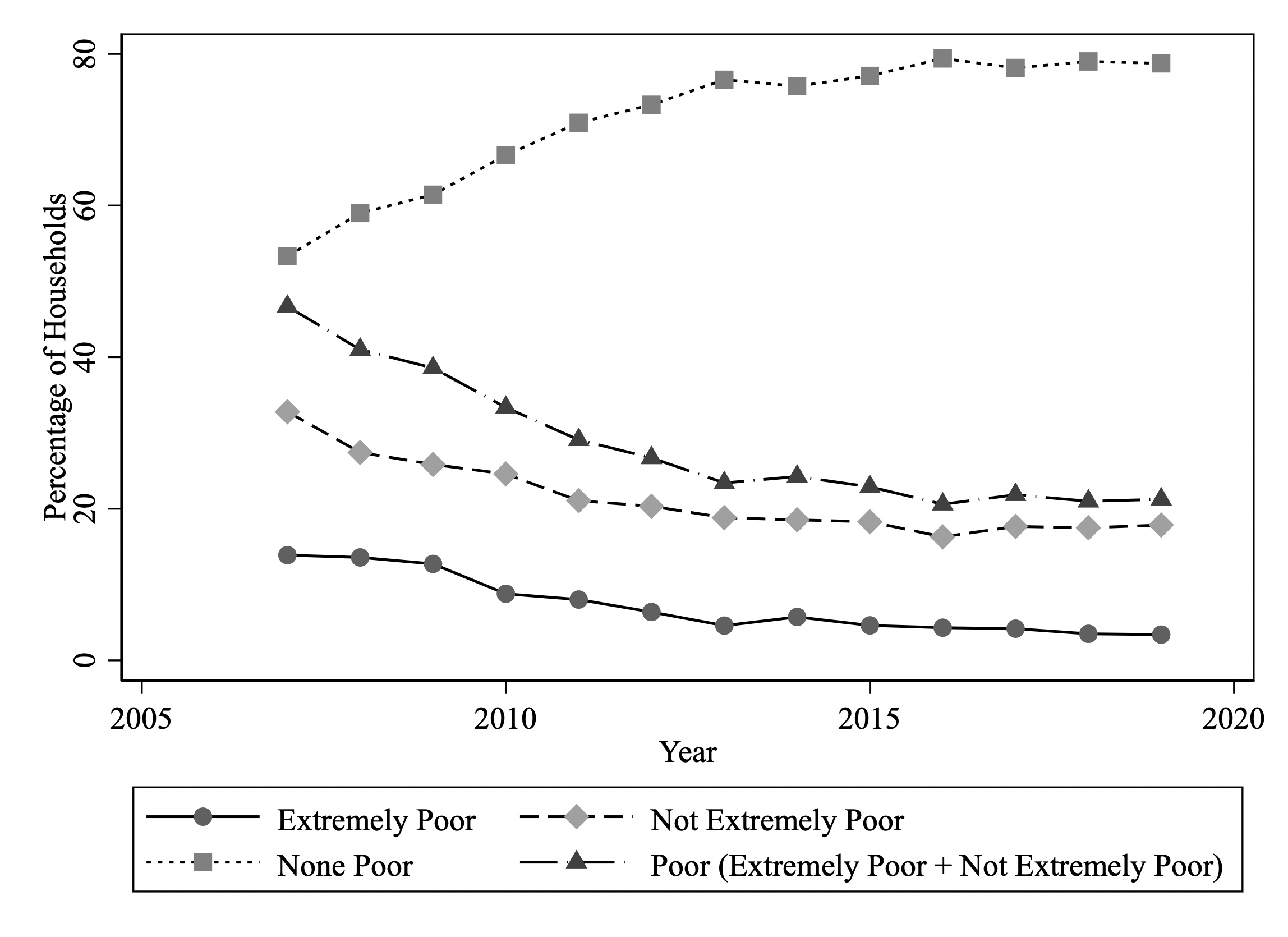}
    \end{center}
    \footnotesize
    {\textit{Source:} Authors’ calculations using ENAHO 2007–2019, restricted to households observed in at least three survey years and to household heads aged 25–55 in all observed years. \\ 
    \textit{Note:} We applied our sample restriction: keep households that stayed in the survey for at three two years, and restricted the sample to household heads between 25 and 55 years old in any given survey years.} \\
\end{figure}

Figure \ref{fig:GFE_poverty_trend_by_year} shows the trend in poverty status in our selected sample. Consistent with Peru's broad poverty reduction over the period, the share of households that are poor (extremely poor plus not extremely poor) declines steadily from 2007 to 2019, while the share of non-poor households rises correspondingly. Both components of poverty fall over time: extremely poverty drops sharply and becomes a relatively small share of the sample by the end of the period, and non-extreme poverty also trends downward. We present the non-restricted (full) sample poverty status in Appendix Figure \ref{fig:GFE_poverty_trend_by_year_allsample}. The overall patterns are very similar, suggesting that our main sample restriction do not change the narrative. However, the levels are slightly different, with the restricted sample exhibiting a higher poverty rate in the earlier years. Requiring repeated observation selects households that can be re-contacted and followed over time, which may differ from households observed only once (e.g., in mobility, location, or demographics). This selection can shift poverty levels relative to the full sample. Importantly, the time trends are very similar across sample, and we further assess sensitivity by reporting robustness checks that relax the age restriction and use the broader sample. In both figures, poverty categories are mutually exclusive within each year, so the shares of poor and non-poor sum to 100 percent.  

Table \ref{tab:GFE.sumstats} reports weighted summary statistics for our restricted ENAHO sample. Panel A summarizes household-year outcomes used in the poverty dynamic analysis. Average monthly per-capita total expenditures are around 445 (in 2007 dollars), and average total and food poverty lines are 235 and 125 (also in 2007 dollars), respectively. We also report the corresponding Inverse Hyperbolic Sine (IHS) transformed measures. Panel B describes baseline household and head characteristics measured in each household's first observed survey year to avoid over-weighting households that appear in multiple waves. In this sample, around 21 percent of the household heads are female, and the average head age is around 41 years old. 30 percent of household heads have primary education, 38 percent of household heads have secondary education, 18 percent of household heads have college education, and the rest of household heads has no primary education. Most households are Spanish-speaking (75 percent) and married or cohabiting (77 percent). Household living conditions are relatively well in our sample: 76 percent live in urban areas, 74 percent have access to water, and 89 percent have electricity.

\input{GFE/tables/main/GFE_sumstats}

\subsection{Model Specification}
\label{sec:model_specification}

Our main welfare measure $y$ is the IHS-transformed monthly household total expenditure per capita (in 2007 dollar). We consider two specifications that differ in the set of covariates $x_{it}$. The first specification is a conservative one which we include characteristics that are more deterministic or unlikely to vary over time. These characteristics include age, gender, educational attainment, and the Spanish spoken language of the head of the household. The second specification adds additional characteristics that are more likely to vary over time, but important in capturing variation in expenditure. These characteristics include marital status, urban/rural residence, and whether the household has water and electricity. 

To examine how much additional characteristics in the second specification capture household expenditure, we regress household expenditure on both specifications. Table \ref{tab:GFE.reg.results.provinceFE} reports estimates of pooled household-year regression of IHS expenditure per capita on the two covariate sets. All specification include province and year fixed effect to net out time-invariant differences across local areas (e.g., geography, prices, and local labor market conditions etc.), apply the ENAHO survey expansion weight, and cluster standard errors clustered at the household level.\footnote{Peru’s administrative divisions are hierarchical: the country is divided into departments (also referred to as regions), which are subdivided into provinces, which in turn are subdivided into districts. In the ENAHO data, the geographic identifier UBIGEO corresponds to the district level. In our selected sample, we have 25 departments, 195 province, and 1121 districts.} The richer specification increases the explanatory power of the model, raising the $R^2$ from 0.49 to 0.54, and we use these two sets of covariates as alternative inputs in our subsequent GFE analysis. In the Appendix, we show that the main qualitative patterns in Table \ref{tab:GFE.reg.results.provinceFE} are not sensitive to using coarser (department) (Table \ref{tab:GFE.reg.results.departmentFE}) or more granular (Table \ref{tab:GFE.reg.results.districtFE}) (district/UBIGEO) location fixed effects. We choose province fixed effects to control for a meaningful amount of local price/labor market/geography differences without fully saturating location. This specification leaves enough within-province variation for our GFE model when grouping the households into different latent type $G$.

\input{GFE/tables/main/GFE_reg_results_provinceFE}

\subsection{Estimation Steps}
\label{sec:estimation_steps}

Because our intended application is to use information observed for a household in earlier survey rounds to predict its welfare in a later (unobserved) round in an unbalanced rotating panel, we therefore evaluate the GFE model using a validation design that mimics this forecasting task: for each household, we hold out its last observed survey year (test data) and estimate the model on all remaining household--year observations (training data). We prefer this test set design to other alternatives: a random subset of households for the entire period or all observations from a specific calendar year. We do not prefer leaving a random subset of households for an entire period, since our primary objective is not to predict outcomes for households that have never been in the survey. Instead, we aim to predict a household’s future welfare using its earlier observed history, which is exactly the application of GFE in rotating panels. In addition, holding out entire households does not change the set of years used for estimation but tests a different notion of generalization (across households rather than forward in time within a household). A further alternative---holding out later calendar years entirely--is also not appropriate in our setting because it would remove the information needed to estimate $\alpha_{g t}$ for those later holdout years, and GFE does not extrapolate unknown future $\alpha_{gt}$ without additional structure.

\paragraph{Step 1: Construct a rotating-panel estimation sample and mask.}

Let $i=1,\dots,N$ index households and $t=1,\dots,T$ index calendar years in our analysis window. We stack the data into household--year observations and define an observation indicator $d_{it}\in\{0,1\}$ equal to one if household $i$ is observed in year $t$ with non-missing outcome and covariates, and zero otherwise. All estimation and objective function evaluations use only observations with $d_{it}=1$, consistent with the unbalanced-panel formulation of the GFE objective.

\paragraph{Step 2: Define training and test data by last-observed-year holdout.}

For each household $i$, let $t_i^{\text{last}}$ denote the last year in which the household is observed (i.e., the largest $t$ such that $d_{it}=1$). We construct a training mask
$$
d^{\text{train}}_{it} = d_{it}\cdot \mathbf{1}\{t \neq t_i^{\text{last}}\},
$$
so that each household contributes all observed years except its last one to the estimation, and this data is our training data. The held out data consists of the observations $\{(i,t_i^{\text{last}}): d_{i,t_i^{\text{last}}}=1\}$, and this is our test data. This split preserves the rotating-panel structure: although a given household’s last year is excluded from estimation, other households observed in the same calendar year remain in the training data, so the year-specific group effects $\alpha_{gt}$ are still identified from the cross-sectional sample in that year.

\paragraph{Step 3: Estimate GFE for a candidate number of groups $G$.}

For each candidate $G$, we estimate the GFE model in Eq.~(\ref{eq:basicmodel}) by minimizing the
masked sum of squared residuals over the training data:
$$
\min_{\theta,\alpha,\gamma} \sum_{i=1}^N \sum_{t=1}^T d^{\text{train}}_{it}
\left(y_{it} - x_{it}'\theta - \alpha_{g_i t} - \mu_{p_i}\right)^2
$$
where $\gamma=\{g_i\}_{i=1}^N$ denotes the household-to-group assignments. We implement Algorithm 2 of BM2015 (VNS with local search) adapted to the unbalanced panel by evaluating objectives and assignment losses only over years with $d^{\text{train}}_{it}=1$. To reduce sensitivity to local minima, we use a multi-start strategy with $\texttt{n\_starts}$ different initializations (different random seeds) and retain the solution with the lowest objective value for each $G$.

\paragraph{Step 4: Model selection over $G$ and predictive validation.}

For each $G$ we compute (i) the BIC on the training data (based on the minimized masked SSE and the corresponding parameter count) and (ii) out-of-sample predictive performance measured using RMSE on the test data. For each held-out test observation $(i,t_i^{\text{last}})$, we form the prediction:
$$
\widehat{y}_{i,t_i^{\text{last}}} = x'_{i,t_i^{\text{last}}}\widehat{\theta} +
\widehat{\alpha}_{\widehat{g}_i,\,t_i^{\text{last}}} + \widehat{\mu}_{p_i}
$$
where $\widehat{g}_i$ is the estimated group assignment for the household $i$ obtained from the training fit. We summarize the prediction accuracy using RMSE
computed over the test data. 

\paragraph{Step 5: Final estimation for the chosen $G$ and stability checks.}

After selecting $G$, we re-estimate the GFE model for the six best-performing candidate values of $G$ using a larger number of random starts (higher $\texttt{n\_starts}$) to verify that the optimal selected latent group $G$ is stable to initialization. We then select a single final $G$ taken into account of both the BIC and RMSE values, and take its estimated group assignments $\widehat{g}$, coefficient vector $\widehat{\theta}$, location fixed effects $\widehat{\mu}$ (when included), and group--time effects $\widehat{\alpha}_{gt}$ as the parameters used in the subsequent analysis of poverty dynamics.

To construct poverty dynamics, we distinguish between two uses of the estimated model. First, for all predictive validation exercises (e.g., one-step-ahead transitions), we use the parameter estimates obtained from the training sample that excludes each household’s last observed year. This ensures that the end-year poverty status used in validation is out-of-sample at the household-year level. Importantly, because other households remain observed in each calendar year, the group--year effects $\widehat{\alpha}_{gt}$ are still identified for all years in the analysis window except for 2019 because our training data does not have 2019 observations and therefore we cannot estimate $\alpha_{g,2019}$.

Second, when we move from validation to describing poverty dynamics and constructing completed welfare paths (i.e., imputing non-survey years), we re-estimate $\widehat{y}_{it}=x'_{it}\widehat{\theta}+\widehat{\alpha}_{\widehat{g}_i t}+\widehat{\mu}_{p_i}$ without leaving the last year observed as a test set, whenever the covariates $x_{it}$ are available (or completed as described in Section \ref{sec:from_estimateG_to_poverty_dynamics}), which allows us to trace model-implied welfare and poverty status over the full set of years covered by the survey.

\section{Results}
\label{sec:results}

\subsection{Selecting Best G using GFE}

We run the GFE model with two sets of different covariates. The first sets of covariates are the ones from Table~\ref{tab:GFE.reg.results.provinceFE} column (1), including head of household gender, age, education level, and if primary language is Spanish. The second sets of covariates are the ones from Table~\ref{tab:GFE.reg.results.provinceFE} column (2) which add additional covariates including marriage status, household living in urban area, has water, and has electricity. We call the first sets of covariates as model specification 1, and the second sets of covariates as model specification 2. 

Running the GFE model with $\texttt{n\_starts}=3$, $\texttt{itermax}=10$, $\texttt{neighmax}=5$, and set the $\texttt{max\_local\_iters}=2$ for latent groups from 1 to 40, we can calculate the BIC for each of the latent groups using the training data, and calculate the RMSE using the test data. 

\begin{figure}[!htbp]
    \begin{center}
        \caption{Model fit (BIC and RMSE) by different sets of covariates ($\texttt{n\_starts}=3$)}
        \label{fig:Fig_bic_vs_rmse_spec1_vs_spec2}
        \includegraphics[width=0.95\linewidth]{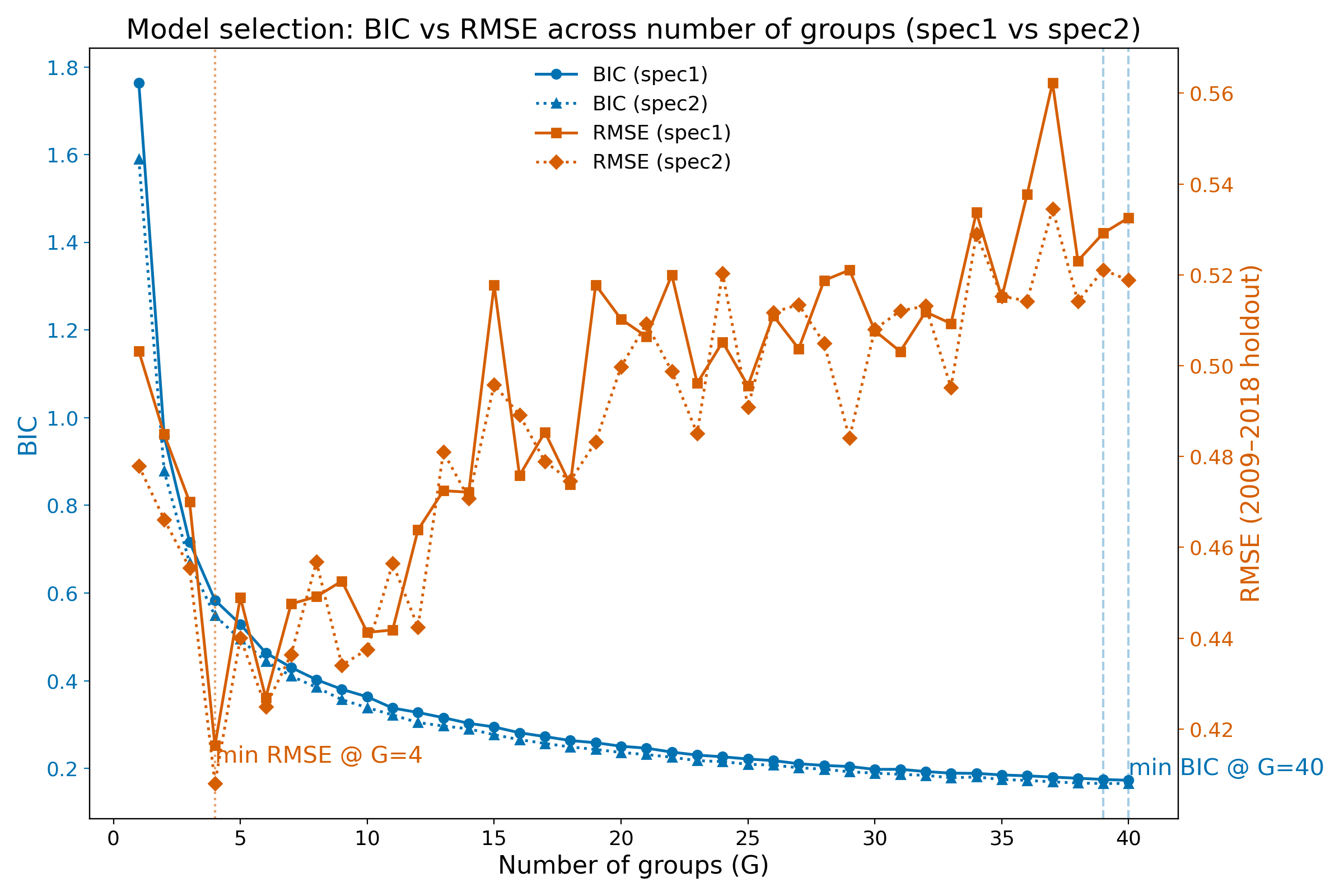}
    \end{center}
    \footnotesize
    {\textit{Source:} Authors’ calculations using ENAHO 2007–2019, restricted to households observed in at least three survey years and to household heads aged 25–55 in all observed years. \\ 
    \textit{Note:} The figure plots model selection diagnostics for the GFE estimator across initial numbers of groups $G\in\{1,\cdots,40\}$ with $\texttt{n\_starts}=3$, $\texttt{itermax}=10$, $\texttt{neighmax}=5$, and $\texttt{max\_local\_iters}=2$ under two covariate specifications (spec1 and spec2). BIC is computed on the training data; RMSE is computed out-of-sample using the last observed household-year held out for each household (years 2009–2018) test data. Lower values indicate better fit/prediction.} \\
\end{figure}

Figure~\ref{fig:Fig_bic_vs_rmse_spec1_vs_spec2} shows the change in the BIC and RMSE values between different latent groups $G$ using two alternative covariate specifications (specification~1 and specification~2). In both specifications, BIC declines steadily as $G$ increases and attains its minimum at the upper end of the grid (e.g., the maximal $G$ we tried, in this case for $G=40$ for both specifications), indicating that adding groups keep improving in-sample fit. This monotone pattern is expected in our setting because increasing $G$ expands the number of group-time intercepts $\alpha_{gt}$, making the model increasingly flexible and able to fit heterogeneity in an unbalanced rotating panel. As $G$ becomes large, however, the model begins to fragment the sample into many small groups. In an unbalanced rotating panel, this can leave some group-year $(g,t)$ supported by only a very small number of observations, so the corresponding $\alpha_{gt}$ is estimated with substantial sampling noise. This ``thin support'' makes group-time path less stable and can also reduce the reliability of prediction, even if the in-sample fit (measured by the BIC) continues to improve. 

Empirically, at $G=40$, we observe group-time cells with as small as two observations only, which implies that the $\alpha_{gt}$ values are identified from only two observations, amplifying estimation variance and also weakening the interpretability of the estimated latent type.

In contrast, predictive accuracy shows a clear interior optimum: RMSE drop sharply when moving from very small $G$ to moderate values, but then flattens and eventually increases as $G$ becomes large. For both specifications, the minimal RMSE on the held-out last-observed household-year (2009--2018) test data is minimized at $G=4$, suggesting that additional groups beyond this points mainly capture idiosyncratic variation rather than improving forecasting performance. Since our primary goal is prediction, we therefore emphasize the RMSE-based criterion for selecting the best $G$, using BIC as a complementary diagnostic about in-sample fit and model complexity.

To ensure the low-RSME results are not just due to a lucky (or unlucky) random starting point that captures specific local minima rather than a truly better grouping structure, we conduct an additional stability check by increasing the number of random starts, as described in Step 5 in Section \ref{sec:estimation_steps}. Specifically, we take the six values of $Gs$ with the lowest holdout RMSE in the initial grid search as the six best-performing candidate values, $G\in\{4,5,6,7,10,11\}$ for specification 1 and $G\in\{4,5,6,7,9,10\}$ for specification 2,  and re-estimate the model with the same algorithm setting but a larger number of initialization (we increase $\texttt{n\_starts}$ from 3 to 10). Using more random starts makes it less likely that the algorithm reaches local minima strictly greater than global minimum for a given $G$, so the comparison of prediction performance across different $G$ value is more reliable.\footnote{We only increase to $\texttt{n\_starts}=10$ after the initial $G\in\{1, \dots, 40\}$ grid search because computation time grows quickly with the number of random starts, so it is more efficient to reserve the higher $\texttt{n\_starts}$ run for a small set of promising $G$ values identified in the initial run.}

\begin{figure}[!htbp]
    \begin{center}
        \caption{Model fit (BIC and RMSE) by different sets of covariates ($\texttt{n\_starts}=10$)}
        \label{fig:Fig_bic_vs_rmse_spec1_vs_spec2_selectedG_nstarts10}
        \includegraphics[width=0.95\linewidth]{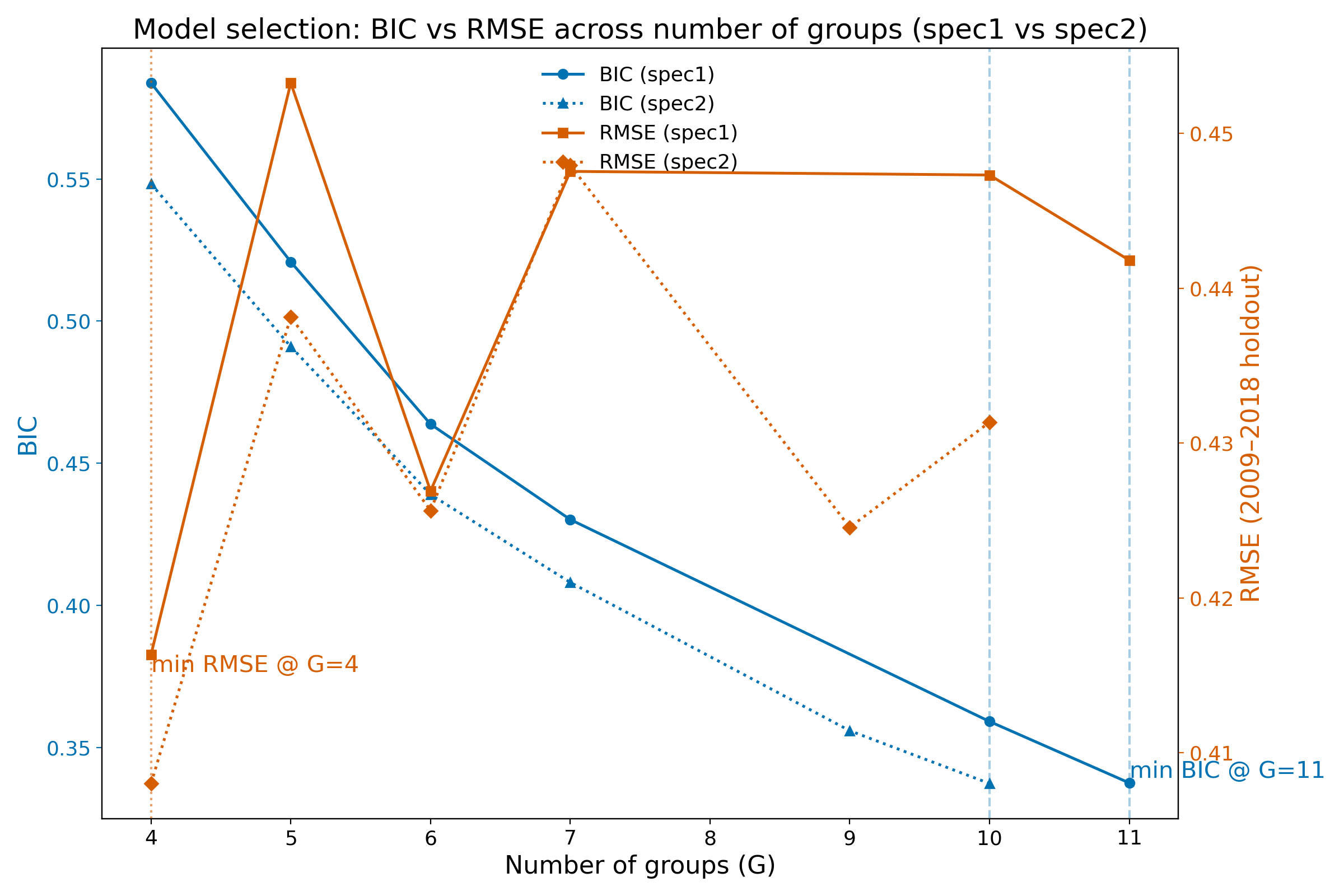}
    \end{center}
    \footnotesize
    {\textit{Source:} Authors’ calculations using ENAHO 2007–2019, restricted to households observed in at least three survey years and to household heads aged 25–55 in all observed years. \\ 
    \textit{Note:} The figure plots model selection diagnostics for the GFE estimator across candidate numbers of groups $G$ under two covariate specifications (spec1 and spec2) with $\texttt{n\_starts}=10$, $\texttt{itermax}=10$, $\texttt{neighmax}=5$, and $\texttt{max\_local\_iters}=2$. BIC is computed on the training data; RMSE is computed out-of-sample using the last observed household-year held out for each household (years 2009–2018) test data. Lower values indicate better fit/prediction.}
\end{figure}

Figure~\ref{fig:Fig_bic_vs_rmse_spec1_vs_spec2_selectedG_nstarts10} summarizes the resulting BIC and RMSE for these re-estimated candidate values of $G$ with $\texttt{n\_starts}=10$. We find that, firstly, the ranking by RMSE is broadly stable. The values of $G$ that performed well in the initial run continues to perform well when we increase the number of starts, and the location of the minimum RMSE remains at a small-to-moderate number of groups rather than shifting towards very large $G$. Secondly, the gap in RMSE among the top candidates is relatively small, indicating that predictive performance is fairly flat once $G$ is in the neighborhood of the RMSE minimum. This pattern reinforces the interpretation that beyond a modest number of groups, additional flexibility primarily increases variance (through thinner support for some group-year cells) rather than improving generalization. 

Based on these results, we select $G=4$ as our preferred baseline choice for both covariates specifications. This value achieves the lowest (or near-lowest) out-of-sample RMSE while avoiding the instability concerns that arise when the model is forced to estimate many group-time effects from very small numbers of observations. 

Our results on model fit across different values of $G$ and specifications point to two practical takeaways. First, moving from specification 1 to specification 2 by adding the extra household-level controls tends to improve fit and forecasting performance at the same $G$: BIC is lower and the test RMSE is generally smaller in specification 2. This suggests that additional covariates capture meaningful variation that would otherwise be absorbed by the latent groups, although the reduction of RMSE by adding additional covariates in our specification is quite small. 

Second, re-running the best-performing set of $G$ values with a larger number of random starts produces almost identical RMSE across most cases, and only modest improvements in a few of them. This pattern indicates that the estimation is generally not being driven by unlucky initializations: with $\texttt{n\_starts}=3$, the algorithm already tends to find solutions with very similar out-of-sample performance, and increasing $\texttt{n\_starts}$ mainly serves as a robustness check against local minima rather than changing the substantive ranking of candidate $G$ values. 
Taken together, these results support our approach of using a low $\texttt{n\_starts}$ grid search to narrow down promising values of $G$, followed by a higher $\texttt{n\_starts}$ rerun on the shortlist to confirm stability before selecting our optimal G.\footnote{We report the full table for BIC and RMSE by specification and the number of random starts in Appendix Table~\ref{tab:Table_main_model_selection_BIC_RMSE}. See Column ``(8)-(6)'' for the first point, and column ``Diff (ns10-ns3)'' for the second point.}

In the remainder of the analysis, unless otherwise noted, we report results using the model estimated at $G=4$ with $\texttt{n\_starts}=10$, and we use the corresponding group assignments, group-time paths, and location effects as the basis for constructing predicted welfare trajectories and poverty transitions. 

\subsection{Welfare Trajectories at Selected G}
\label{sec:welfare_trajectory}

\begin{figure}[!htbp]
    \begin{center}
        \caption{Estimated group--year intercept paths $\hat{\alpha}_{gt}$ for $G=4$}
        \label{fig:Fig_alpha_paths_G4_overlay}
        \includegraphics[width=0.90\linewidth]{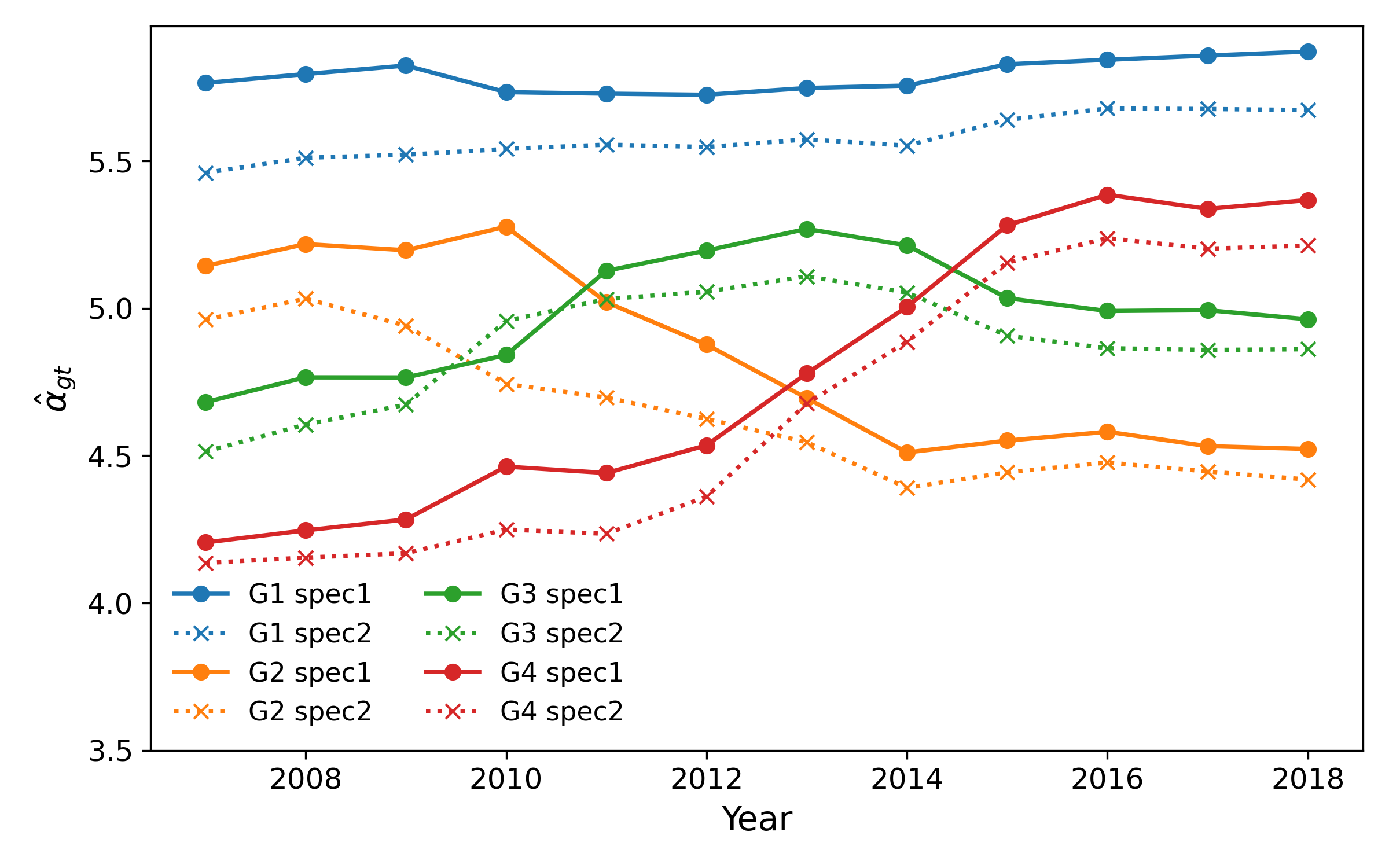}
    \end{center}
    \footnotesize
    {\textit{Data:} Estimated group-year intercept paths for G=4 using training data. \\\textit{Note:} Colors indicate groups. Solid lines are specification 1 and dotted lines are specification 2. $\hat{\alpha}_{gt}$ is the estimated group$\times$year intercept from the GFE model (net of covariates). Interpretation should focuses on relative differences and dynamics across groups rather than the value. }
\end{figure}

Figure~\ref{fig:Fig_alpha_paths_G4_overlay} shows the estimated group-year intercepts $\hat{\alpha}_{gt}$ for our selected model with $G=4$, overlaying specification 1 (solid lines) and specification 2 (dotted lines). These intercept paths summarize the evolution of latent welfare components after netting out the contribution of observed covariates and the province fixed effects, so difference in $\hat{\alpha}_{gt}$ reflect time-varying unobserved heterogeneity that is common to households assigned to the same latent group. 

As shown in the figure, we can tell that it reveals four distinct welfare trajectories. Group 1 is consistently at the top of the distribution throughout the sample with only modest movements over time, suggesting a persistently better-off latent type. Group 3 experiences an increase in the early years followed by a gradual decline after year 2013, while Group 2 is comparatively stable and keep declining in the later years. In contract, Group 4 shows the most pronounced upward movement. The similarities in welfare trajectories between the two specifications indicate that adding the richer set of controls does not significantly alter the overall pattern of latent welfare dynamics captured by the GFE. 

\begin{figure}[!htbp]
    \begin{center}
        \caption{Poverty rate trends by GFE group (G=4): Specification 1 vs. Specification 2.}
        \label{fig:Fig_poverty_rate_G4_overlay}
        \includegraphics[width=0.90\linewidth]{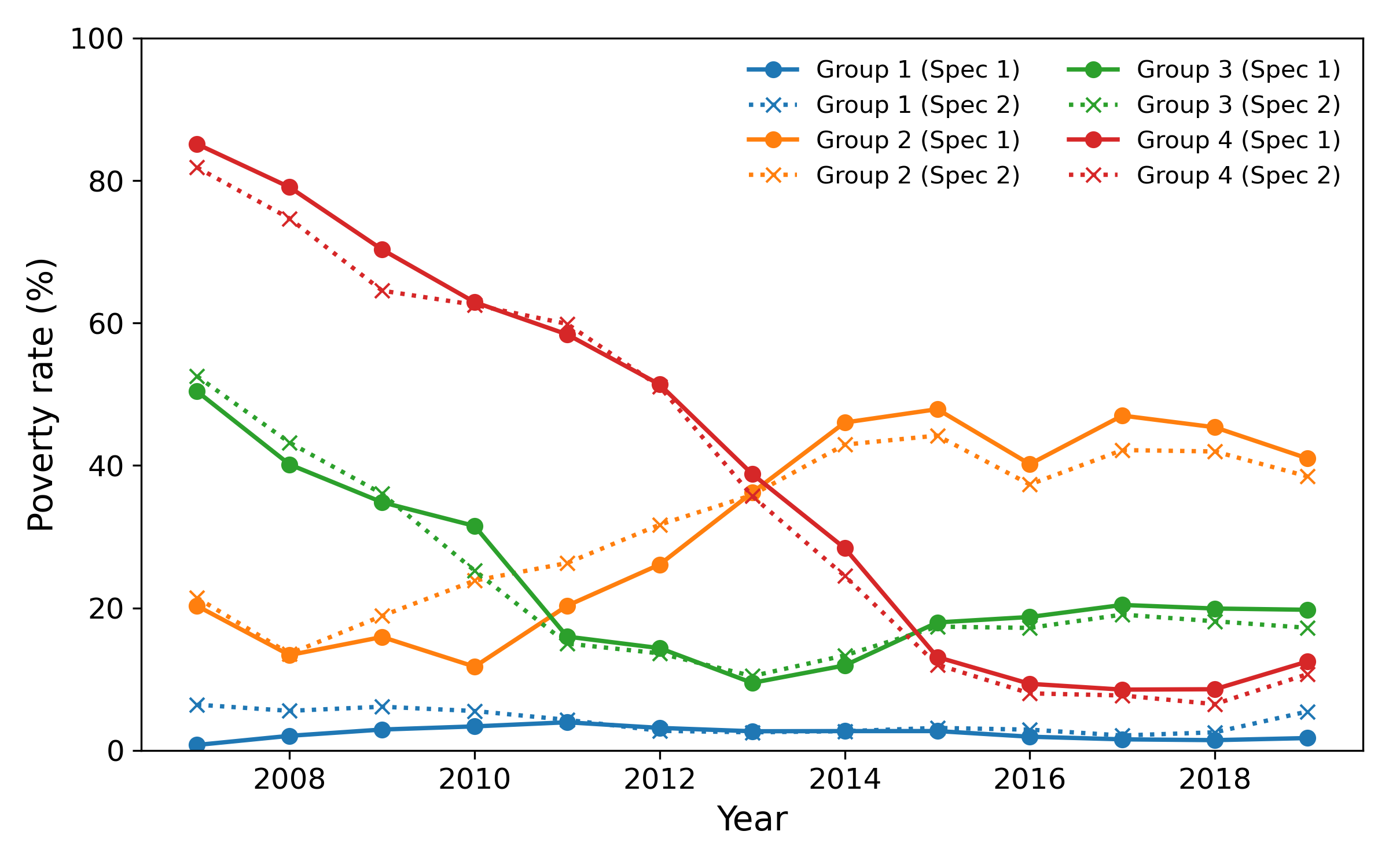}
    \end{center}
    \footnotesize
    {\textit{Data:} All available selected sample (training + test). \\ \textit{Note:} Values are survey-weighted poverty rates by group-year (ENAHO expansion weights). Solid lines are for specification 1 and dotted lines are for specification 2. Colors denote the same group across specifications. Groups use household-level GFE assignments from the selected G=4 model.}
\end{figure}

To complement Figure~\ref{fig:Fig_alpha_paths_G4_overlay}, Figure~\ref{fig:Fig_poverty_rate_G4_overlay} translates the latent welfare trajectories into intuitive welfare metrics: the share of households living below the poverty line. For each year $t$ and group $g$, we compute the survey-weighted poverty rate among household-year observations assigned to group $g$ under the selected $G=4$, overlaying specification 1 (solid lines) and specification 2 (dotted lines). As we can tell from these two figures, the resulting patterns of Figure \ref{fig:Fig_poverty_rate_G4_overlay} closely track the estimated $\hat{\alpha}_{gt}$ paths in the opposite direction, consistent with the interpretation of $\hat{\alpha}_{gt}$ as a latent component of welfare. Group 1, which is the consistently high $\hat{\alpha}_{gt}$ group, exhibits the lowest poverty rates throughout the sample and remains largely stable over time. Group 3, which shows an early improvement of $\hat{\alpha}_{gt}$ followed by a post-2013 declines, similarly begins with very high poverty rates that falls in the early years and then slowly increase the poverty rate in the later period. Group 2 displays initial relatively low poverty levels with increase of poverty rate later, while Group 4 shows the most significant improvement over time: poverty declines a lot as the $\hat{\alpha}_{gt}$ rises. As the intercept paths, the poverty trajectories are highly similar across the two specifications, indicating that the added households controls in specification 2 do not materially change the qualitative welfare ordering or the timing of improvements across latent groups.

This close similarity between $\hat{\alpha}_{gt}$ dynamics and poverty rates helps us validate that the selected $G=4$ grouping captures economically meaningful differences in welfare trajectories rather than purely statistical variation. This is because in our current specifications, if we treat our covariates as time-invariant, and both the coefficient vector $\hat{\theta}$ and the location effect $\hat{\mu}$ are time-constant, then the only flexible component that can capture systematic changes in welfare over time within each latent type is the group--year intercept $\hat{\alpha}_{gt}$, making its dynamics a direct summary of the estimated welfare trajectory for group $g$.

\subsection{Model Validation at the Aggregate Level}
\label{sec:model_validation_agglevel}

\begin{figure}[!htbp]
    \begin{center}
        \caption{Actual vs. predicted poverty rate over time in the test data (G=4)}
        \label{fig:Fig_poverty_actual_vs_pred_df_test_two_panels_spec1_spec2_t}
        \includegraphics[width=\linewidth]{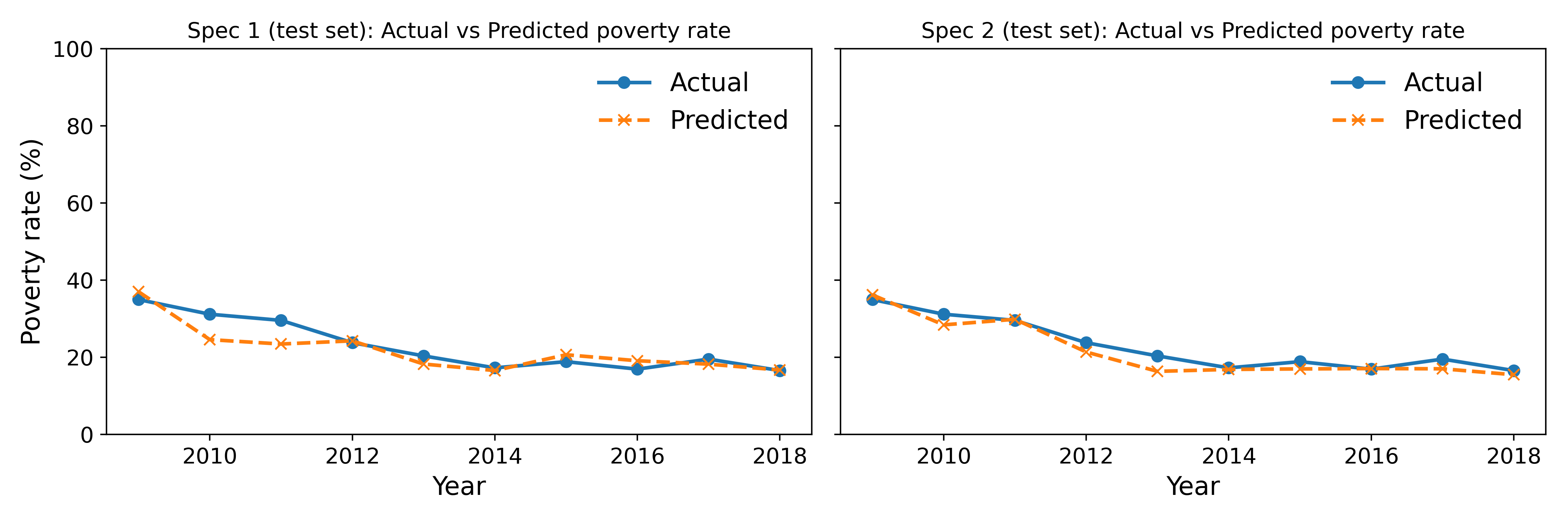}
    \end{center}
    \footnotesize
    {\textit{Data:} Selected sample, test data only. \\ \textit{Note:} Poverty rates are survey-weighted (ENAHO expansion weights). The test data consists of each household’s last observed survey year. Predicted poverty is based on model-predicted welfare from the GFE model and the IHS poverty line (both in IHS scale).}
\end{figure}

Before turning to poverty transitions, we assess whether the GFE model reproduces the overall time pattern of poverty at the aggregate level when evaluated out of sample. Figure~\ref{fig:Fig_poverty_actual_vs_pred_df_test_two_panels_spec1_spec2_t} compares the survey-weighted poverty rate computed from observed expenditures to the poverty rate implied by GFE-predicted welfare in the test data, where each household contributes its last observed survey year.\footnote{Predicted poverty is defined as $\hat{P}_{i,t_i^{\text{last}}}=1\{\hat{y}_{i,t_i^{\text{last}}}<z_{t_i^{\text{last}}}\}$, with $\hat{y}_{it}=x'_{it}\hat{\theta}+\hat{\alpha}_{\hat{g}_i,t}+\hat{\mu}_{p_i}$, and all year-level rates are aggregated using ENAHO expansion weights.}
In both specifications, the predicted poverty rates are very close to the observed poverty rate across years, capturing the broad decline in poverty over the study period and remaining near the realized poverty rate in most years. This result indicates that the estimated group--year components, together with observables and location fixed effects, capture the main time variation in poverty at the population level even when predictions are formed on held-out household-years (test data).

As a reference, Appendix Figure~\ref{fig:Fig_poverty_actual_vs_pred_df_all_two_panels_spec1_spec2_t} repeats the same comparison using all available observations (training plus test) and therefore shows an almost perfect overlap between actual and predicted poverty rates, as expected for an in-sample fit.

\subsection{Poverty Transitions: One-step-ahead Validation}
\label{sec:one-step-ahead-pred}

We next examine whether the GFE model can reproduce observed patterns of poverty transitions (poor-poor, poor-non-poor, non-poor-poor, and non-poor-non-poor). To provide an out-of-sample check that is feasible in our setting, we implement a one-step-ahead validation that mirrors how the model would be used in practice. For each household, we take its final observed survey year $t$ as a held-out outcome and use the observed poverty status in the preceding year $t-1$ together with the GFE-predicted welfare in year $t$ to construct predicted transitions. Specifically, we compare the realized transition based on $(P_{i,t-1},P_{it})$ to the predicted transition $(P_{i,t-1},\hat P_{it})$, where $\hat P_{it}=1\{\hat y_{it}<z_t\}$. This design avoids using the household’s realized welfare in the target year when evaluating transition patterns, while still allowing us to assess whether the model accurately captures changes in poverty status from one year to the next.\footnote{We also validated our results using all available household-year observations for which the model-implied welfare $\hat y_{it}$ can be constructed. Figures~\ref{fig:Fig_pov_transitions_actual_vs_pred_spec1_G4} and \ref{fig:Fig_pov_transitions_actual_vs_pred_spec2_G4} show the results. Because these figures draw on the same observations used to estimate the model (and therefore include substantial training-sample information), they should be interpreted as showing that the estimated GFE structure delivers a transition pattern that is consistent with the data in sample, rather than as evidence of forecasting performance.}

\begin{figure}[!htbp]
    \begin{center}
        \caption{One-step-ahead poverty transition shares into held-out final observations: actual vs.\ GFE-predicted ($G=4$), for specification 1.}
        \label{fig:Fig_pov_transitions_onestep_spec1_G4}
        \includegraphics[width=\linewidth]{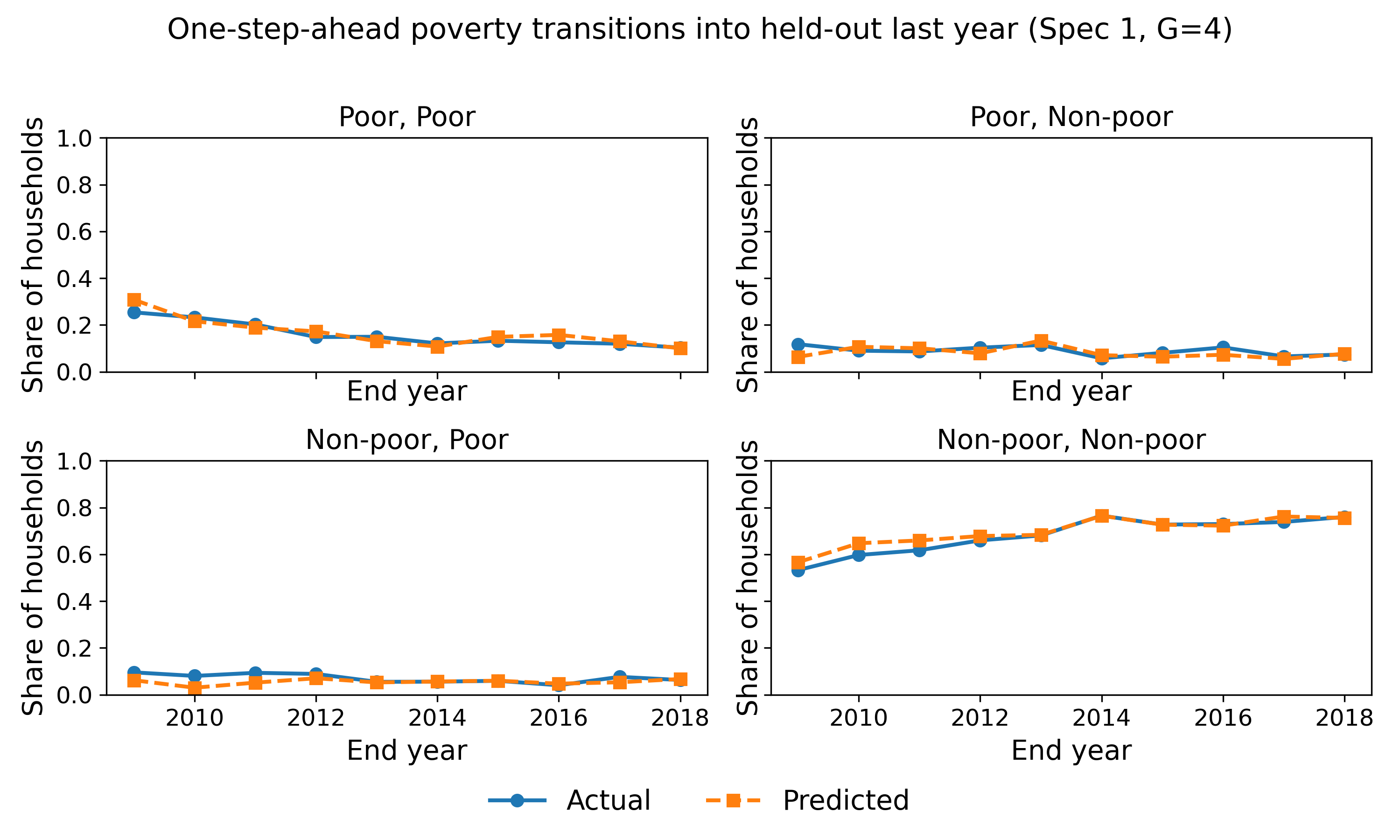}
    \end{center}
    \footnotesize
    {\textit{Data:} Last year of training data + all test data. \\ \textit{Note:} The figure reports survey-weighted shares of households transitioning between poverty states from $t-1$ to $t$, where $t$ is each household's held-out last observed year and $t-1$ is its preceding observed year (restricted to consecutive years; one transition per household). ``Actual'' transitions are computed using observed (IHS-transformed) per-capita expenditure relative to the (IHS-transformed) poverty line in both years. ``GFE-predicted'' transitions use actual poverty at $t-1$ and predicted poverty at $t$, where predicted poverty is $1\{\hat y_{it}<z_{it}\}$ based on model-predicted welfare $\hat y_{it}=x'_{it}\hat\theta+\hat\alpha_{\hat g_i,t}+\hat\mu_{p_i}$. Transition shares are weighted using ENAHO expansion weights in year $t$. In this one-step-ahead sample, the expansion-weighted misclassification rate of $\widehat{\mathrm{poor}}_{it}$ in year $t$ is 17.6\%.}
\end{figure}

Figure~\ref{fig:Fig_pov_transitions_onestep_spec1_G4} reports the results of our one-step-ahead exercise for Specification 1 (For Specification 2, see Appendix Figure~\ref{fig:Fig_pov_transitions_onestep_spec2_G4}). Each point corresponds to an end year $t$ and summarizes the distribution of two-year transitions among households whose final observed survey year is $t$ and whose previous observed year is $t-1$ (restricting to consecutive pairs). The solid line in each panel shows the observed transition share computed from $(P_{i,t-1},P_{it})$, while the dashed line replaces the end-year status with the model-implied classification, $(P_{i,t-1},\hat P_{it})$. Across the four panels, the predicted series closely tracks the observed series, indicating that the GFE model reproduces not only the overall level of persistence in poverty (poor--poor and non-poor--non-poor), but also the relative frequency of entries into poverty (non-poor--poor) and exits from poverty (poor--non-poor) in held-out end years. This provides an out-of-sample check on the model’s ability to capture short-run poverty dynamics, using information available in real time (poverty status in $t-1$ and predicted welfare in $t$). Overall, this one-step-ahead validation model obtains an expansion-weighted poverty-status prediction accuracy of 83.3\% / 83.1\% in the held-out year for Specification 1 and 2 respectively.

\subsection{Comparing GFE to Synthetic Panels}
\label{sec:GFE_vs_synthetic_panel}

\begin{figure}[!htbp]
    \begin{center}
        \caption{Poverty transition shares: actual vs. synthetic panel predictions}
        \label{fig:Fig_pov_transitions_synthpanel_actual_vs_pred_spec1}
        \includegraphics[width=\linewidth]{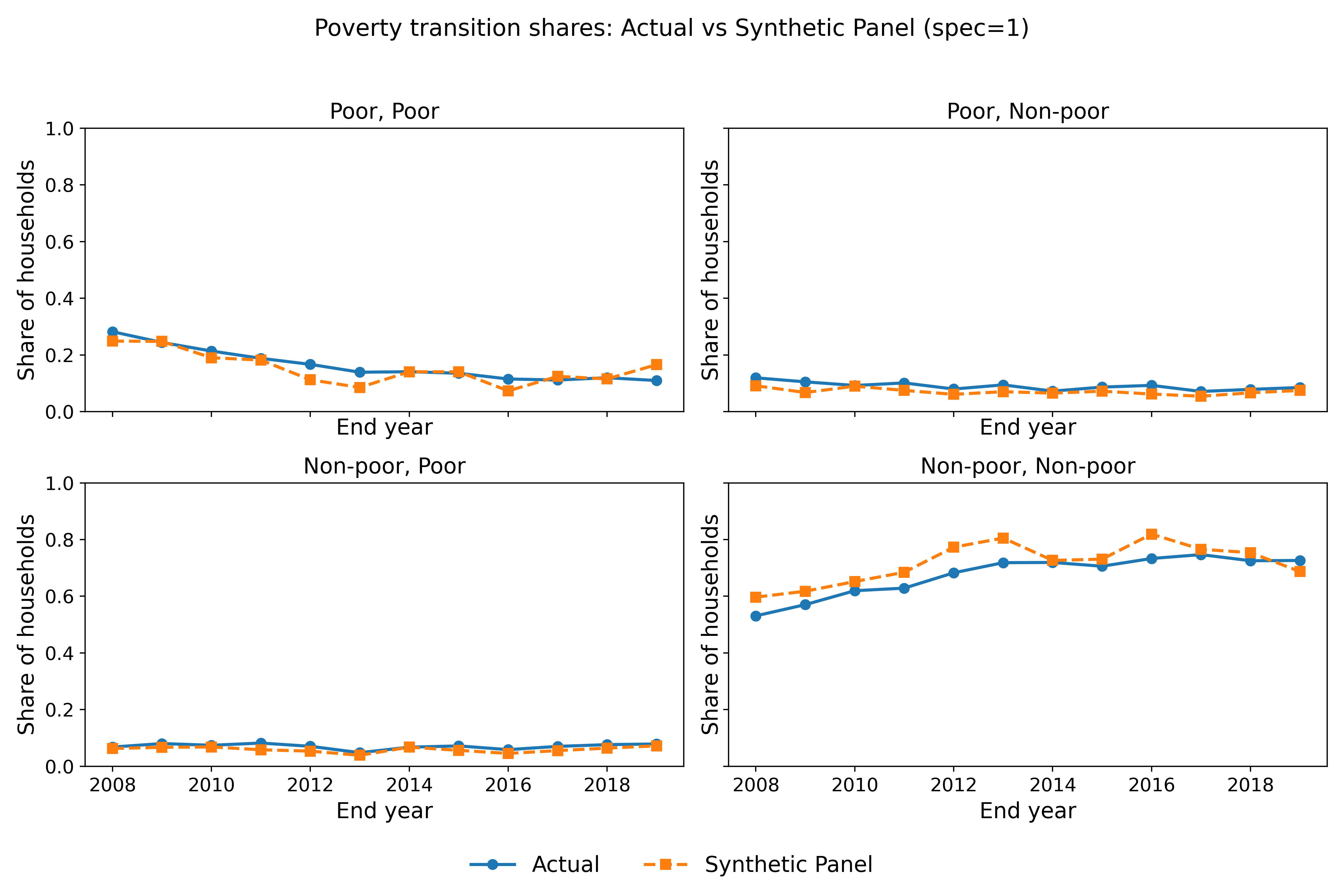}
    \end{center}
    \footnotesize
    {\textit{Note:} Each panel plots the share of households in one of four poverty transition states between $t-1$ and $t$: poor--poor, poor--non-poor, non-poor--poor, and non-poor--non-poor. The solid line (``Actual'') reports transitions computed directly from observed survey data using per-capita expenditure relative to the poverty line in both years. The dashed line reports the corresponding transition shares predicted by the synthetic panel method, which is estimated using repeated cross-sections rather than true panel links. The horizontal axis uses the end year $t$ for each two-year pair. Because of the set up and use of repeated cross-sections rather than true panel, the actual and prediction poverty transition can go up to $t = 2019$. The sample only use household head between 25-55 years old as well, the same as our sample.}
\end{figure}

The most well-known approach to estimate poverty dynamics is the synthetic panel method proposed by \citet{dang_using_2014} and \citet{dang2023measuring}, which use repeated cross-sections to recover poverty transition rates. To benchmark our results against their approach, we replicate the synthetic panel point estimates using the same ENAHO data and plot the four transition shares over time: poor--poor, poor--non-poor, non-poor--poor, and non-poor--non-poor. Figure~\ref{fig:Fig_pov_transitions_synthpanel_actual_vs_pred_spec1} compares the synthetic panel point estimates to the corresponding transition shares computed directly from the data for Specification 1 (see Appendix Figure~\ref{fig:Fig_pov_transitions_synthpanel_actual_vs_pred_spec2} for Specification 2 comparison). The synthetic panel method tracks the broad time patterns in the transition shares reasonably well, but it also exhibits noticeable deviations in several years, especially for the non-poor--non-poor share, where small level differences in predicted transitions can cumulate into visible gaps.

\begin{figure}[!htbp]
    \begin{center}
        \caption{Prediction error in poverty transition shares: GFE one-step versus synthetic panel.}
        \label{fig:Fig_transition_fit_compare_RMSE_yearly_spec1}
        \includegraphics[width=\linewidth]{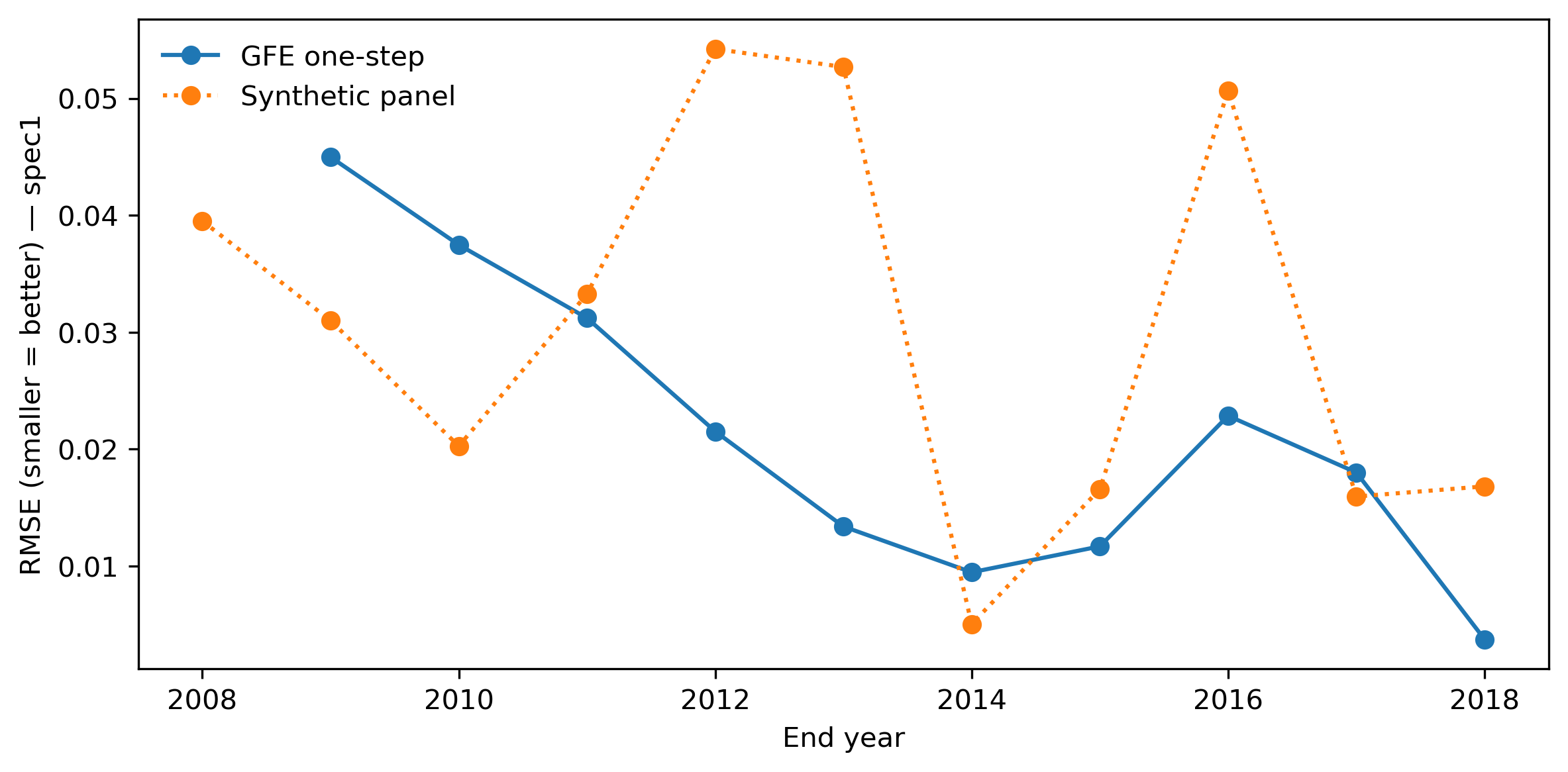}
    \end{center}
    \footnotesize
    {\textit{Note:} The figure plots the root mean squared error (RMSE) between predicted and observed two-year poverty transition shares for each end year $t$. For a given end year, the RMSE is computed across the four transition states (poor--poor, poor--non-poor, non-poor--poor, non-poor--non-poor), comparing the predicted transition-share vector to the observed transition-share vector; smaller values indicate closer fit. The solid line shows the GFE one-step approach, which uses observed poverty status in $t\!-\!1$ and predicts poverty status in $t$. The dotted line shows the synthetic panel approach, which produces transition-share predictions from repeated cross-sections. Because the underlying samples used to compute the transition shares may differ across approaches, the figure is intended as a descriptive comparison of fit within each method. Additionally, the GFE one-step RMSE begins in 2009 instead of 2008 because the one-step-ahead validation requires observing each household in both $t-1$ and its held-out last observed year $t$. Under our analysis sample restriction (at least three observed survey years per household), no household has $t=2008$ as its last observed year.}
\end{figure}

We then evaluate how closely each method reproduces observed transition shares using a simple error metric. For each end year $t$, we treat the four transition shares as a probability distribution and compute the total variation distance between predicted and observed shares per each share $s$, $TV_t = \tfrac{1}{2}\sum_s |\hat p_{s,t} - p_{s,t}|$, along with Mean Absolute Error (MAE) and RMSE averaged across the four transition states. Table~\ref{tab:transition_fit_compare} summarizes these metrics for (i) the synthetic panel approach and (ii) our GFE one-step-ahead prediction exercise we had in earlier Section \ref{sec:one-step-ahead-pred}, where we use observed poverty status in $t-1$ and predict poverty in $t$. GFE one-step yields smaller errors ($RMSE \approx 0.021$ for Specification 1 and $RMSE \approx 0.020$ for Specification 2) than the synthetic panel estimates ($RMSE \approx 0.031$ for Specification 1 and $RMSE \approx 0.028$ for Specification 2), although the difference is modest. Figure~\ref{fig:Fig_transition_fit_compare_RMSE_yearly_spec1} plots the end-year RMSE in the four transition shares for the GFE one-step approach and the synthetic panel point estimates under Specification 1. The relative performance is not driven by a single year: across end years, the two methods exhibits errors of comparable magnitude, with the GFE one-step RMSE being generally smaller than the synthetic panel approach in several years. While synthetic-panel-based prediction performs slightly better than the GFE-based prediction in some years, there are some years where the former performs poorly, consistent with occasional deviation shown in Figure~\ref{fig:Fig_pov_transitions_synthpanel_actual_vs_pred_spec1}. Meanwhile, GFE-based predictions are stable across years. For Specification 2, Figure~\ref{fig:Fig_transition_fit_compare_RMSE_yearly_spec2} shows a qualitatively similar pattern. Note that the GFE one-step RMSE begins in 2009 rather than 2008 because the one-step-ahead validation requires observing households in both $t-1$ and their held-out last-observed year $t$.

Beyond these fit comparisons, the two approaches are different in what they can deliver for more analysis. The synthetic panel method is designed to recover transition shares from repeated cross-sections, but it does not assign households to latent types and does not produce a full set of predicted welfare outcomes at the household-year level. By contrast, GFE estimates group-specific welfare components $\hat{\alpha}_{g,t}$ and assigns each household to a latent group, which allows us to summarize heterogeneity in welfare dynamics and to construct completed welfare paths for households even when some survey years are missing (under explicit assumptions about how covariates evolve between survey rounds). For researchers that plan and would benefit from describing heterogeneous welfare trajectories and using the estimated latent structure to fill missing outcomes, GFE provides additional structure that is not available in the synthetic panel framework, while achieving comparable (even better) accuracy in reproducing aggregate poverty transitions. In the next Section \ref{sec:gfe_fill_all_missing}, we show the results of using GFE to fill all missing years information in the rotating panel data.

\input{GFE/tables/main/Table_transition_fit_compare}

\subsection{GFE for Filling All Missing Years Information}
\label{sec:gfe_fill_all_missing}
One advantage for the GFE is that we can predict all the missing year's results using the group-year $\alpha_{gt}$ as long as we find ways to fill the missing covariates information. We discussed about the ways that researchers can fill the missing covariates information in Section \ref{sec:from_estimateG_to_poverty_dynamics}. 

To study welfare dynamics over time, we construct a completed outcome that fills in years when a household is not observed in the survey. For each household $i$ and year $t$ from 2007--2018, we set $y^{\mathrm{comp}}_{it}$ equal to the observed welfare measure $y_{it}$ whenever it is available (i.e., $d_{it}=1$). When $y_{it}$ is missing (i.e., $d_{it}=0$), we replace it with the GFE-predicted value $\hat y_{it}=x'_{it}\hat\theta+\hat{\alpha}_{\hat g_i,t}+\hat{\mu}_{p_i}$, which combines (i) the contribution of observed household characteristics, (ii) the estimated group--year component for the household's assigned group, and (iii) province fixed effect estimates. 

To compute $\hat y_{it}$ in non-survey years, we also need values of the covariates $x_{it}$ in those years. There is no single ``correct'' way to do this, and researchers should choose an imputation rule that matches their setting and the variables they use. For example, characteristics that are truly fixed (such as gender or language) can be carried forward and backward within household, while other variables may reasonably evolve over time (such as age) and can be updated mechanically. In richer specifications, some covariates may change in ways that are not mechanical (such as education, marital status, or assets), and different assumptions---e.g., holding them constant, carrying forward the last observed value, interpolating between observed values, or using external information---will change the predicted component $x'_{it}\hat\theta$ and therefore can shift the level and shape of the group-average paths.

To make this possible in non-survey years in our setting, we fill in missing household characteristics using simple rules: characteristics that are effectively time-invariant in both specifications are carried forward/backward within the same household, and age is updated when missing so that it moves with the calendar year. 

\begin{figure}[!htbp]
    \begin{center}
        \caption{Mean completed welfare by GFE group over time ($G=4$), overlaying specifications.}
        \label{fig:Fig_mean_y_completed_by_group_G4_overlay}
        \includegraphics[width=\linewidth]{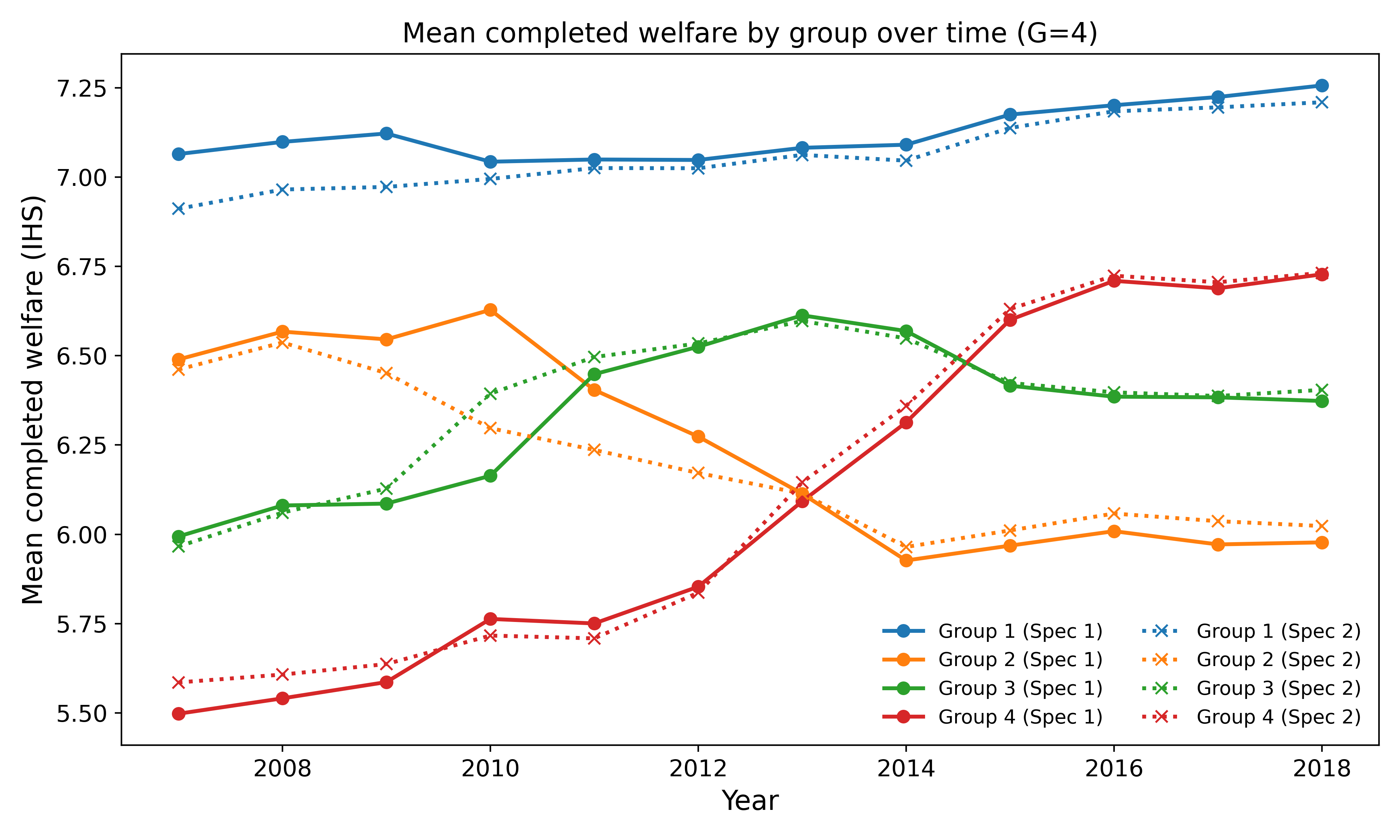}
    \end{center}
    \footnotesize
    {\textit{Data:} All available selected sample (training + test). \\ \textit{Note:} The figure plots the mean of the completed welfare measure $y^{\mathrm{comp}}_{it}$ by latent group $g$ and year $t$ for 2007--2018. The completed outcome equals observed welfare when available and is imputed using the GFE prediction when missing: $y^{\mathrm{comp}}_{it}=y_{it}$ if observed and $y^{\mathrm{comp}}_{it}=\hat y_{it}$ otherwise, where $\hat y_{it}=x'_{it}\hat\theta+\hat{\alpha}_{\hat g_i,t}+\hat{\mu}_{p_i}$. Solid lines correspond to Specification~1 and dotted lines correspond to Specification~2; colors identify groups. Imputation for non-survey years relies on within-household filling rules for covariates (time-invariant characteristics are carried forward/backward within household; age is updated by one year when missing. The welfare measure that we use is the IHS-transformed household expenditure per capita.}
\end{figure}

Figure~\ref{fig:Fig_mean_y_completed_by_group_G4_overlay} shows the average $y^{\mathrm{comp}}_{it}$ by group and year, overlaying Specification 1 and 2. The completed series shows us very clear and stable differences across groups: Group~1 remains the highest throughout the period, Group~2 keep declining in the later years, Group~3 shows an increase in the early years followed by a gradual decline after year 2013, and Group~4 exhibits the largest improvement. The patterns are very similar to our estimated group-year $\alpha_{gt}$ path as shown in Figure \ref{fig:Fig_alpha_paths_G4_overlay}.

\subsection{Group Composition by Baseline Characteristics}
\label{sec:group-composition}

To understand the baseline profile by each latent group, we prepared two summary Tables~\ref{tab:Table_g4_baseline_spec1} and \ref{tab:Table_g4_baseline_spec2} to summarize group size and baseline characteristics for the selected $G=4$ model under Specifications 1 and 2, respectively. Baseline characteristics are measured in each household’s first observed survey year, and all statistics are weighted using the ENAHO expansion factors so that the reported shares and means are population-representative. To make the group labels interpretable, we order groups by the mean estimated group effect $\bar{\alpha}_g$ and refer to them as \emph{Top}, \emph{Upper-mid}, \emph{Lower-mid}, and \emph{Bottom}.

The two tables show clear differences in baseline welfare levels across groups. In Specification 1 (Table~\ref{tab:Table_g4_baseline_spec1}), the Top group accounts for about 11\% of the weighted population and has the highest baseline welfare (mean IHS expenditure of 7.32), while the Bottom group accounts for about 23\% and has the lowest baseline welfare (5.61). The middle groups together comprise roughly two thirds of the population, with baseline welfare levels that fall between these extremes (6.58 for the Lower-mid group and 6.09 for the Upper-mid group). The education profile also varies a lot across groups: the share with college education is much higher in the Top group (38.6\%) than in the Bottom group (9.7\%), and the Top group is more likely to report Spanish as the household language (84.9\% versus 67.9\%). In contrast, average age is very similar across groups (around 41 years), suggesting that the group ordering primarily reflects persistent welfare and human-capital differences rather than age composition.

Table~\ref{tab:Table_g4_baseline_spec2} extends this comparison by adding additional household characteristics. The same broad patterns remain: the Top group is smaller (13\% of the population) and has higher baseline welfare (7.15), while the Bottom group is larger (24\%) and has lower baseline welfare (5.71). The expanded covariates also reveal meaningful differences in living conditions. In particular, access to basic services is substantially higher in the Top group than in the Bottom group (water access: 74.6\% vs.\ 58.0\%; electricity: 85.0\% vs.\ 78.7\%). Language differences are also pronounced (Spanish: 85.5\% in the Top group vs.\ 68.7\% in the Bottom group). 

Overall, the baseline profiles in both specifications help clarify which kinds of households are classified into each latent group. The ordering by $\bar{\alpha}_g$ aligns closely with a drop in baseline living standards: households in the Top group enter the sample with higher per-capita expenditure, substantially higher schooling attainment, and a higher likelihood of speaking Spanish, while households in the Bottom group start with lower expenditure, much lower rates of college education, and a lower prevalence of Spanish. In Specification~2, this decline extends to housing and local amenities---the Top group is more likely to have access to water and electricity than the Bottom group. These systematic baseline differences suggest that the GFE classification is not arbitrary: it sorts households into groups that are already meaningfully distinct in observable socioeconomic conditions at baseline. At the same time, the groups are not determined by any single observed characteristic alone. For example, average age is very similar across groups, and the middle groups have comparable age profiles but differ in education and baseline welfare. This reinforces the interpretation that the latent grouping captures broader ``types'' of households defined by a bundle of correlated living-standard indicators, which then map into different latent welfare trajectories over time through the estimated group--year components.

The baseline characteristics by GFE grouping provide researchers very useful information to understand poverty dynamics. It moves the analysis beyond a single snapshot of who is poor in a given year and instead organizes households into a small number of economically meaningful ``types'' that evolve differently over time. When groups reflect a bundle of correlated living-standard indicators (education, language, and basic services), they provide a transparent way to summarize heterogeneity in both the level of welfare and the path of welfare. In practice, this helps researchers distinguish households that are persistently better-off from those that remain structurally disadvantaged, and it also helps identify intermediate groups that may experience upward mobility or vulnerability to downturns. Because the group--year components trace how each type’s latent welfare changes over time, researchers can connect poverty transitions to macroeconomic shocks and policy periods in a more structured way, and can report dynamic patterns (e.g., catch-up, stagnation, or decline) that are difficult to see from year-by-year poverty rates alone. Finally, once households are classified into these types, the framework can be used to describe who is most likely to enter or exit poverty and to target subsequent analysis (and policy discussion) toward the groups whose trajectories suggest persistent deprivation or heightened vulnerability.

\begin{landscape}
    \input{GFE/tables/main/Table_g4_baseline_spec1}
    \input{GFE/tables/main/Table_g4_baseline_spec2}
\end{landscape}

\section{Robustness Check}
\label{sec:robustness_check}

\subsection{No Age Restriction}
\label{sec:r1_NoAgeLimit}

As the first robustness check, we re-estimate the GFE model on a broader sample that does not impose any age restriction on the household head for the ENAHO data. We keep the same rotating-panel requirement used in the main analysis (households observed in at least three survey years) and the same estimation and evaluation set up. We end up having a sample of 104,304 household-year observation from 27,034 unique households. The average age of head of households for the sample is around 52, with a standard deviation of 15 based on the household head's first survey year age. Approximately 31\% of household heads are aged 60 or older. 

\begin{figure}[!htbp]
    \begin{center}
        \caption{BIC vs. RMSE by different sets of covariates (no age restriction)}
        \label{fig_r1:Fig_bic_vs_rmse_spec1_vs_spec2}
        \includegraphics[width=0.95\linewidth]{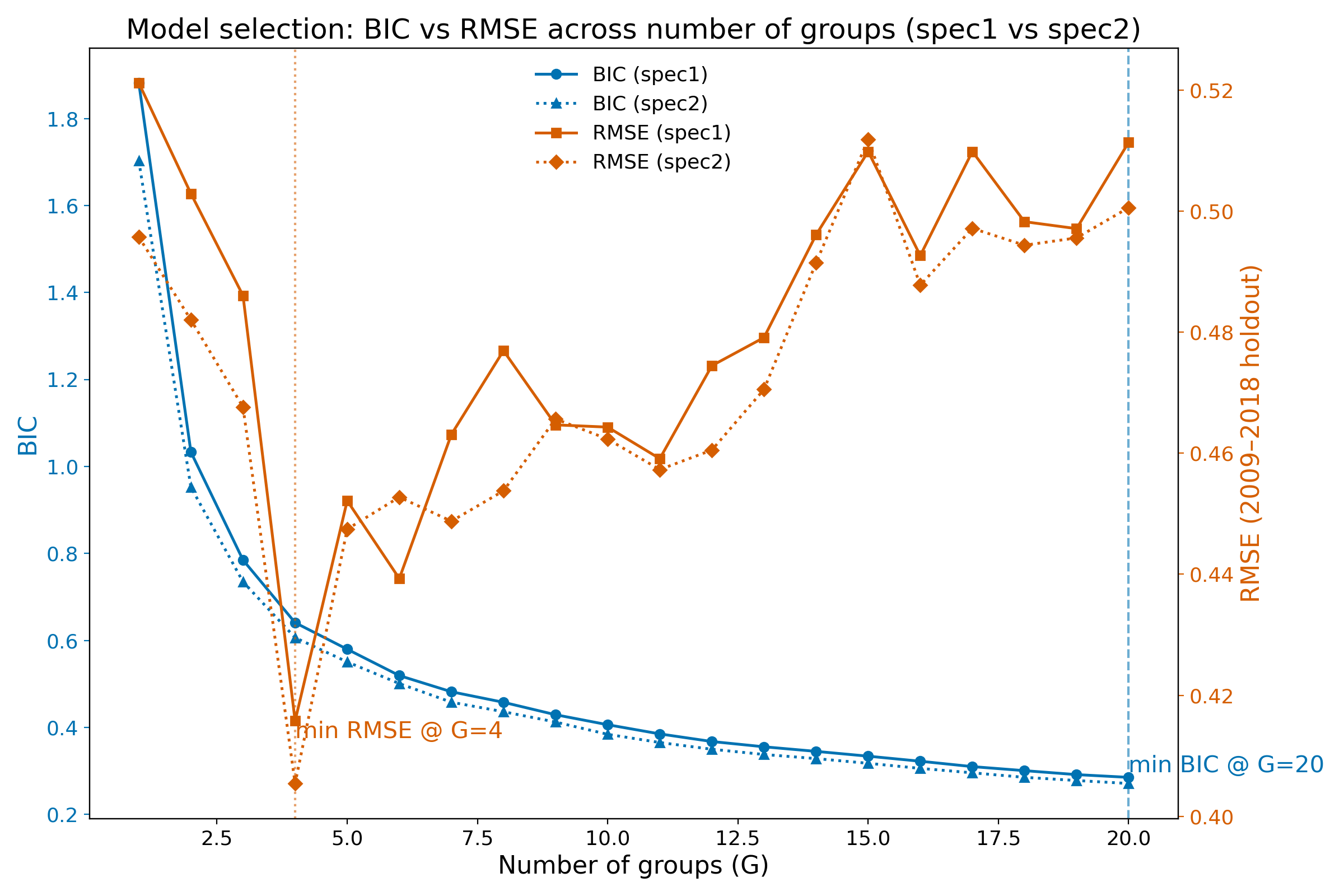}
    \end{center}
    \footnotesize
    {\textit{Source:} Authors’ calculations using ENAHO 2007–2019, restricted to households observed in at least three survey years but does not apply age restrictions for household heads. \\ 
    \textit{Note:} The figure plots model selection diagnostics for the GFE estimator across initial numbers of groups $G\in\{1,\cdots,40\}$ with $\texttt{n\_starts}=3$, $\texttt{itermax}=10$, $\texttt{neighmax}=5$, and $\texttt{max\_local\_iters}=2$ under two covariate specifications (spec1 and spec2). BIC is computed on the training data; RMSE is computed out-of-sample using the last observed household-year held out for each household (years 2009–2018) test data. Lower values indicate better fit/prediction.} \\
\end{figure}

We first run the GFE model with a grid over $G = 1,\dots,20$ with $\texttt{n\_starts}=3$ random initialization for each G. Similarly to Figure~\ref{fig:Fig_bic_vs_rmse_spec1_vs_spec2} that shows the change in the BIC and RMSE values between different latent groups $G$ using two alternative covariate specifications (specification~1 and specification~2), our result with the sample of no age restriction (Figure~\ref{fig_r1:Fig_bic_vs_rmse_spec1_vs_spec2}) continue to favor a small number of latent groups: the holdout RMSE is minimized at $G=4$ under both specifications, while the BIC decreases mechanically with increase of $G$ and thus prefer much larger $G$. 

\begin{figure}[!htbp]
    \begin{center}
        \caption{BIC vs. RMSE by different sets of covariates for selected G (no age restriction)}
        \label{fig_r1:Fig_bic_vs_rmse_spec1_vs_spec2_selectedG_nstarts10}
        \includegraphics[width=0.95\linewidth]{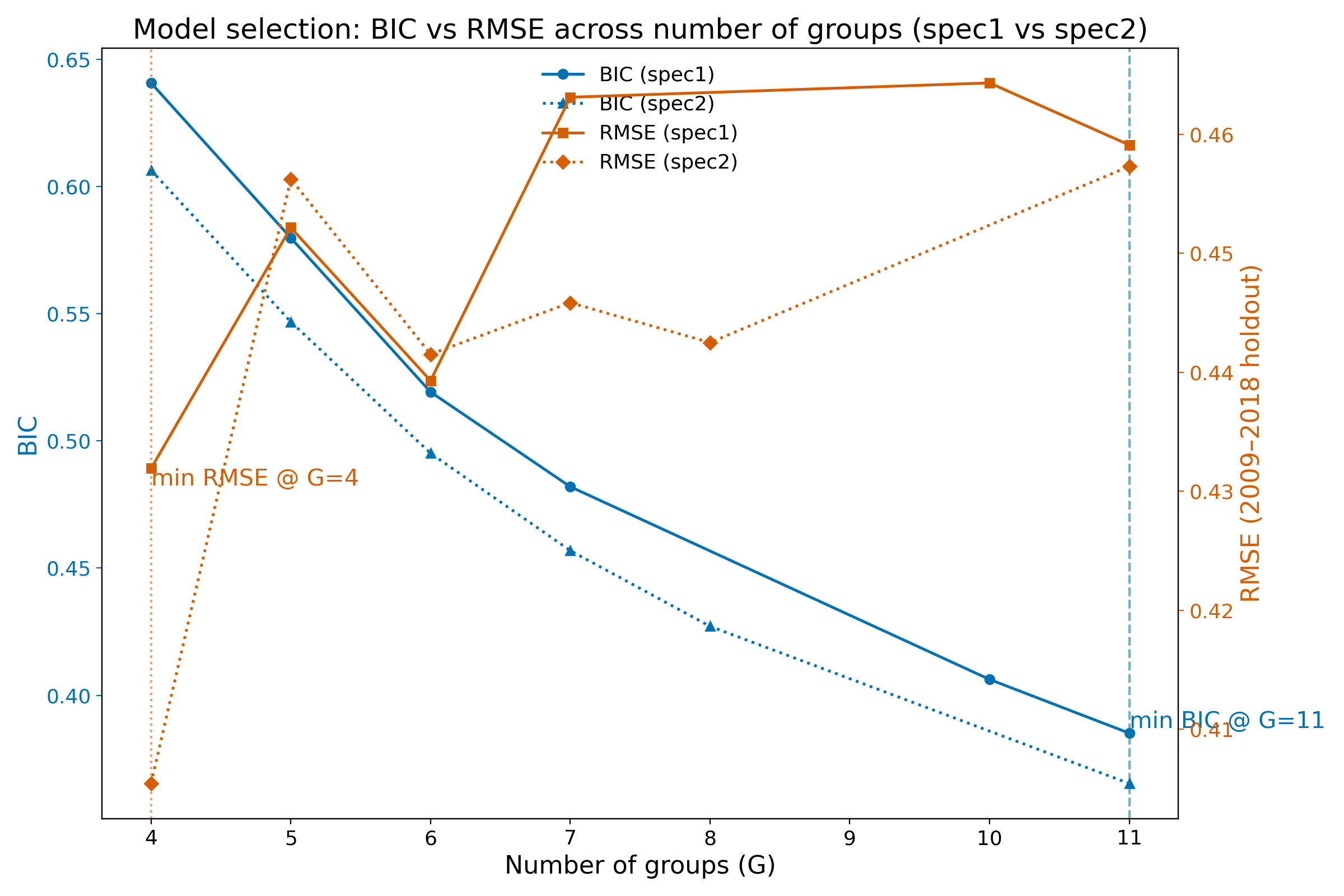}
    \end{center}
    \footnotesize
    {\textit{Source:} Authors’ calculations using ENAHO 2007–2019, restricted to households observed in at least three survey years but does not apply age restrictions for household heads. \\ 
    \textit{Note:} The figure plots model selection diagnostics for the GFE estimator across candidate numbers of groups $G$ under two covariate specifications (spec1 and spec2) with $\texttt{n\_starts}=10$, $\texttt{itermax}=10$, $\texttt{neighmax}=5$, and $\texttt{max\_local\_iters}=2$. BIC is computed on the training data; RMSE is computed out-of-sample using the last observed household-year held out for each household (years 2009–2018) test data. Lower values indicate better fit/prediction.}
\end{figure}

After rerun the model with a higher random initialization ($\texttt{n\_starts}=3$) for the six $Gs$ that have the lowest RMSE, the model still shows that the best $G$ is 4 (Figure~\ref{fig_r1:Fig_bic_vs_rmse_spec1_vs_spec2_selectedG_nstarts10}). Overall, dropping the age restriction does not materially affect the model-selection conclusion: the preferred number of latent groups remains as $G=4$, consistent with the results using the sample in the main analysis. This suggests that the main structural patterns identified by the GFE model are not driven by the head-age restriction. Expanding the sample changes the composition of households but does not change the degree of latent heterogeneity favored by out-of-sample fit.

\begin{figure}[!htbp]
    \begin{center}
        \caption{Estimated group--year intercept paths $\hat{\alpha}_{gt}$ for $G=4$ (no age restriction)}
        \label{fig_r1:Fig_alpha_paths_G4_overlay}
        \includegraphics[width=0.90\linewidth]{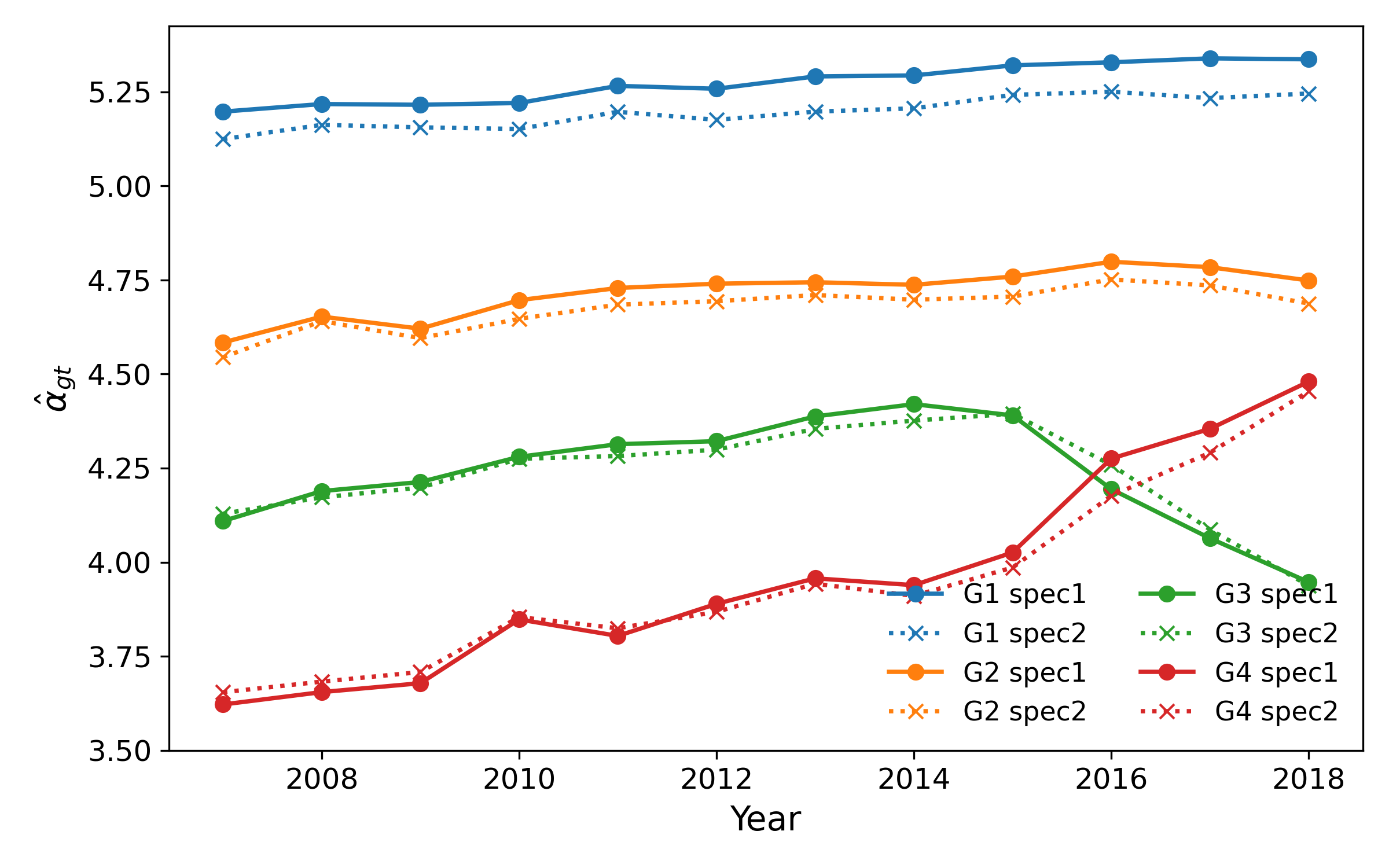}
    \end{center}
    \footnotesize
    {\textit{Data:} Estimated group-year intercept paths for G=4 using training data of sample without age restriction. \\\textit{Note:} Colors indicate groups. Solid lines are specification 1 and dotted lines are specification 2. $\hat{\alpha}_{gt}$ is the estimated group$\times$year intercept from the GFE model (net of covariates). Interpretation should focuses on relative differences and dynamics across groups rather than the value. }
\end{figure}

Figure~\ref{fig_r1:Fig_alpha_paths_G4_overlay} reproduces the Figure~\ref{fig:Fig_alpha_paths_G4_overlay} exercise using the sample without any head-age restriction, plotting the estimated group-year intercepts $\hat{\alpha}_{gt}$ for $G=4$ under Specification 1 and 2. The qualitative welfare dynamics remain very similar to the sample used for the main results. We can tell that there is a clear and persistent separation between the top and bottom groups throughout the period. Importantly, the estimated $\hat{\alpha}_{gt}$ for $G=4$ paths are closer to each other with Specification 1 and 2 (solid vs. dotted lines) than that of the results in the main results (Figure~\ref{fig:Fig_alpha_paths_G4_overlay}). A natural explanation is that removing the age restriction increase the effective sample size and improves support within each group--year cell, which stabilizes the estimation of group-year intercepts and reduces the sensitivity of $\hat{\alpha}_{gt}$ to a particular set of covariates included in $X_{it}$. With more observations per $(g,t)$, the within-cell averages that construct the updated group effects are less noisy, so the additional covariates in Specification 2 primarily reallocate the variation between the $X_{it} \hat{\theta}$ component and the fixed-effect component without changing the implied group-level trajectory.  This finding suggest that when the sample provides sufficiently rich support - especially enough observations in each group-year cell - the specific choice of covariates in the GFE model becomes less critical for the estimated group trajectories. In that case, researchers can often work with a parsimonious set of largely time-invariant covariates, which has the practical advantage of making covariate completion over missing survey years more credible and less assumption intensive. By contrast, when the effective sample size is smaller (or group-year cells are thin), the allocation of variation between the covariate component and the group effects is less tightly pinned down, and the estimated $\hat{\alpha}_{gt}$ paths can become more sensitive to the covariates set.

Table~\ref{tab_r1:transition_fit_compare} summarizes the fit between predicted and observed two-year poverty transition shares. Across both specifications, the GFE one-step approach matches the actual transition distribution more closely than synthetic panel method. Removing the age restriction does not overturn the main conclusion from our main results: our one-step GFE prediction provides a better match to the actual poverty transitions observed in the data than synthetic panel point estimates. Appendix Figure~\ref{fig_r1:Fig_transition_fit_compare_RMSE_yearly} plot the end-year RMSE in the four transition shares for the GFE one-step approach and the synthetic panel point estimates under Specification 1 and 2. Similarly to our conclusion in the main results (Figure~\ref{fig:Fig_transition_fit_compare_RMSE_yearly_spec1} and Figure~\ref{fig:Fig_transition_fit_compare_RMSE_yearly_spec2}), the relative performance is not driven by a single year and the GFE one-step RMSEs are generally smaller than the synthetic panel approach. 

\input{GFE/tables/RobustnessCheck_NoAgeLimit/Table_transition_fit_compare}

Tables~\ref{tab_r1:Table_g4_baseline_spec1} and~\ref{tab_r1:Table_g4_baseline_spec2} replicate the baseline group-profile exercise using the sample without age restriction for Specification 1 and 2, respectively. The resulting group composition and baseline characteristics are consistent with Tables~\ref{tab:Table_g4_baseline_spec1} and~\ref{tab:Table_g4_baseline_spec2} in the main results. The Top group has the highest baseline IHS expenditure per capita and the largest share with college education, while the Bottom group has the lowest baseline welfare and substantially lower educational attainment. The Top group is more likely to have access to water and electricity than the Bottom group in Specification 2. Overall, dropping the age restriction does not change the interpretation of the latent groups as distinct welfare types.

\begin{landscape}
\input{GFE/tables/RobustnessCheck_NoAgeLimit/Table_g4_baseline_spec1}    \input{GFE/tables/RobustnessCheck_NoAgeLimit/Table_g4_baseline_spec2}
\end{landscape}

\subsection{Household Surveyed At Least Five Years}
\label{sec:r1_atleast5yrs}

We also re-estimate the GFE model that apply the age restriction but keep only households surveyed at least five years in the ENAHO data. This restriction gives us a sample of 15,529 household-year observation from 2,946 unique households, which only account for around 7.8 \%  (2,946 / 375,897) of unique households in the sample that we use for our main analysis, a significant sample decreases as compared to the sample used in the main analysis. 

\begin{figure}[!htbp]
    \begin{center}
        \caption{BIC vs. RMSE by different sets of covariates for selected G (households surveyed at least five years)}
        \label{fig_r2:Fig_bic_vs_rmse_spec1_vs_spec2_selectedG_nstarts10}
        \includegraphics[width=0.95\linewidth]{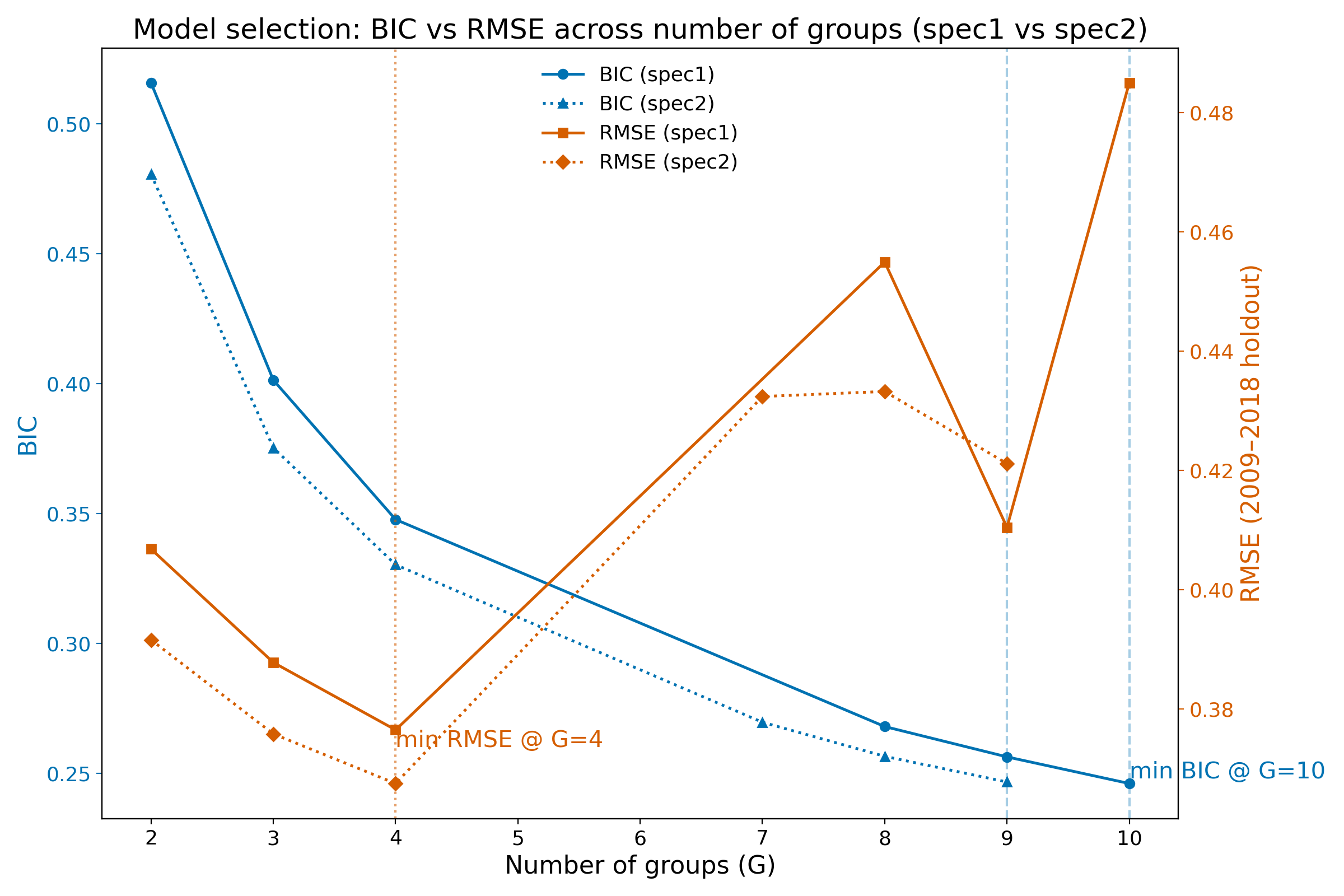}
    \end{center}
    \footnotesize
    {\textit{Source:} Authors’ calculations using ENAHO 2007–2019, restricted to households observed in at least five survey years and to household heads aged 25-55 in all observed years. \\ 
    \textit{Note:} The figure plots model selection diagnostics for the GFE estimator across candidate numbers of groups $G$ under two covariate specifications (spec1 and spec2) with $\texttt{n\_starts}=10$, $\texttt{itermax}=10$, $\texttt{neighmax}=5$, and $\texttt{max\_local\_iters}=2$. BIC is computed on the training data; RMSE is computed out-of-sample using the last observed household-year held out for each household (years 2009–2018) test data. Lower values indicate better fit/prediction.}
\end{figure}

After running the GFE model with a grid over $G = 1,\dots,20$ with $\texttt{n\_starts}=3$ random initialization for each G, and then select the best six Gs that with the lowest RMSEs to re-run the model with $\texttt{n\_starts}=10$, Figure~\ref{fig_r2:Fig_bic_vs_rmse_spec1_vs_spec2_selectedG_nstarts10} shows that the best $G$ with the lowest RMSE is still 4. This result is consistent with our main result. 

\begin{figure}[!htbp]
    \begin{center}
        \caption{Estimated group--year intercept paths $\hat{\alpha}_{gt}$ for $G=4$ (households surveyed at least five years)}
        \label{fig_r2:Fig_alpha_paths_G4_overlay}
        \includegraphics[width=0.90\linewidth]{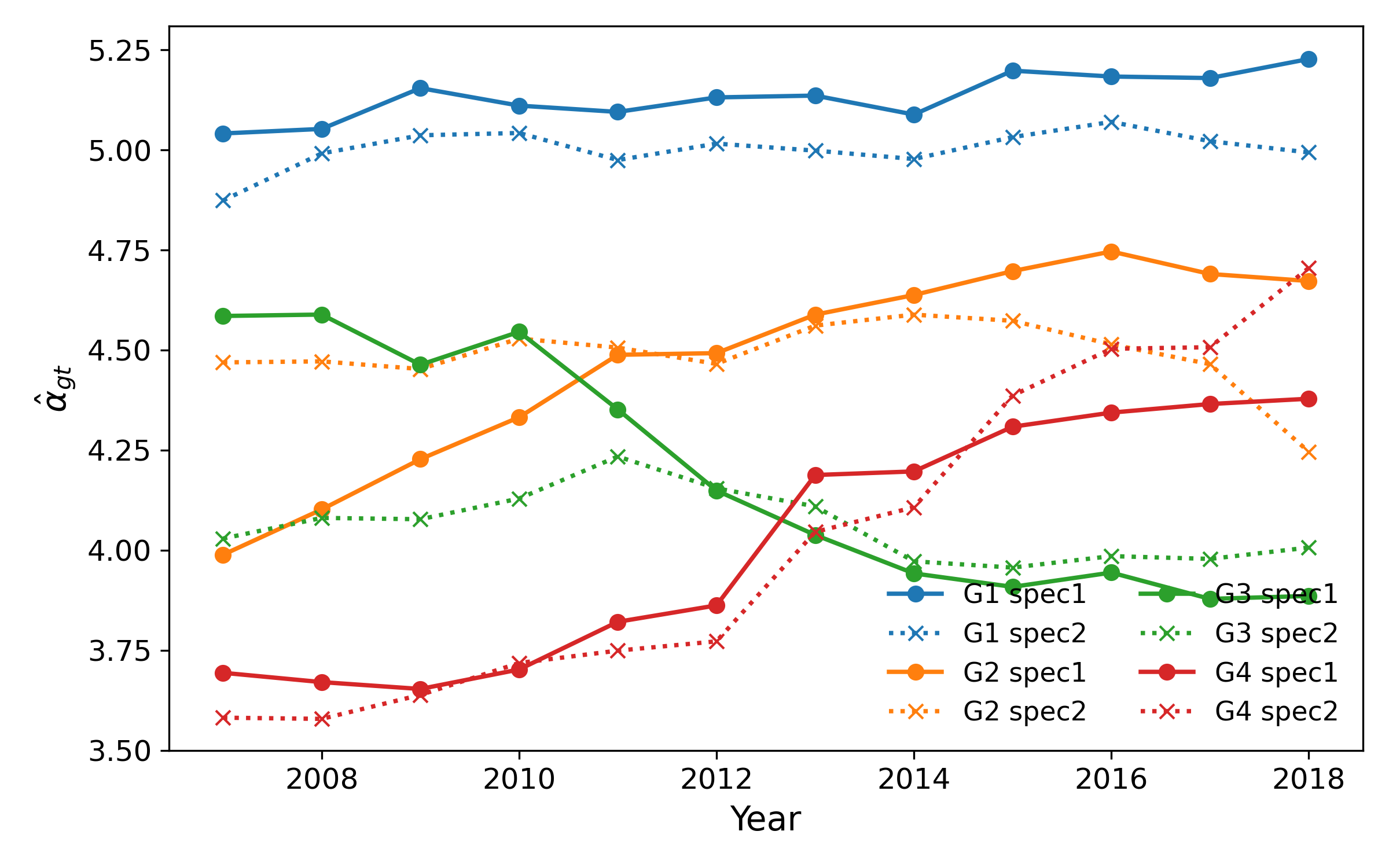}
    \end{center}
    \footnotesize
    {\textit{Data:} Estimated group-year intercept paths for G=4 using training data of households stayed in the survey for at least five years. \\\textit{Note:} Colors indicate groups. Solid lines are specification 1 and dotted lines are specification 2. $\hat{\alpha}_{gt}$ is the estimated group$\times$year intercept from the GFE model (net of covariates). Interpretation should focuses on relative differences and dynamics across groups rather than the value. }
\end{figure}

Figure~\ref{fig_r2:Fig_alpha_paths_G4_overlay} reproduces the Figure~\ref{fig:Fig_alpha_paths_G4_overlay} exercise using the sample with household surveyed for at least five years, plotting the estimated group-year intercepts $\hat{\alpha}_{gt}$ for $G=4$ under Specification 1 and 2. Relative to the sample we used in our main results, the two specifications produce noticeably less aligned group patterns and more divergent $\hat{\alpha}_{gt}$ trajectories (solid vs.\ dotted lines), especially for Group 2 and Group 3. A plausible explanation is that the $\geq 5$-wave requirement sharply reduces effective sample size and thins support within some group--year cells. With fewer observations per $(g,t)$, the estimated group--time effects are less precisely pinned down and the decomposition of welfare dynamics into the covariate component $X_{it}\hat{\theta}$ versus the fixed-effect component $\hat{\alpha}_{gt}$ becomes more sensitive to the covariate set. In addition, conditioning on long survey participation may select a non-random subset of households (e.g., households that are less likely to move), changing the composition of the sample and potentially the nature of latent heterogeneity. 

In conclusion, this robustness exercise suggests that while the four-group structure remains stable, inference about fine differences in group trajectories is less stable when the rotating panel is restricted to a very small set of long-duration households. More broadly, the results highlight a practical trade-off faced by researchers working with rotating panels: restricting to households observed for many years increases the within-household information, but it can sharply reduce the effective sample size and thin support in group-year cells, making the estimated trajectories more sensitive to model specification choices and sampling noise. In our ENAHO application, this trade-off leans more towards sample size, and maintaining broader coverage with shorter household participation yields more stable and comparable group patterns than focusing on the smaller subset of long-duration households.

\subsection{Medium-run Poverty Prediction}
\label{sec:r1_atleast4yrs}

As a robustness check, we restrict the sample to households surveyed for at least four years and reserve the final two observed years for a validation. We therefore estimate the GFE model using the earlier observed years. This provides us an opportunity to demonstrate how well the model is in predicting medium-run outcomes based on the alpha trajectories since we emphasize earlier that one of the advantages of the GFE model is its ability to fill all the missing years information. This restriction gives us a sample of 35,693 household-year observation from 7,987 unique households.


\input{GFE/tables/RobustnessCheck_atleast4yrs_holdout2yrs/holdout2_classification_accuracy}

Table~\ref{tab:holdout2_classification_accuracy} reports weighted poverty classification results using households observed for at least four years. The final two observed years are reserved for testing, and the GFE model is estimated using only earlier observed years. We can tell that poverty classification accuracy remains high throughout the held-out period for both specifications. Averaging across all eligible held-out observations, the weighted classification accuracy is 82.89\% for Specification 1 and 82.79\% for Specification 2. Year-specific accuracy generally lies between about 79 and 86 percent. These prediction accuracy suggest that the model captures medium-run poverty reasonably well.



\section{Discussion and Implications}
\label{sec:conclusion}

This paper implements a GFE approach to measuring poverty dynamics in rotating panel surveys, where households are observed for only a few years but there is partial overlap across waves. Using Peru's ENAHO rotating panel as an application, we show that the GFE framework can leverage short-panel overlap to recover interpretable latent welfare trajectories and to construct poverty transition measures over a long period of time. The model classifies households into a number of latent types and estimates group--year welfare trajectories that summarize how these types evolve over time, after accounting for observed covariates and province fixed effects.

We show that empirically, the estimated groups corresponding to meaningful baseline differences in living conditions: higher-ranked groups have higher baseline expenditure, higher education level, and better access to water and electricity, while lower-ranked groups are disadvantaged in multiple margins. The estimated group--year trajectories show distinct welfare paths over the periods of 2007--2018, and these patterns are robust in two different covariate specifications we have. In addition, the model-implied poverty rates are closely aligned with the true poverty rates in the test data, and our one-step-ahead validation shows that GFE can reproduce observed two-year transition shares in out-of-sample end years very nicely. 

Our findings show the advantage of applying GFE to rotating panels which contains more information about welfare and poverty dynamics than repeated cross-section data. An important advantage of GFE is that it provides us two outputs that are both useful for applied work: (i) a summary of heterogeneous welfare dynamics through a number of group trajectories, and (ii) a very practical group-year parameter for constructing completed welfare paths in years when households are not observed, conditional on researchers makes transparent assumptions about how covariates evolve overtime. This can be very helpful when the researchers want to understand and describe medium to long run poverty persistence and mobility and to relate these patterns to policy changes. 

One caveat is important for the interpretation. The completed welfare paths rely on assumptions about missing covariates in non-survey years. In our application, we use a simple completion rule (forward/backward filling for time-invariant characteristics and deterministic updating for age), but other reasonable choices are possible, and they can change the level and slope of the completed welfare trajectories. We therefore recommend that researchers treat covariate completion as part of the empirical design and assess robustness to alternative imputation rules whenever completed panels are used for substantive conclusions.

A second practical lesson concerns sample design in rotating panels. Our robustness checks show that removing the household head's age restriction leaves the optimal number of latent group selection and preferred group structure conclusion unchanged, consistent with the idea that richer group-year support stabilizes $\hat{\alpha}_{gt}$ and reduces sensitivity to the covariate set. By contrast, restricting samples to households observed for at least 5 years sharply reduces the effective sample size and yields greater divergence in estimated trajectories across specifications, even though the RMSE-based choice of optimal $G$ remains the same. This divergence may not be an issue for most rotating panel designs, where the same unit is tracked for up to 2 years (e.g., the Current Population Survey (CPS) and the Vietnam Household Living Standards Survey (VHLSS)). However, for a rotating panel design where the same units are tracked more than two rounds, our robustness check suggests that maintaining coverage with shorter household participation can be more valuable for having stable trajectory inference than for focusing on the small subset of long-duration households.

While our implementation follows the discrete GFE framework of BM2015, extensions that allow for richer forms of heterogeneity may be valuable in some settings, but the discrete grouping approach has the practical benefit of transparency and interpretability, which is central for describing poverty dynamics in applied work.

Looking forward, our findings suggest that GFE can serve as a practical bridge between short rotating panels and the long-run welfare dynamics that researchers and policymakers often cares about. By combining interpretable latent types with a transparent completion strategy to fill all the missing years dynamics, the approach can produce transition predictions that align closely with observed mobility patterns. Our results motivate the use of GFE model for studying poverty dynamics when true long panels are unavailable, and they point the value of developing diagnostics and model-selection criteria that explicitly account for cell support, imputation assumptions, and out-of-sample performance.


\pagebreak


\clearpage	
\bibliographystyle{apalike} 
\bibliography{Peru}


\newpage 

\begin{appendices}

\begin{center}
\section*{\LARGE Appendix} 
\end{center}

\thispagestyle{empty}
\pagenumbering{arabic}
\renewcommand*{\thepage}{\arabic{page}}
\setcounter{equation}{0}
\renewcommand\theequation{A\arabic{equation}}
\setcounter{table}{0}
\renewcommand{\thetable}{A\arabic{table}}
\setcounter{figure}{0}
\renewcommand{\thefigure}{A\arabic{figure}}
\renewcommand{\thesubsection}{Appendix \Alph{subsection}}

\renewcommand{\thefigure}{A\arabic{figure}} \setcounter{figure}{0}

\section{Pseudocode}
\label{appendix:pseudocode}
\begin{algorithm}[!htbp]
\scriptsize
\RemoveAlgoNumber 
\caption{GFE Estimation via VNS + Local Search}
\label{alg:gfe_vns}
\DontPrintSemicolon
\SetKwInOut{Input}{Input}\SetKwInOut{Output}{Output}
\Input{$\{y_{it},x_{it},d_{it}\}$, number of groups $G$ to explore, tuning: \texttt{n\_starts}, \texttt{neighmax}, \texttt{itermax}, \texttt{max\_local\_iters}}
\Output{$(\hat{\theta},\hat{\alpha},\hat{\gamma},\hat{\mu})$}

\BlankLine

\Fn{$Q(\theta,\alpha,\gamma,\mu)$}{\Return $\sum_{i,t} d_{it}(y_{it}-x'_{it}\theta-\alpha_{g_i t}- \mu_{p_i})^2 $}

\Fn{Assign$(\theta,\alpha,\mu)$}{For each $i$: $g_i\leftarrow \arg\min_g \sum_t d_{it}(y_{it}-x'_{it}\theta-\alpha_{gt}-\mu_{p_i})^2$;\; \Return $\gamma$}

\Fn{OLSUpdate$(\gamma)$}{OLS of $y_{it}$ on $x_{it}$ and group$\times$time dummies and province fixed effects using $d_{it}=1$;\; \Return $(\theta,\alpha,\mu)$}

\Fn{LocalSearch$(\gamma,\theta,\alpha,\mu,\texttt{max\_local\_iters})$}{
    \tcp*[l]{Greedy single-unit moves holding $(\theta,\alpha,\mu)$ fixed}
    perform at most \texttt{max\_local\_iters} greedy reassignment passes, stopping earlier if no improving move exists\;
    \Return $\gamma$\;
}

\BlankLine
$Q^\ast \leftarrow +\infty$\;

\For{$r=1$ \KwTo \texttt{n\_starts}}{

    \tcp*[l]{Step 1: Initialization}
    $\theta^{(0)},\mu^{(0)} \leftarrow$ pooled OLS using $d_{it}=1$\;
    $\alpha^{(0)}_{t} \leftarrow \overline{y_{i t}-x_{i t}^{\prime} \theta^{(0)}-\mu_{p(i)}^{(0)}}$ (means over $d_{it}=1$)\;
    $\alpha^{(0)}_{gt}\leftarrow \alpha^{(0)}_{t}$ for all $g,t$\;

    $\gamma \leftarrow$ Assign$(\theta^{(0)},\alpha^{(0)},\mu^{(0)})$\;
    $(\theta,\alpha,\mu)\leftarrow$ OLSUpdate$(\gamma)$\;

    $\gamma^\star\leftarrow \gamma$;\;
    $(\theta^\star,\alpha^\star,\mu^\star)\leftarrow (\theta,\alpha,\mu)$;\;
    $Q_r^\star\leftarrow Q(\theta^\star,\alpha^\star,\gamma^\star,\mu^\star)$\;

    \tcp*[l]{Outer VNS cycles}
    \For{$j=1$ \KwTo \texttt{itermax}}{

        \tcp*[l]{Step 2: Start from neighborhood size $n=1$ each cycle}
        $n\leftarrow 1$\;

        \While{$n \le \texttt{neighmax}$}{

            \tcp*[l]{Step 3: Neighborhood jump ("shake") from current best $\gamma^\star$}
            Select random set $\mathcal{I}_n$ with $|\mathcal{I}_n|=n$\;
            $\gamma' \leftarrow \gamma^\star$\;
            Reassign each $i\in\mathcal{I}_n$ to a uniformly random group in $\{1,\ldots,G\}$\;

            \tcp*[l]{Step 4: Refinement after the neighborhood jump}
            $(\theta',\alpha',\mu')\leftarrow$ OLSUpdate$(\gamma')$\;
            
            \tcp*[l]{Step 5: Greedy local search over single-unit moves}
            $\gamma''\leftarrow$ LocalSearch$(\gamma',\theta',\alpha',\mu',\texttt{max\_local\_iters})$\;
            $(\theta'',\alpha'',\mu'')\leftarrow$ OLSUpdate$(\gamma'')$\;
            $Q''\leftarrow Q(\theta'',\alpha'',\gamma'',\mu'')$\;

            \tcp*[l]{Step 6: Acceptance rule}
            \eIf{$Q'' < Q_r^\star$}{
                $\gamma^\star\leftarrow \gamma''$;\;
                $(\theta^\star,\alpha^\star,\mu^\star)\leftarrow (\theta'',\alpha'',\mu'')$;\;
                $Q_r^\star\leftarrow Q''$\;
                $n\leftarrow 1$\; \tcp*[l]{restart neighborhood search around improved solution}
            }{
                $n\leftarrow n+1$\; \tcp*[l]{Step 7: increase neighborhood size}
            }
        }
    }

    \tcp*[l]{Keep best start}
    \If{$Q_r^\star < Q^\ast$}{
        $(\hat{\theta},\hat{\alpha},\hat{\gamma},\hat{\mu})\leftarrow (\theta^\star,\alpha^\star,\gamma^\star,\mu^\star)$;\;
        $Q^\ast\leftarrow Q_r^\star$\;
    }
}

\Return $(\hat{\theta},\hat{\alpha},\hat{\gamma},\hat{\mu})$\;
\end{algorithm}

\section{Tables}
\label{appendix:tables}

\begin{landscape}
\input{GFE/tables/main/Table_Rotating_Panel_List}
\end{landscape}

\clearpage

\input{GFE/tables/main/GFE_reg_results_departmentFE}
\input{GFE/tables/main/GFE_reg_results_districtFE}

\input{GFE/tables/main/Table_main_model_selection_BIC_RMSE}

\section{Figures}
\label{appendix:figures}

\begin{figure}[H]
    \begin{center}
        \caption{The trend of poverty status by year - Unrestricted full sample}
        \label{fig:GFE_poverty_trend_by_year_allsample}
        \includegraphics[width=0.95\linewidth]{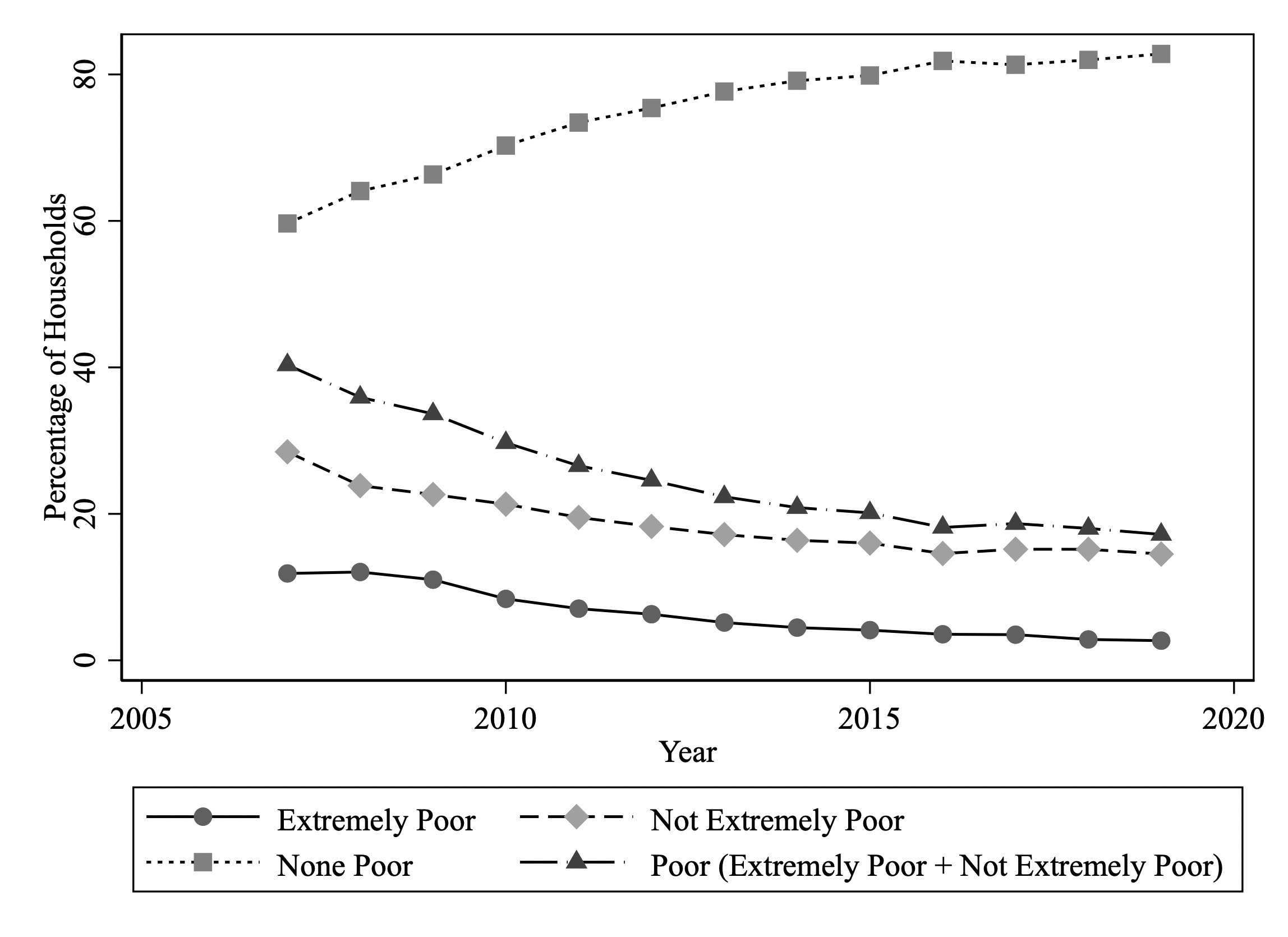}
    \end{center}
    \footnotesize
    {\textit{Source:} Author's calculation using full ENAHO dataset. \\ 
    \textit{Note:} This figure present the trend of poverty status using the original sample without years and age restriction.} \\
\end{figure}

\begin{figure}[!htbp]
    \begin{center}
        \caption{Actual vs. predicted poverty rate over time in all available data (G=4)}
        \label{fig:Fig_poverty_actual_vs_pred_df_all_two_panels_spec1_spec2_t}
        \includegraphics[width=\linewidth]{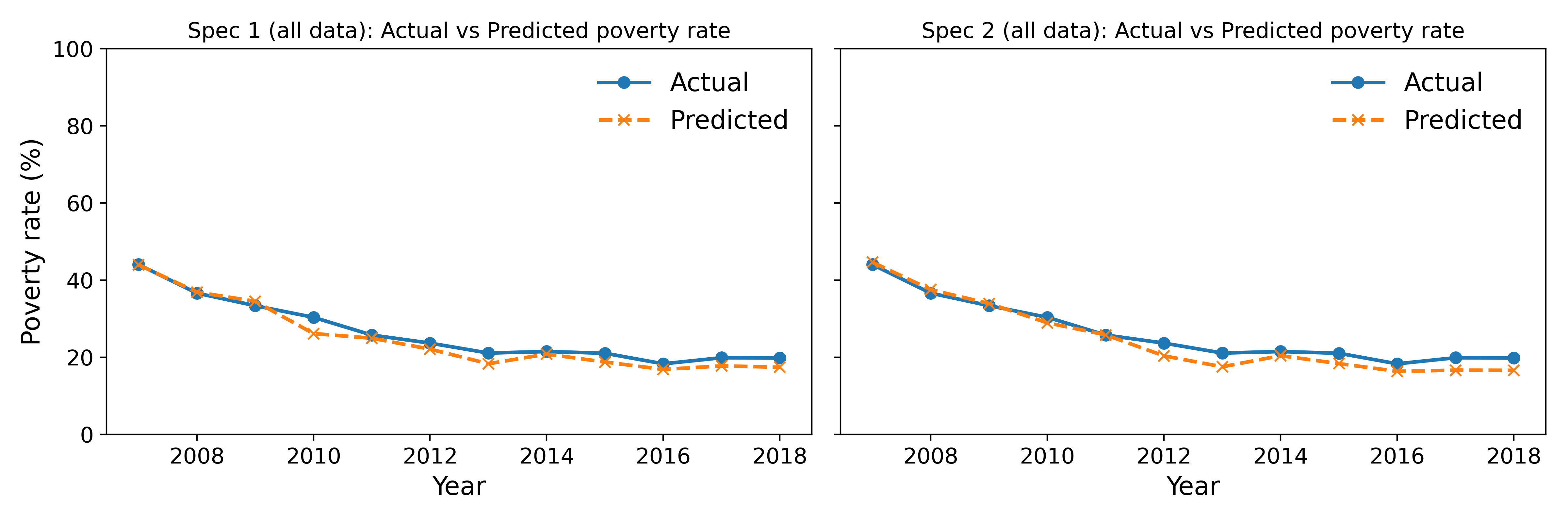}
    \end{center}
    \footnotesize
    {\textit{Data:} All available selected sample (training + test). \\ \textit{Note:} Poverty rates are survey-weighted (ENAHO expansion weights). This figure is prepared using all the available data (training + test). Predicted poverty is based on model-predicted welfare from the GFE model and the IHS poverty line (both in IHS scale).}
\end{figure}

\begin{figure}[!htbp]
    \begin{center}
        \caption{Two-year poverty transition shares over time: actual vs.\ GFE-predicted (all observed household--year pairs, $G=4$), for specification 1.}
        \label{fig:Fig_pov_transitions_actual_vs_pred_spec1_G4}
        \includegraphics[width=\linewidth]{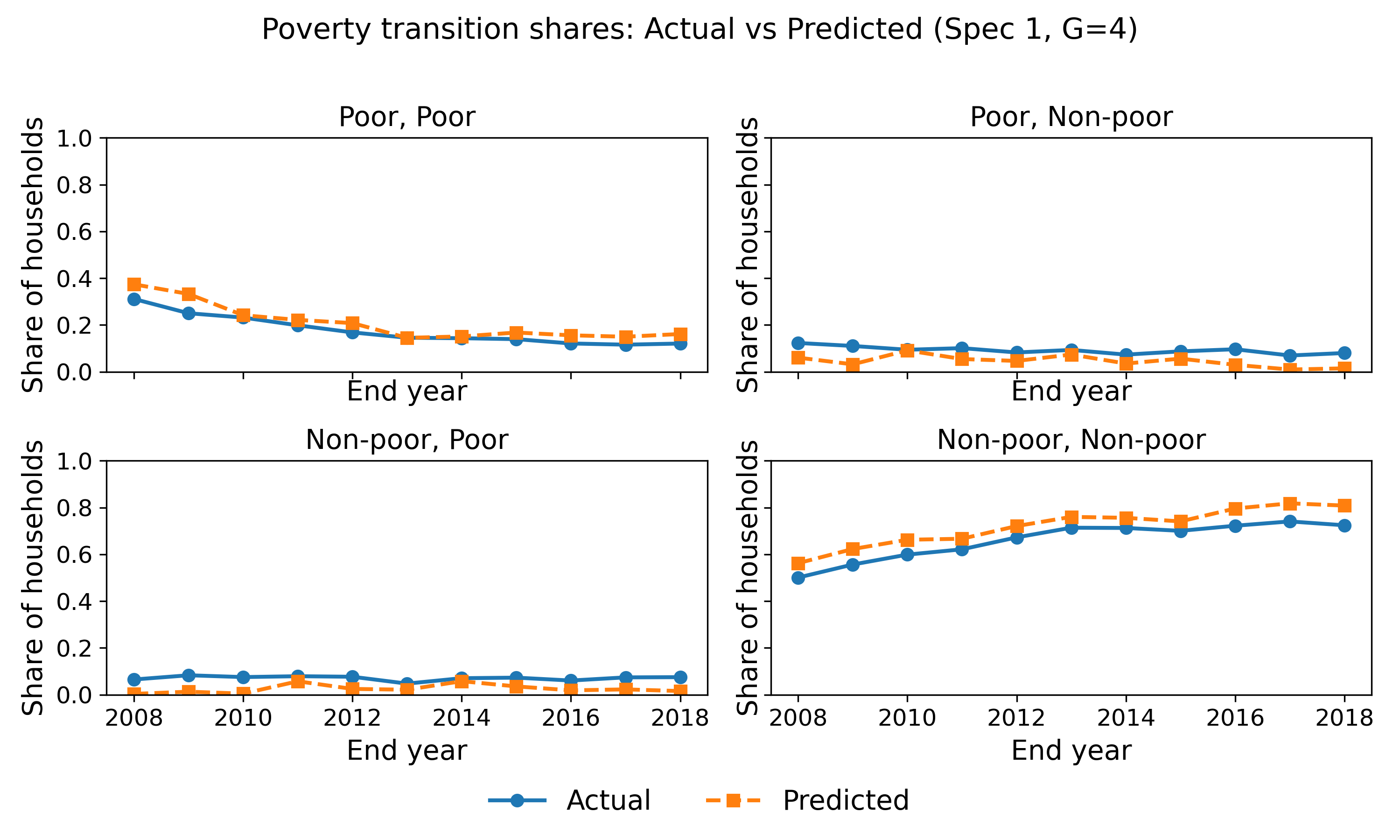}
    \end{center}
    \footnotesize
    {\textit{Data:} All available selected sample (training + test). \\ \textit{Note:} Each panel plots the survey-weighted share of households in one of four poverty transition states between consecutive survey years $(t-1,t)$: Poor$\rightarrow$Poor, Poor$\rightarrow$Non-poor, Non-poor$\rightarrow$Poor, and Non-poor$\rightarrow$Non-poor. ``Actual'' transitions are computed from observed (IHS-transformed) per-capita expenditure relative to the (IHS-transformed) total poverty line. ``Predicted'' transitions replace observed expenditure with GFE-predicted welfare $\hat y_{it}=x'_{it}\hat\theta+\hat\alpha_{\hat g_i,t}+\hat\mu_{p_i}$ and classify predicted poverty as $1\{\hat y_{it}<z_{it}\}$. Only households observed in both years of a given pair are included; transition shares are weighted using ENAHO expansion weights (measured in the end year $t$). The final year-pair is omitted when group--year intercepts are unavailable (e.g., no $\hat\alpha_{g,2019}$).}
\end{figure}

\begin{figure}[!htbp]
    \begin{center}
        \caption{Two-year poverty transition shares over time: actual vs.\ GFE-predicted (all observed household--year pairs, $G=4$), for specification 2.}
        \label{fig:Fig_pov_transitions_actual_vs_pred_spec2_G4}
        \includegraphics[width=\linewidth]{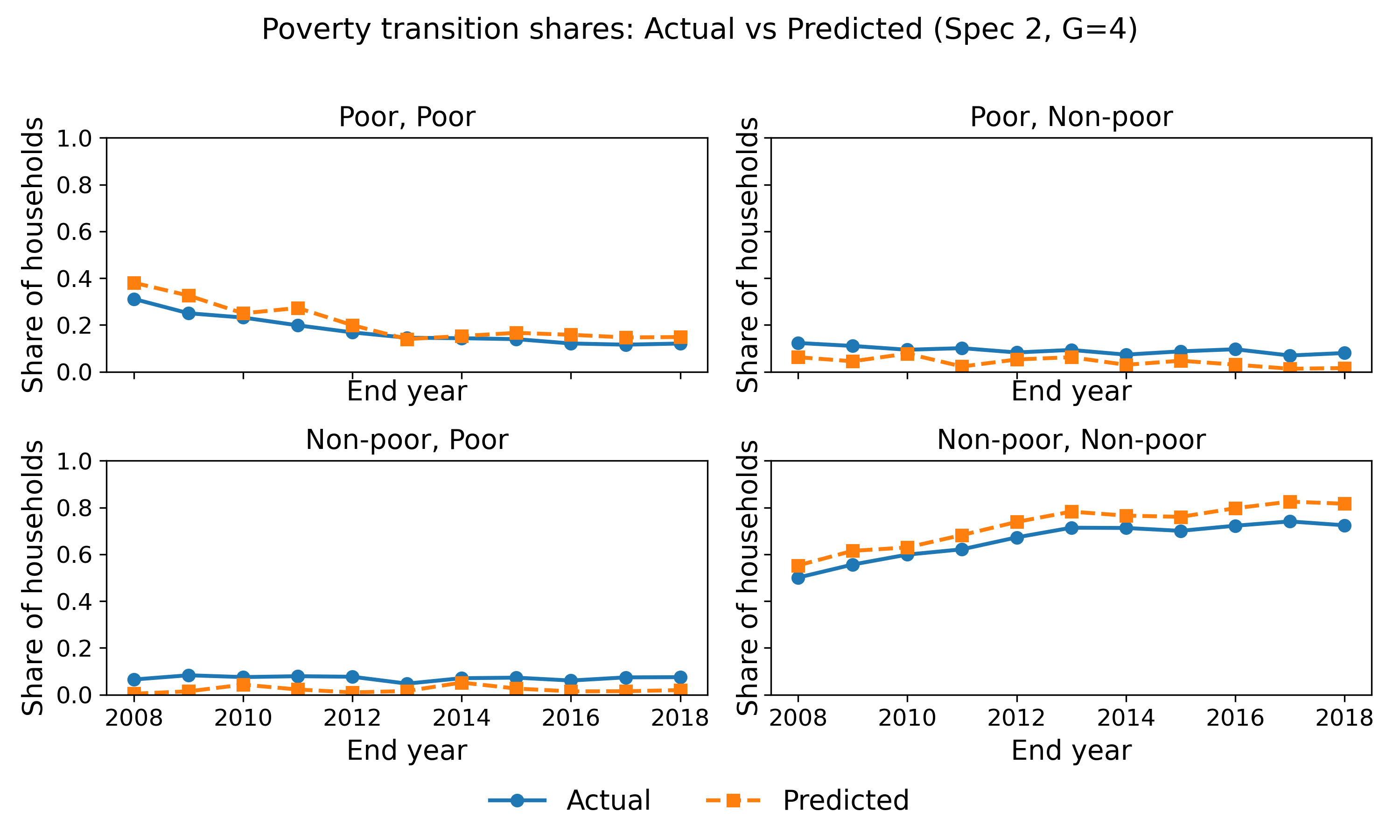}
    \end{center}
    \footnotesize
    {\textit{Data:} All available selected sample (training + test). \\ \textit{Note:} Each panel plots the survey-weighted share of households in one of four poverty transition states between consecutive survey years $(t-1,t)$: Poor$\rightarrow$Poor, Poor$\rightarrow$Non-poor, Non-poor$\rightarrow$Poor, and Non-poor$\rightarrow$Non-poor. ``Actual'' transitions are computed from observed (IHS-transformed) per-capita expenditure relative to the (IHS-transformed) total poverty line. ``Predicted'' transitions replace observed expenditure with GFE-predicted welfare $\hat y_{it}=x'_{it}\hat\theta+\hat\alpha_{\hat g_i,t}+\hat\mu_{p_i}$ and classify predicted poverty as $1\{\hat y_{it}<z_{it}\}$. Only households observed in both years of a given pair are included; transition shares are weighted using ENAHO expansion weights (measured in the end year $t$). The final year-pair is omitted when group--year intercepts are unavailable (e.g., no $\hat\alpha_{g,2019}$).}
\end{figure}

\begin{figure}[!htbp]
    \begin{center}
        \caption{One-step-ahead poverty transition shares into held-out final observations: actual vs.\ GFE-predicted ($G=4$), for specification 2.}
        \label{fig:Fig_pov_transitions_onestep_spec2_G4}
        \includegraphics[width=\linewidth]{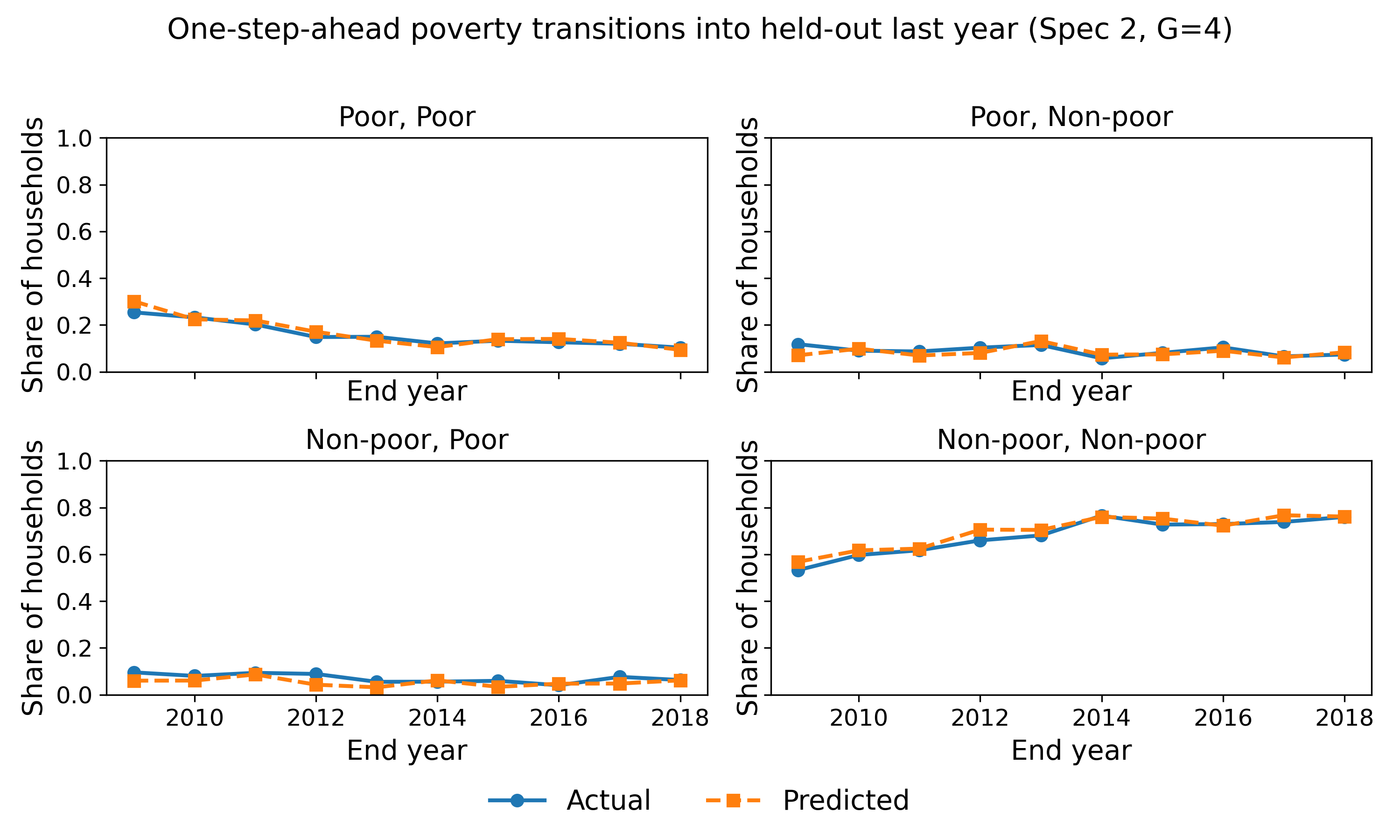}
    \end{center}
    \footnotesize
    {\textit{Data:} Last year of training data + all test data. \\ \textit{Note:} The figure reports survey-weighted shares of households transitioning between poverty states from $t-1$ to $t$, where $t$ is each household's held-out last observed year and $t-1$ is its preceding observed year (restricted to consecutive years; one transition per household). ``Actual'' transitions are computed using observed (IHS-transformed) per-capita expenditure relative to the (IHS-transformed) poverty line in both years. ``GFE-predicted'' transitions use actual poverty at $t-1$ and predicted poverty at $t$, where predicted poverty is $1\{\hat y_{it}<z_{it}\}$ based on model-predicted welfare $\hat y_{it}=x'_{it}\hat\theta+\hat\alpha_{\hat g_i,t}+\hat\mu_{p_i}$. Transition shares are weighted using ENAHO expansion weights in year $t$. In this one-step-ahead sample, the expansion-weighted misclassification rate of $\widehat{\mathrm{poor}}_{it}$ in year $t$ is 17.6\%.}
\end{figure}

\begin{figure}[!htbp]
    \begin{center}
        \caption{Poverty transition shares: actual vs. synthetic panel predictions, for specification 2}
        \label{fig:Fig_pov_transitions_synthpanel_actual_vs_pred_spec2}
        \includegraphics[width=\linewidth]{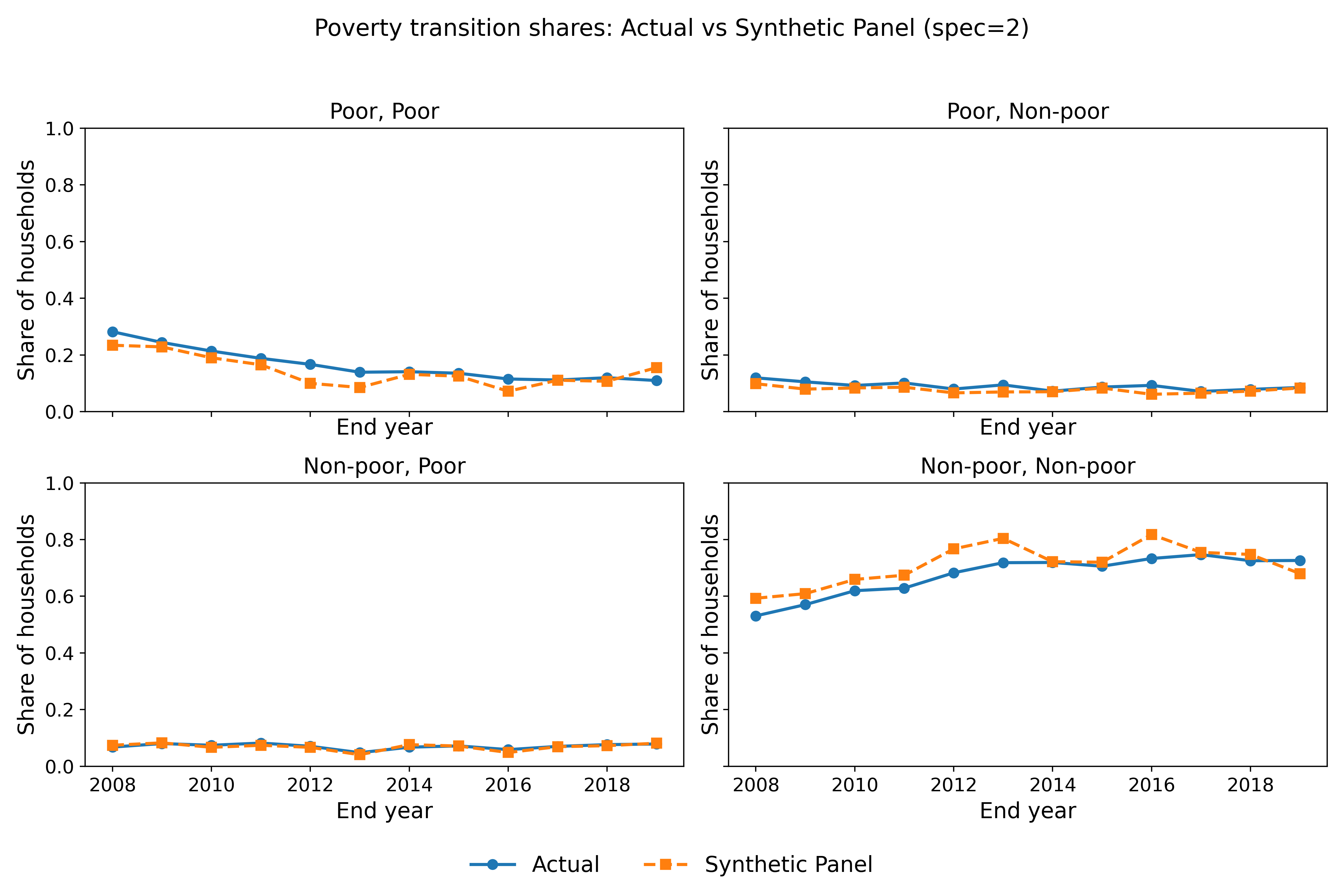}
    \end{center}
    \footnotesize
    {\textit{Note:} Each panel plots the share of households in one of four poverty transition states between $t-1$ and $t$: poor--poor, poor--non-poor, non-poor--poor, and non-poor--non-poor. The solid line (``Actual'') reports transitions computed directly from observed survey data using per-capita expenditure relative to the poverty line in both years. The dashed line reports the corresponding transition shares predicted by the synthetic panel method, which is estimated using repeated cross-sections rather than true panel links. The horizontal axis uses the end year $t$ for each two-year pair. Because of the set up and use of repeated cross-sections rather than true panel, the actual and prediction poverty transition can go up to $t = 2019$. The sample only use household head between 25-55 years old as well, the same as our sample.}
\end{figure}

\begin{figure}[!htbp]
    \begin{center}
        \caption{Prediction error in poverty transition shares: GFE one-step versus synthetic panel.}
        \label{fig:Fig_transition_fit_compare_RMSE_yearly_spec2}
        \includegraphics[width=\linewidth]{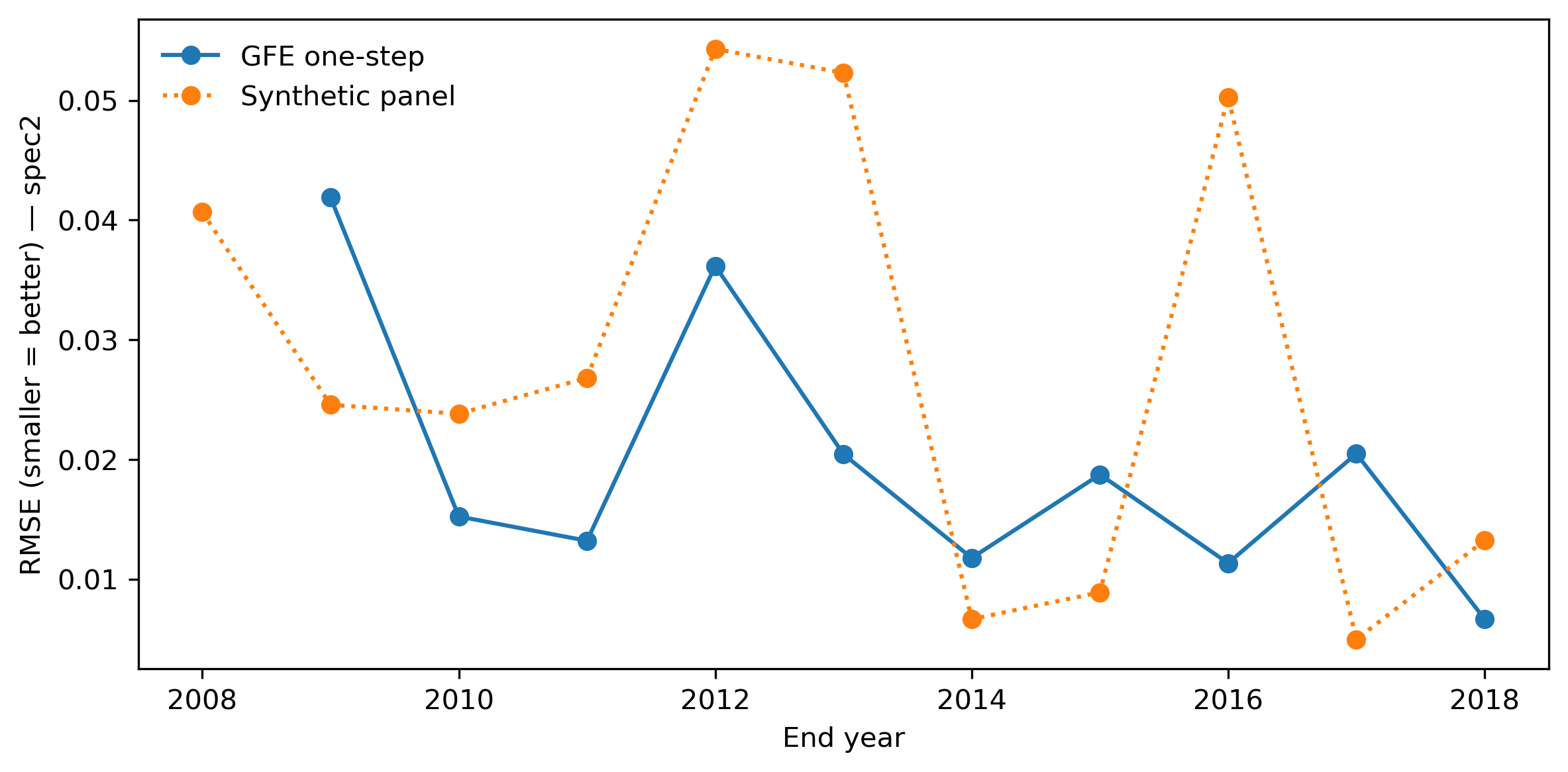}
    \end{center}
    \footnotesize
    {\textit{Note:} The figure plots the root mean squared error (RMSE) between predicted and observed two-year poverty transition shares for each end year $t$. For a given end year, the RMSE is computed across the four transition states (poor--poor, poor--non-poor, non-poor--poor, non-poor--non-poor), comparing the predicted transition-share vector to the observed transition-share vector; smaller values indicate closer fit. The solid line shows the GFE one-step approach, which uses observed poverty status in $t\!-\!1$ and predicts poverty status in $t$. The dotted line shows the synthetic panel approach, which produces transition-share predictions from repeated cross-sections. Because the underlying samples used to compute the transition shares may differ across approaches, the figure is intended as a descriptive comparison of fit within each method. Additionally, the GFE one-step RMSE begins in 2009 instead of 2008 because the one-step-ahead validation requires observing each household in both $t-1$ and its held-out last observed year $t$. Under our analysis sample restriction (at least three observed survey years per household), no household has $t=2008$ as its last observed year.}
\end{figure}

\begin{figure}[!htbp]
\caption{Prediction error in poverty transition shares: GFE one-step versus synthetic panel using data without age restriction}
\label{fig_r1:Fig_transition_fit_compare_RMSE_yearly}
 \begin{subfigure}{\textwidth}
     \caption{Specification 1}
     \includegraphics[width=\textwidth]{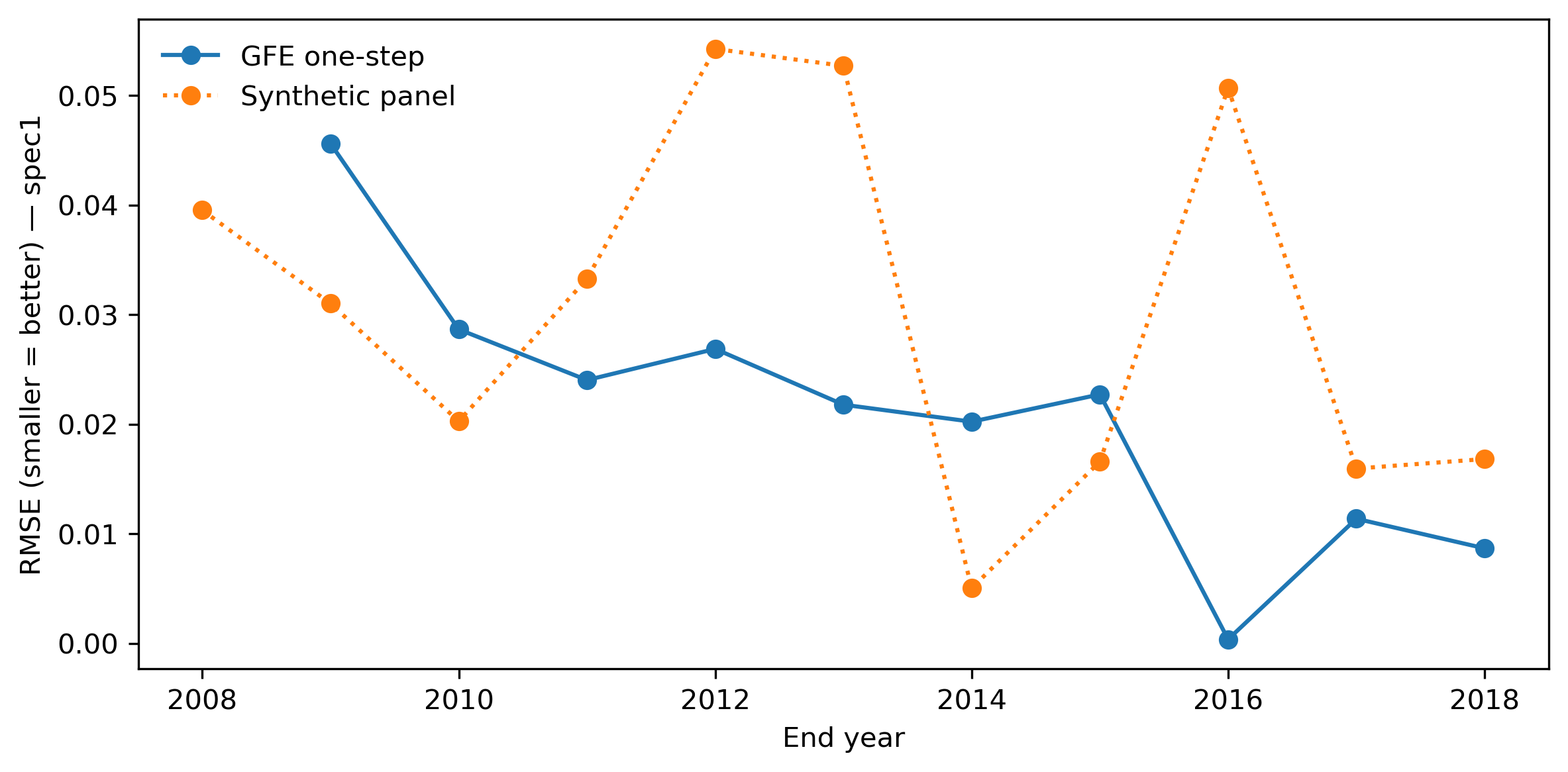}
     \label{fig_r1:Fig_transition_fit_compare_RMSE_yearly_spec1}
 \end{subfigure}
 \hfill
 \begin{subfigure}{\textwidth}
      \caption{Specification 2}
     \includegraphics[width=\textwidth]{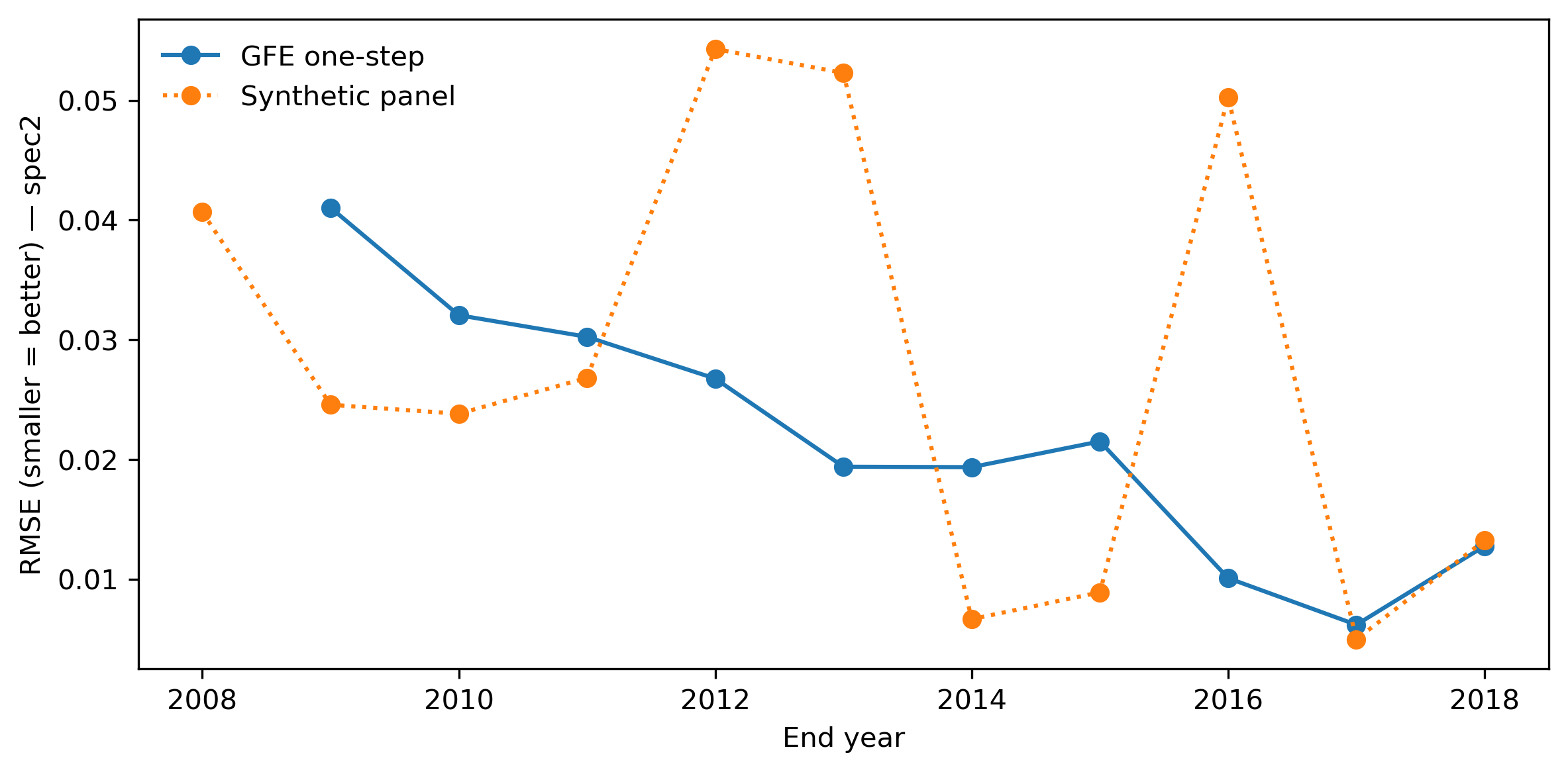}
     \label{fig_r1:Fig_transition_fit_compare_RMSE_yearly_spec2}
 \end{subfigure}
    \footnotesize
    {\textit{Note:} The figures plot the root mean squared error (RMSE) between predicted and observed two-year poverty transition shares for each end year $t$. For a given end year, the RMSE is computed across the four transition states (poor--poor, poor--non-poor, non-poor--poor, non-poor--non-poor), comparing the predicted transition-share vector to the observed transition-share vector; smaller values indicate closer fit. The solid line shows the GFE one-step approach, which uses observed poverty status in $t\!-\!1$ and predicts poverty status in $t$. The dotted line shows the synthetic panel approach, which produces transition-share predictions from repeated cross-sections. Because the underlying samples used to compute the transition shares may differ across approaches, the figures are intended as a descriptive comparison of fit within each method. Additionally, the GFE one-step RMSE begins in 2009 instead of 2008 because the one-step-ahead validation requires observing each household in both $t-1$ and its held-out last observed year $t$. Under our analysis sample restriction (at least three observed survey years per household), no household has $t=2008$ as its last observed year.}
\end{figure}

\end{appendices}

\end{document}

%% file: GFE/tables/main/GFE_sumstats.tex
\begin{table}[!htbp]
\centering
\footnotesize
\caption{Summary Statistics (Weighted)}
\label{tab:GFE.sumstats}
\renewcommand{\arraystretch}{0.95}

\begin{tabular}{p{9cm}
    >{\centering\arraybackslash}p{1.7cm}
    >{\centering\arraybackslash}p{1.7cm}}
\toprule
 & Mean & SD  \\
\midrule
\multicolumn{3}{l}{\textbf{Panel A: Household-year level (N=56,390)}} \\
\hspace{0.4cm} Total expenditures per capita (monthly) - 2007 dollars &  444.92 &  418.24 \\
\hspace{0.4cm} Total poverty line - 2007 dollars &  235.42 &   48.44 \\
\hspace{0.4cm} Food poverty line - 2007 dollars &  124.73 &   18.81 \\
\hspace{0.4cm} IHS (Total expenditures per capita (monthly) - 2007 dollars) &    6.53 &    0.70 \\
\hspace{0.4cm} IHS (Total Poverty Line - 2007 dollars) &    6.13 &    0.20  \\
\hspace{0.4cm} IHS (Food Poverty Line - 2007 dollars) &    5.51 &    0.15  \\
\addlinespace

\multicolumn{3}{l}{\textbf{Panel B: Household-level (first observed year) (N=14,886)}} \\
\multicolumn{3}{l}{\hspace{0.2cm}\textit{Head of household}} \\
\hspace{0.4cm} Female &    0.21 &    0.41 \\
\hspace{0.4cm} Age &   41.22 &    7.60  \\
\hspace{0.4cm} Primary school &    0.30 &    0.46 \\
\hspace{0.4cm} Secondary school &    0.38 &    0.48  \\
\hspace{0.4cm} College &    0.18 &    0.38  \\
\hspace{0.4cm} Language - Spanish &    0.75 &    0.43 \\
\hspace{0.4cm} Married or cohabiting &    0.77 &    0.42  \\
\addlinespace

\multicolumn{3}{l}{\hspace{0.2cm}\textit{Household}} \\
\hspace{0.4cm} Urban &    0.76 &    0.43  \\
\hspace{0.4cm} Has water &    0.74 &    0.44 \\
\hspace{0.4cm} Has electricity &    0.89 &    0.31  \\
\bottomrule
\end{tabular}

\vspace{0.6em}
\begin{minipage}{0.95\textwidth}
\footnotesize
\textit{Source:} Authors’ calculations using ENAHO 2007–2019, restricted to households observed in at least three survey years and to household heads aged 25–55 in all observed years. \\
\textit{Note:} The table reports weighted means, standard deviations, and counts. The sample includes households observed at least three times over 2007-2019, with head age 25--55 in each survey year. Panel A uses household-year observations; Panel B uses the household’s first observed year. All the summary statistics are weighted by the Annual Expansion Factor CPV-2007 projections from the ENAHO data.
\end{minipage}
\end{table}

%% file: GFE/tables/main/GFE_reg_results_provinceFE.tex
\begin{table}[!htbp]\centering\footnotesize
\caption{Household Total Expenditure Per Capita Estimates}
\label{tab:GFE.reg.results.provinceFE}
\begin{tabularx}{0.85\textwidth}{X>{\centering\arraybackslash}p{2cm}>{\centering\arraybackslash}p{2cm}}
\toprule
                    &\multicolumn{1}{c}{(1)}         &\multicolumn{1}{c}{(2)}         \\
\midrule
Head of Household - Female&      0.1248\sym{***}&     -0.1075\sym{***}\\
                    &      0.0120         &      0.0164         \\
Head of Household - Age&     -0.0241\sym{***}&     -0.0168\sym{***}\\
                    &      0.0066         &      0.0060         \\
Head of Household - Age squared&      0.0004\sym{***}&      0.0003\sym{***}\\
                    &      0.0001         &      0.0001         \\
Head of Household - Primary school&      0.1547\sym{***}&      0.1143\sym{***}\\
                    &      0.0114         &      0.0108         \\
Head of Household - Secondary school&      0.3849\sym{***}&      0.2907\sym{***}\\
                    &      0.0134         &      0.0130         \\
Head of Household - College&      0.8049\sym{***}&      0.6792\sym{***}\\
                    &      0.0178         &      0.0173         \\
Language - Spanish  &      0.1582\sym{***}&      0.1258\sym{***}\\
                    &      0.0127         &      0.0121         \\
Head of Household - Married or Cohabitant&                     &     -0.3029\sym{***}\\
                    &                     &      0.0160         \\
Urban               &                     &      0.2340\sym{***}\\
                    &                     &      0.0121         \\
Has water           &                     &      0.1325\sym{***}\\
                    &                     &      0.0102         \\
Has electricity     &                     &      0.1656\sym{***}\\
                    &                     &      0.0150         \\
\midrule
R-squared           &       0.492         &       0.538         \\
N                   &       56390         &       56390         \\
\bottomrule
\end{tabularx}
\vspace{0.5em}
\begin{minipage}{0.85\textwidth}\footnotesize
\textit{Source:} Authors’ calculations using ENAHO 2007–2019, restricted to households observed in at least three survey years and to household heads aged 25–55 in all observed years. \\                                         \textit{Note:} ***, **, * indicate significance at the 1\%, 5\%, and 10\% levels. The dependent variable is monthly household total expenditures per capita (2007 dollars), inverse hyperbolic sine transformed. The sample includes households observed at least three times and with head age 25--55 in each survey year, 2007--2019. All regressions include province and year fixed effects and are weighted by the annual expansion factor (CPV-2007 projections). Standard errors (in parentheses) are clustered at the household level.
\end{minipage}
\end{table}

%% file: GFE/tables/main/Table_transition_fit_compare.tex
\begin{table}
\centering
\caption{Fit of predicted vs. observed poverty transition shares (lower values indicate closer match).}
\label{tab:transition_fit_compare}
\begin{tabular}{lrrrrr}
\toprule
        dataset &  spec & MAE\_avg & RMSE\_avg & TV\_avg & TV\_max \\
\midrule
   GFE one-step &     1 &   0.019 &    0.021 &  0.038 &  0.088 \\
   GFE one-step &     2 &   0.018 &    0.020 &  0.036 &  0.083 \\
Synthetic panel &     1 &   0.026 &    0.031 &  0.051 &  0.090 \\
Synthetic panel &     2 &   0.023 &    0.028 &  0.046 &  0.086 \\
\bottomrule
\end{tabular}
\begin{minipage}{0.95\linewidth}\footnotesize \emph{Notes:} The table compares predicted and observed two-year poverty transition shares across the four transition states (poor--poor, poor--non-poor, non-poor--poor, and non-poor--non-poor). MAE and RMSE are averaged across the four states within each end year. TV denotes the total variation distance, $TV_t=\tfrac{1}{2}\sum_s|\hat p_{s,t}-p_{s,t}|$, computed for each end year $t$ and then averaged (TV\_avg) or maximized (TV\_max) over years. ``GFE one-step'' uses observed poverty status in $t\!-\!1$ and predicts poverty in $t$, while ``Synthetic panel'' uses synthetic panel point estimates (Dang et al., 2014). Lower values indicate closer fit.\end{minipage}
\end{table}

%% file: GFE/tables/main/Table_g4_baseline_spec1.tex
\begin{table}[!htbp]
\centering
\footnotesize
\caption{Group size and baseline characteristics (Specification 1, $G=4$).}
\label{tab:Table_g4_baseline_spec1}
\begin{tabular}{lcccccccccc}
\toprule
{} & Type & Households & Pop.\ share & IHS exp.\ pc & Female head & Age & Primary edu & Secondary edu & College+ & Spanish \\
Group & (by $\bar\alpha_g$) & $N$ & (\%) & (baseline) & (\%) & (years) & (\%) & (\%) & (\%) & (\%) \\
\midrule
1     &        Top &      2,614 &      10.58 &        7.32 &        20.7 &  41.43 &        17.5 &          32.8 &     38.6 &    84.9 \\
2     &  Lower-mid &      3,380 &      29.87 &        6.58 &        15.4 &  40.79 &        31.8 &          38.2 &     16.1 &    76.0 \\
3     &  Upper-mid &      5,056 &      36.87 &        6.09 &        19.1 &  40.72 &        34.9 &          33.0 &     12.5 &    70.7 \\
4     &     Bottom &      3,836 &      22.68 &        5.61 &        16.5 &  40.43 &        33.7 &          36.3 &      9.7 &    67.9 \\
\bottomrule
\end{tabular}
\begin{flushleft}
\footnotesize\emph{Notes:} Baseline covariates are measured in each household’s first observed survey year. All entries are weighted using ENAHO expansion factors. Binary variables are reported as percentages. Groups are ordered by mean estimated group effect ($\bar\alpha_g$), from Top to Bottom. Population share is calculated by summing the sample weight for each G and dividing by the sum of weights for the entire sample.
\end{flushleft}
\end{table}

%% file: GFE/tables/main/Table_g4_baseline_spec2.tex
\begin{table}[!htbp]
\centering
\footnotesize
\caption{Group size and baseline characteristics (Specification 2, $G=4$).}
\label{tab:Table_g4_baseline_spec2}
\begin{tabular}{lcccccccccccccc}
\toprule
{} & Type & Households & Pop.\ share & IHS exp.\ pc & Female head & Age & Primary edu & Secondary edu & College+ & Spanish & Married & Urban & Water & Elec \\
Group & (by $\bar\alpha_g$) & $N$ & (\%) & (baseline) & (\%) & (years) & (\%) & (\%) & (\%) & (\%) & (\%) & (\%) & (\%) & (\%) \\
\midrule
1     &        Top &      2,569 &      13.48 &        7.15 &        17.9 &  41.20 &        21.9 &          32.3 &     33.7 &    85.5 &    71.4 &  72.0 &  74.6 &  85.0 \\
2     &  Lower-mid &      3,722 &      29.72 &        6.53 &        14.9 &  40.88 &        30.2 &          37.6 &     16.7 &    73.6 &    82.5 &  71.8 &  68.6 &  81.4 \\
3     &  Upper-mid &      5,039 &      33.08 &        6.04 &        18.3 &  40.28 &        36.5 &          33.9 &     11.5 &    71.0 &    80.1 &  70.4 &  62.8 &  78.4 \\
4     &     Bottom &      3,556 &      23.73 &        5.71 &        19.7 &  40.99 &        33.2 &          35.8 &     10.1 &    68.7 &    77.4 &  71.5 &  58.0 &  78.7 \\
\bottomrule
\end{tabular}
\begin{flushleft}
\footnotesize\emph{Notes:} Baseline covariates are measured in each household’s first observed survey year. All entries are weighted using ENAHO expansion factors. Binary variables are reported as percentages. Groups are ordered by mean estimated group effect ($\bar\alpha_g$), from Top to Bottom. Population share is calculated by summing the sample weight for each G and dividing by the sum of weights for the entire sample.
\end{flushleft}
\end{table}

%% file: GFE/tables/RobustnessCheck_NoAgeLimit/Table_transition_fit_compare.tex
\begin{table}
\centering
\caption{Fit of predicted vs. observed poverty transition shares using data without age restriction (lower values indicate closer match).}
\label{tab_r1:transition_fit_compare}
\begin{tabular}{lrrrrr}
\toprule
        dataset &  spec & $MAE\_avg$ & $RMSE\_avg$ & $TV\_avg$ & $TV_max$ \\
\midrule
   GFE one-step &     1 &   0.019 &    0.021 &  0.038 &  0.090 \\
   GFE one-step &     2 &   0.020 &    0.022 &  0.040 &  0.081 \\
Synthetic panel &     1 &   0.026 &    0.031 &  0.051 &  0.090 \\
Synthetic panel &     2 &   0.023 &    0.028 &  0.046 &  0.086 \\
\bottomrule
\end{tabular}
\begin{minipage}{0.95\linewidth}\footnotesize \emph{Notes:} The table compares predicted and observed two-year poverty transition shares across the four transition states (poor--poor, poor--non-poor, non-poor--poor, and non-poor--non-poor). MAE and RMSE are averaged across the four states within each end year. TV denotes the total variation distance, $TV_t=\tfrac{1}{2}\sum_s|\hat p_{s,t}-p_{s,t}|$, computed for each end year $t$ and then averaged (TV\_avg) or maximized (TV\_max) over years. ``GFE one-step'' uses observed poverty status in $t\!-\!1$ and predicts poverty in $t$, while ``Synthetic panel'' uses synthetic panel point estimates (Dang et al., 2014). Lower values indicate closer fit.\end{minipage}
\end{table}

%% file: GFE/tables/RobustnessCheck_NoAgeLimit/Table_g4_baseline_spec1.tex
\begin{table}[!htbp]
\centering
\footnotesize
\caption{Group size and baseline characteristics using sample without age restriction (Specification 1, $G=4$).}
\label{tab_r1:Table_g4_baseline_spec1}
\begin{tabular}{lcccccccccc}
\toprule
{} & Type & Households & Pop.\ share & IHS exp.\ pc & Female head & Age & Primary edu & Secondary edu & College+ & Spanish \\
Group & (by $\bar\alpha_g$) & $N$ & (\%) & (baseline) & (\%) & (years) & (\%) & (\%) & (\%) & (\%) \\
\midrule
1     &        Top &      4,495 &      13.03 &        7.34 &        27.5 &  53.14 &        24.4 &          25.6 &     29.7 &    83.3 \\
2     &  Upper-mid &      9,662 &      32.99 &        6.62 &        21.8 &  51.25 &        30.2 &          29.5 &     15.8 &    75.6 \\
3     &  Lower-mid &      7,257 &      34.60 &        6.07 &        21.9 &  49.02 &        32.2 &          27.6 &      8.9 &    66.3 \\
4     &     Bottom &      5,620 &      19.39 &        5.51 &        22.7 &  49.82 &        29.7 &          25.8 &      8.2 &    61.9 \\
\bottomrule
\end{tabular}
\begin{flushleft}
\footnotesize\emph{Notes:} Baseline covariates are measured in each household’s first observed survey year. All entries are weighted using ENAHO expansion factors. Binary variables are reported as percentages. Groups are ordered by mean estimated group effect ($\bar\alpha_g$), from Top to Bottom. Population share is calculated by summing the sample weight for each G and dividing by the sum of weights for the entire sample.
\end{flushleft}
\end{table}

%% file: GFE/tables/RobustnessCheck_NoAgeLimit/Table_g4_baseline_spec2.tex
\begin{table}[!htbp]
\centering
\footnotesize
\caption{Group size and baseline characteristics using sample without age restriction (Specification 2, $G=4$).}
\label{tab_r1:Table_g4_baseline_spec2}
\begin{tabular}{lcccccccccccccc}
\toprule
{} & Type & Households & Pop.\ share & IHS exp.\ pc & Female head & Age & Primary edu & Secondary edu & College+ & Spanish & Married & Urban & Water & Elec \\
Group & (by $\bar\alpha_g$) & $N$ & (\%) & (baseline) & (\%) & (years) & (\%) & (\%) & (\%) & (\%) & (\%) & (\%) & (\%) & (\%) \\
\midrule
1     &        Top &      4,547 &      12.98 &        7.30 &        24.0 &  53.10 &        22.2 &          26.1 &     30.5 &    83.9 &    66.8 &  72.6 &  78.3 &  84.1 \\
2     &  Upper-mid &      9,761 &      32.29 &        6.57 &        21.8 &  50.58 &        31.5 &          28.3 &     15.4 &    74.2 &    73.2 &  71.8 &  72.9 &  82.3 \\
3     &  Lower-mid &      7,469 &      35.50 &        6.09 &        21.1 &  49.26 &        31.1 &          28.1 &      9.0 &    68.1 &    75.0 &  70.4 &  65.4 &  79.2 \\
4     &     Bottom &      5,257 &      19.22 &        5.60 &        26.5 &  50.61 &        30.8 &          26.5 &      8.4 &    60.7 &    66.3 &  68.9 &  57.7 &  74.3 \\
\bottomrule
\end{tabular}
\begin{flushleft}
\footnotesize\emph{Notes:} Baseline covariates are measured in each household’s first observed survey year. All entries are weighted using ENAHO expansion factors. Binary variables are reported as percentages. Groups are ordered by mean estimated group effect ($\bar\alpha_g$), from Top to Bottom. Population share is calculated by summing the sample weight for each G and dividing by the sum of weights for the entire sample.
\end{flushleft}
\end{table}

%% file: GFE/tables/RobustnessCheck_atleast4yrs_holdout2yrs/holdout2_classification_accuracy.tex
\begin{table}[htbp]
\centering
\caption{Poverty classification accuracy in held-out years}
\label{tab:holdout2_classification_accuracy}
\begin{threeparttable}
\footnotesize
\begin{tabular}{llrrrrr}
\toprule
Specification & Year & Households & Obs. & Actual poverty & Predicted poverty & Classification \\ &  &  &  & rate (\%) & rate (\%) & accuracy (\%) \\
\midrule
Spec 1 & 2009 & 856 & 856 & 33.99 & 31.80 & 83.68 \\
 & 2010 & 1,391 & 1,391 & 30.66 & 26.53 & 81.18 \\
 & 2011 & 674 & 674 & 29.34 & 24.39 & 78.60 \\
 & 2012 & 1,054 & 1,054 & 29.10 & 26.03 & 80.46 \\
 & 2013 & 1,790 & 1,790 & 20.26 & 18.02 & 83.91 \\
 & 2014 & 1,531 & 1,531 & 19.94 & 17.92 & 85.42 \\
 & 2015 & 1,062 & 1,062 & 22.86 & 19.29 & 82.68 \\
 & 2016 & 1,234 & 1,234 & 16.17 & 16.48 & 84.63 \\
 & 2017 & 1,101 & 1,101 & 17.39 & 17.72 & 82.61 \\
 & Overall & 5,829 & 10,693 & 23.85 & 21.53 & 82.89 \\
\addlinespace
Spec 2 & 2009 & 856 & 856 & 33.99 & 31.04 & 81.72 \\
 & 2010 & 1,391 & 1,391 & 30.66 & 23.29 & 79.49 \\
 & 2011 & 674 & 674 & 29.34 & 21.33 & 79.40 \\
 & 2012 & 1,054 & 1,054 & 29.10 & 26.48 & 80.38 \\
 & 2013 & 1,790 & 1,790 & 20.26 & 17.88 & 84.25 \\
 & 2014 & 1,531 & 1,531 & 19.94 & 17.10 & 86.05 \\
 & 2015 & 1,062 & 1,062 & 22.86 & 15.53 & 81.74 \\
 & 2016 & 1,234 & 1,234 & 16.17 & 14.06 & 85.72 \\
 & 2017 & 1,101 & 1,101 & 17.39 & 16.51 & 83.63 \\
 & Overall & 5,829 & 10,693 & 23.85 & 19.96 & 82.79 \\
\bottomrule
\end{tabular}
\begin{tablenotes}[flushleft]
\footnotesize
\item \textit{Notes:} For each held-out household-year observation within the supported period of the estimated group-time effects, poverty status is predicted and compared with realized poverty status. Classification accuracy is the weighted share of correctly classified poor/non-poor observations, using ENAHO household survey weights.
\end{tablenotes}
\end{threeparttable}
\end{table}

%% file: GFE/tables/main/Table_Rotating_Panel_List.tex
\begin{table}[htbp]\centering
\def\sym#1{\ifmmode^{#1}\else\(^{#1}\)\fi}
\small
\caption{Rotating Panel Design Surveys}
\begin{threeparttable}
\begin{tabular}{>{\raggedright\arraybackslash}p{2cm}
    >{\raggedright\arraybackslash}p{8cm}
    >{\raggedright\arraybackslash}p{10cm}}
\hline\hline
 Country & Name      &          Rotation Design \\
 \hline
 Argentina & Permanent Household Survey (EPH) & 2-2-2 rotation scheme: Households are interviewed for two consecutive quarters, dropped for the next two, and interviewed again for the following two. \\ 
European Union & EU-SILC (European Union Statistics on Income and Living Conditions) &  Each year, 25\% of the sample is rotated out and replaced by a new sub-sample. \\
Kyrgyz Republic & Kyrgyz Integrated Household Survey (KIHS) &  25\% of the sampled households are replaced each year.\\ 
Mexico & National Survey of Occupation and Employment (ENOE) & 5-quarter rotating panel where 20\% of the sample is replaced each quarter.\\
Peru & National Household Survey (ENAHO) & Each year, 30\% of households are selected as panel households, surveyed from 2 to 5 years. \\
South Africa & Quarterly Labour Force Survey (QLFS) & One-quarter of the sample being refreshed each quarter. \\ 
United States & Current Population Survey (CPS) &  4-8-4:  Households are interviewed for 4 consecutive months, excluded for 8, and re-interviewed for 4 more before being retired. This means 50\% of households are in the survey during the same month one year earlier. \\
Vietnam & Vietnam Household Living Standards Surveys (VHLSS) & 50\% of households are retained, 50\% are newly surveyed \\
\hline\hline
\end{tabular}
\end{threeparttable}
\label{Tab:RP_list}
\end{table}

%% file: GFE/tables/main/GFE_reg_results_departmentFE.tex
\begin{table}[!htbp]\centering\footnotesize
\caption{Household Total Expenditure Per Capita Estimates}
\label{tab:GFE.reg.results.departmentFE}
\begin{tabularx}{0.85\textwidth}{X>{\centering\arraybackslash}p{2cm}>{\centering\arraybackslash}p{2cm}}
\toprule
                    &\multicolumn{1}{c}{(1)}         &\multicolumn{1}{c}{(2)}         \\
\midrule
Head of Household - Female&      0.1542\sym{***}&     -0.1063\sym{***}\\
                    &      0.0123         &      0.0167         \\
Head of Household - Age&     -0.0244\sym{***}&     -0.0175\sym{***}\\
                    &      0.0068         &      0.0061         \\
Head of Household - Age squared&      0.0004\sym{***}&      0.0003\sym{***}\\
                    &      0.0001         &      0.0001         \\
Head of Household - Primary school&      0.1910\sym{***}&      0.1287\sym{***}\\
                    &      0.0125         &      0.0114         \\
Head of Household - Secondary school&      0.4579\sym{***}&      0.3147\sym{***}\\
                    &      0.0142         &      0.0135         \\
Head of Household - College&      0.8820\sym{***}&      0.6995\sym{***}\\
                    &      0.0182         &      0.0177         \\
Language - Spanish  &      0.1911\sym{***}&      0.1368\sym{***}\\
                    &      0.0128         &      0.0119         \\
Head of Household - Married or Cohabitant&                     &     -0.3141\sym{***}\\
                    &                     &      0.0163         \\
Urban               &                     &      0.3018\sym{***}\\
                    &                     &      0.0121         \\
Has water           &                     &      0.1257\sym{***}\\
                    &                     &      0.0105         \\
Has electricity     &                     &      0.1776\sym{***}\\
                    &                     &      0.0158         \\
\midrule
R-squared           &       0.443         &       0.508         \\
N                   &       56390         &       56390         \\
\bottomrule
\end{tabularx}
\vspace{0.5em}
\begin{minipage}{0.85\textwidth}\footnotesize
\textit{Source:} Authors’ calculations using ENAHO 2007–2019, restricted to households observed in at least three survey years and to household heads aged 25–55 in all observed years. \\                                         \textit{Note:} ***, **, * indicate significance at the 1\%, 5\%, and 10\% levels. The dependent variable is monthly household total expenditures per capita (2007 dollars), inverse hyperbolic sine transformed. The sample includes households observed at least three times and with head age 25--55 in each survey year, 2007--2019. All regressions include department (Peru region) and year fixed effects and are weighted by the annual expansion factor (CPV-2007 projections). Standard errors (in parentheses) are clustered at the household level.
\end{minipage}
\end{table}

%% file: GFE/tables/main/GFE_reg_results_districtFE.tex
\begin{table}[!htbp]\centering\footnotesize
\caption{Household Total Expenditure Per Capita Estimates}
\label{tab:GFE.reg.results.districtFE}
\begin{tabularx}{0.85\textwidth}{X>{\centering\arraybackslash}p{2cm}>{\centering\arraybackslash}p{2cm}}
\toprule
                    &\multicolumn{1}{c}{(1)}         &\multicolumn{1}{c}{(2)}         \\
\midrule
Head of Household - Female&      0.0967\sym{***}&     -0.1020\sym{***}\\
                    &      0.0116         &      0.0149         \\
Head of Household - Age&     -0.0179\sym{***}&     -0.0111\sym{*}  \\
                    &      0.0062         &      0.0058         \\
Head of Household - Age squared&      0.0003\sym{***}&      0.0002\sym{***}\\
                    &      0.0001         &      0.0001         \\
Head of Household - Primary school&      0.1202\sym{***}&      0.0977\sym{***}\\
                    &      0.0113         &      0.0107         \\
Head of Household - Secondary school&      0.3106\sym{***}&      0.2582\sym{***}\\
                    &      0.0133         &      0.0128         \\
Head of Household - College&      0.6621\sym{***}&      0.5898\sym{***}\\
                    &      0.0167         &      0.0161         \\
Language - Spanish  &      0.0994\sym{***}&      0.0870\sym{***}\\
                    &      0.0130         &      0.0125         \\
Head of Household - Married or Cohabitant&                     &     -0.2759\sym{***}\\
                    &                     &      0.0140         \\
Urban               &                     &      0.2045\sym{***}\\
                    &                     &      0.0149         \\
Has water           &                     &      0.1321\sym{***}\\
                    &                     &      0.0107         \\
Has electricity     &                     &      0.1638\sym{***}\\
                    &                     &      0.0159         \\
\midrule
R-squared           &       0.567         &       0.595         \\
N                   &       56390         &       56390         \\
\bottomrule
\end{tabularx}
\vspace{0.5em}
\begin{minipage}{0.85\textwidth}\footnotesize
\textit{Source:} Authors’ calculations using ENAHO 2007–2019, restricted to households observed in at least three survey years and to household heads aged 25–55 in all observed years. \\                                         \textit{Note:} ***, **, * indicate significance at the 1\%, 5\%, and 10\% levels. The dependent variable is monthly household total expenditures per capita (2007 dollars), inverse hyperbolic sine transformed. The sample includes households observed at least three times and with head age 25--55 in each survey year, 2007--2019. All regressions include UBIGEO (department--province--district level location) and year fixed effects and are weighted by the annual expansion factor (CPV-2007 projections). Standard errors (in parentheses) are clustered at the household level.
\end{minipage}
\end{table}

%% file: GFE/tables/main/Table_main_model_selection_BIC_RMSE.tex
\begin{table}
\footnotesize
\centering
\caption{BICs and RMSEs by specification and number of random starts}
\label{tab:Table_main_model_selection_BIC_RMSE}
\begin{tabular}{lcccccccccccc}
\toprule
& \multicolumn{5}{c}{n\_start=3} & \multicolumn{5}{c}{n\_start=10} & \multicolumn{2}{c}{Diff (ns10-ns3)} \\
\cmidrule(l{2pt}r{2pt}){2-6}
\cmidrule(l{2pt}r{2pt}){7-11}
\cmidrule(l{2pt}r{2pt}){12-13}
& \multicolumn{2}{c}{spec1} & \multicolumn{2}{c}{spec2} & \multicolumn{1}{c}{s2-s1} & \multicolumn{2}{c}{spec1} & \multicolumn{2}{c}{spec2} & \multicolumn{1}{c}{s2-s1} & \multicolumn{1}{c}{spec1} & \multicolumn{1}{c}{spec2} \\ 
\cmidrule(l{2pt}r{2pt}){2-3} 
\cmidrule(l{2pt}r{2pt}){4-5}
\cmidrule(l{2pt}r{2pt}){6-6}
\cmidrule(l{2pt}r{2pt}){7-8} 
\cmidrule(l{2pt}r{2pt}){9-10}
\cmidrule(l{2pt}r{2pt}){11-11}
\cmidrule(l{2pt}r{2pt}){12-12}
\cmidrule(l{2pt}r{2pt}){13-13}
G & BIC & RMSE & BIC & RMSE & Diff & BIC & RMSE & BIC & RMSE & Diff & RMSE & RMSE \\
& (1) & (2) & (3) & (4) & (4)-(2) & (5) & (6) & (7) & (8) & (8)-(6) & (6)-(2) & (8)-(4) \\
\midrule
\toprule
 1 & 1.764 & 0.503 & 1.590 & 0.478 & -0.025 &       &       &       &       &        &       &        \\
 2 & 0.958 & 0.485 & 0.878 & 0.466 & -0.019 &       &       &       &       &        &       &        \\
 3 & 0.717 & 0.470 & 0.669 & 0.455 & -0.015 &       &       &       &       &        &       &        \\
 4 & 0.584 & 0.416 & 0.548 & 0.408 & -0.008 & 0.584 & 0.416 & 0.548 & 0.408 & -0.008 & 0.000 &  0.000 \\
 5 & 0.529 & 0.449 & 0.495 & 0.440 & -0.009 & 0.521 & 0.453 & 0.491 & 0.438 & -0.015 & 0.004 & -0.002 \\
 6 & 0.464 & 0.427 & 0.444 & 0.425 & -0.002 & 0.464 & 0.427 & 0.439 & 0.426 & -0.001 & 0.000 &  0.001 \\
 7 & 0.430 & 0.448 & 0.411 & 0.436 & -0.011 & 0.430 & 0.448 & 0.408 & 0.448 &  0.000 & 0.000 &  0.012 \\
 8 & 0.402 & 0.449 & 0.385 & 0.457 &  0.008 &       &       &       &       &        &       &        \\
 9 & 0.380 & 0.453 & 0.357 & 0.434 & -0.019 &       &       & 0.356 & 0.425 &        &       & -0.010 \\
10 & 0.363 & 0.441 & 0.339 & 0.438 & -0.004 & 0.359 & 0.447 & 0.337 & 0.431 & -0.016 & 0.006 & -0.006 \\
11 & 0.338 & 0.442 & 0.322 & 0.456 &  0.015 & 0.338 & 0.442 &       &       &        & 0.000 &        \\
12 & 0.328 & 0.464 & 0.304 & 0.442 & -0.021 &       &       &       &       &        &       &        \\
13 & 0.316 & 0.472 & 0.297 & 0.481 &  0.009 &       &       &       &       &        &       &        \\
14 & 0.303 & 0.472 & 0.290 & 0.471 & -0.001 &       &       &       &       &        &       &        \\
15 & 0.295 & 0.518 & 0.277 & 0.496 & -0.022 &       &       &       &       &        &       &        \\
16 & 0.281 & 0.476 & 0.266 & 0.489 &  0.013 &       &       &       &       &        &       &        \\
17 & 0.273 & 0.485 & 0.257 & 0.479 & -0.006 &       &       &       &       &        &       &        \\
18 & 0.264 & 0.474 & 0.249 & 0.475 &  0.001 &       &       &       &       &        &       &        \\
19 & 0.259 & 0.518 & 0.243 & 0.483 & -0.035 &       &       &       &       &        &       &        \\
20 & 0.251 & 0.510 & 0.236 & 0.500 & -0.010 &       &       &       &       &        &       &        \\
21 & 0.246 & 0.506 & 0.231 & 0.509 &  0.003 &       &       &       &       &        &       &        \\
22 & 0.237 & 0.520 & 0.225 & 0.499 & -0.021 &       &       &       &       &        &       &        \\
23 & 0.230 & 0.496 & 0.218 & 0.485 & -0.011 &       &       &       &       &        &       &        \\
24 & 0.226 & 0.505 & 0.215 & 0.520 &  0.015 &       &       &       &       &        &       &        \\
25 & 0.221 & 0.496 & 0.209 & 0.491 & -0.005 &       &       &       &       &        &       &        \\
26 & 0.218 & 0.511 & 0.207 & 0.512 &  0.001 &       &       &       &       &        &       &        \\
27 & 0.211 & 0.504 & 0.201 & 0.513 &  0.010 &       &       &       &       &        &       &        \\
28 & 0.207 & 0.519 & 0.198 & 0.505 & -0.014 &       &       &       &       &        &       &        \\
29 & 0.204 & 0.521 & 0.192 & 0.484 & -0.037 &       &       &       &       &        &       &        \\
30 & 0.198 & 0.508 & 0.189 & 0.508 &  0.000 &       &       &       &       &        &       &        \\
31 & 0.198 & 0.503 & 0.186 & 0.512 &  0.009 &       &       &       &       &        &       &        \\
32 & 0.193 & 0.512 & 0.184 & 0.513 &  0.001 &       &       &       &       &        &       &        \\
33 & 0.189 & 0.509 & 0.179 & 0.495 & -0.014 &       &       &       &       &        &       &        \\
34 & 0.189 & 0.534 & 0.180 & 0.529 & -0.005 &       &       &       &       &        &       &        \\
35 & 0.185 & 0.515 & 0.175 & 0.515 &  0.000 &       &       &       &       &        &       &        \\
36 & 0.183 & 0.538 & 0.173 & 0.514 & -0.023 &       &       &       &       &        &       &        \\
37 & 0.180 & 0.562 & 0.169 & 0.535 & -0.028 &       &       &       &       &        &       &        \\
38 & 0.178 & 0.523 & 0.167 & 0.514 & -0.009 &       &       &       &       &        &       &        \\
39 & 0.175 & 0.529 & 0.166 & 0.521 & -0.008 &       &       &       &       &        &       &        \\
40 & 0.173 & 0.533 & 0.166 & 0.519 & -0.014 &       &       &       &       &        &       &        \\
\bottomrule
\end{tabular}
\par\vspace*{6pt}
\begin{minipage}{\linewidth}
\footnotesize
\textit{Notes:} BIC is computed on the training data. RMSE is computed on the test data. For each specification, we estimate the model over the grid of $G$ with $\texttt{n\_starts}=3$ and rerun selected $G$ values with $\texttt{n\_starts}=10$ to assess sensitivity to initialization. “Diff” columns report spec2--spec1 and $\texttt{n\_starts}=10 - \texttt{n\_starts}=3$ RMSE differences. Blank entries indicate $G$ values not rerun with $\texttt{n\_starts}=10$.
\end{minipage}
\end{table}